\documentclass[a4paper, 11pt]{article}
\usepackage{graphicx}
\usepackage{dcolumn}
\usepackage{bm}
\usepackage[utf8]{inputenc}
\usepackage[T1]{fontenc}
\usepackage[dvipsnames]{xcolor}
\usepackage{amsmath}
\usepackage{subfig}
\usepackage{stfloats}
\usepackage{lipsum}
\let\oldAA\AA
\renewcommand{\AA}{\text{\normalfont\oldAA}}
\usepackage{hyperref}
\usepackage[margin=0.5in]{geometry}
\usepackage[separate-uncertainty = true,multi-part-units = repeat]{siunitx}

\usepackage{soul}
\usepackage{authblk}

\newsavebox\affbox

\author[1*]{L. Klochko*}
\author[1,2]{V. Mandrolko}
\author[1]{G. Castanet}
\author[1]{G. Pernot}
\author[1]{F. Lemoine}
\author[3]{ K. Termentzidis}
\author[1]{D. Lacroix}
\author[1]{and M. Isaiev*}
\affil[1]{Universit\'e de Lorraine, CNRS, LEMTA, 54000 Nancy, France} 
\affil[2]{Faculty of Physics, Taras Shevchenko National University of Kyiv, 64/13, Volodymyrs’ka St., Kyiv 01601,Ukraine}
\affil[3]{Universit\'e de Lyon, CNRS, INSA-Lyon, CETHIL UMR5008, F-69621, Villeurbanne, France}

\date{}    

\usepackage{ragged2e}
\justifying

\begin{document}

\pagenumbering{arabic}
\setcounter{page}{1}

\title{Molecular dynamics simulation of thermal transport \\
across solid/liquid interface created by meniscus}

\maketitle


\begin{abstract}

Understandings heat transfer across a solid/liquid interface is important to develop new pathways to improve thermal management in various energy applications.
One of the important questions that arises in this context is the impact of three-phase contact line between solid, liquid and gas on the perturbations of the heat fluxes at the nanoscale.
Therefore, this paper is devoted to the investigations of features of thermal transport across nanosized meniscus constrained between two solid walls.
Different wetting states of the meniscus were considered with molecular dynamics approach by the variation of the interactional potential between atoms of the substrate and the liquid.
The effect of the size of the meniscus on the exchange of energy  between two solid walls was also invetigated.
It was shown that the presence of  a three phase contact line leads to a decrease of the interfacial boundary resistance between solid and liquid.
Further, investigations with the finite element method were used to link atomistic simulations with the continuum mechanics.
We demonstrate that the wetting angle and the interfacial boundary resistance are the required key-parameters  to perform multiscale simulations of such engineering problems with an accurate microscale parametrization.

\textbf{keywords}: thermal transport,  molecular dynamics, finite element method, interfacial thermal resistance, meniscus
\end{abstract}

\section{Introduction}

Current miniaturization of devices  is often related to different thermal management issues such as overheating, hotspots and thermal interface effects. The latters constitute bottlenecks for improving their stability, reliability, and lifetime. Thus, tuning heat transfer at small scales, have a significant importance for a wide range of applied fields. For instance, thermal transport across solid/liquid interface is crucial for: energy applications, materials elaboration, solidification/melting processes, cooling applications, etc. In addition, such interfaces play an increasingly important role at the nanoscale when the surface-to-volume ratio becomes significant. As some examples of such role, the outstanding enhancement of energy transfer and thermal conductivity in nanofluids \cite{GoncalvesSmall2020, Mahian2021} or in  porous system filled by liquids \cite{Lishchuk2015, Isaiev2020} can be mentioned. Such remarkable properties of the systems cause significant applied interest for theirs application in the field of energy harvesting, dissipation, and storage. However, the morphology features of the contact between two different species separating a heat source and heat sinks result in the presence of interfacial thermal resistance (ITR), which lead to the degradation of the heat dissipation performance~\cite{Kim2021sml}.

Despite the significant applied and fundamental interest of this issue, mechanisms that rule the thermal resistance close to a solid/liquid interface remain unclear. This is particularly the case while considering thermal transport in nanofluids with significant amount of interfaces. In the latter case, physic of transport is still under debates. Specifically, numerous models have been already developed for the description of thermal conductivity enhancement in nanofluids, however each of the model works well only for specific nanofluids and concentration level of nanoparticles~\cite{Ali2018}. 

Going back to the basics, heat transfer without mass flow, at the macroscopic level, can be described in the frame of Fourier's law. In the case of multiphase systems, a temperature jump ($\Delta T$) appears at the interface between the two media. The latter is defined in terms of an ITR,  $R = \Delta T / J$, where  $J$ is the normal heat flux across the interface. 

From a microscopic point of view, the heat transfer from the solid to the liquid depends on the characteristics of the vibrational modes presented in the solid phase. For example,  C. Ulises Gonzalez-Valle and B. Ramos-Alvarado~\cite{GonzalezValle2019} showed that out-of-plane modes (perpendicular to the surface) of the substrate contribute more significantly than in-plane-modes (parallel to the surface) to the thermal transport across the interface. However, the dependence of the ITR with the interaction parameters (wetting angle, penetration length adhesion work) is the genuine difference with the solid/solid ITR (Kapitza resistance) which mostly depends on bulk properties of materials~\cite{Zhang2018, PhysRev.60.354}.

The ITR can significantly affect the thermal transport in systems where there is an important surface-to-volume fraction of liquid. Its value depends on the interactions between atoms of solid and fluid in the vicinity of the interface~\cite{Merabia2009,Frank2018}. In solid/liquid systems, interfacial properties also affect: the wetting angle and the adhesion strength, while both parameters influence the heat transfer. For instance, a well-pronounced dependence of the ITR with the wetting angle was found~\cite{Acharya2011,Shenogina2009}. 
Based on the spectral analysis of the heat flow in silicon-based systems with solid/liquid interface, it was shown that the scaling law of ITR dependence on wetting angle is not unique~\cite{Alosious2019,RamosAlvarado2017} while liquid density depletion layer thickness~\cite{GonzalezValle2019,GonzalezValle2019_1} complies with a more generic behaviour. However, this parameter is not universal for all materials \cite{Gonzalez-Valle2021}, and further investigation is required to go deeper in the understanding of interfacial thermal transport. 

The three-phase contact line is the line separating three different phases for instance solid, liquid and gas. The line often appears in the physical processes like nucleate boiling~\cite{Basavaraja2021}, meniscus characteristics~\cite{Basavaraja2021}, crystallization occurance~\cite{AzimiYancheshme2020} etc. The energy transport close to the three-phase contact line is important for understanding: the heat transfer in phase change materials~\cite{Qiu2020, Jilte2021}, the contact photothermally-induced bubbles grow~\cite{Yan2020}, or the various physical-based theranostic modalities~\cite{Li2021}. It is clear that the efficiency of energy transfer in this case depends on the physical properties of all species as well as interactions between them. However, the features of this interaction is even more poorly understood then the interfacial thermal transport. Eventually, three-phase contact line can also significantly contribute to the thermal transport at the nanoscale, especially where phase change occurs.  As an example, it was previously reported~\cite{KUNKELMANN20121896,Pruthvik_thesis} that the contribution of the contact line on the thermal transport in the nucleation process is crucial. 

Thus, in order to describe this complex behaviour more accurately, one needs to find scalable approaches, as the characteristic size of a nucleation site of the water is in micron range. Thus, physical insight regarding features of thermal transport close to the three-phase contact line is essential for further improvement of the approaches in the energy field. 

In this framework, the main goal of this work is to provide a detailed analysis of the thermal transport close to the triple line. For this, we considered heat transfer across a meniscus connecting two solid walls as a model system. Firstly, molecular dynamics (MD) approach was adopted to mine insight regarding features of energy propagation across the meniscus. The dependence of the wetting angle as well as on the meniscus size was analyzed. In a second step, results of MD simulations were compared with ones performed by finite elements method (FEM) to find appropriate parametrization required to build links between nanoscale and micro/macroscale approaches.

\section{Model and Methods}

\subsection{Simulations details}

\begin{figure*}
 	\centering
 	\includegraphics[width = 3.1in]{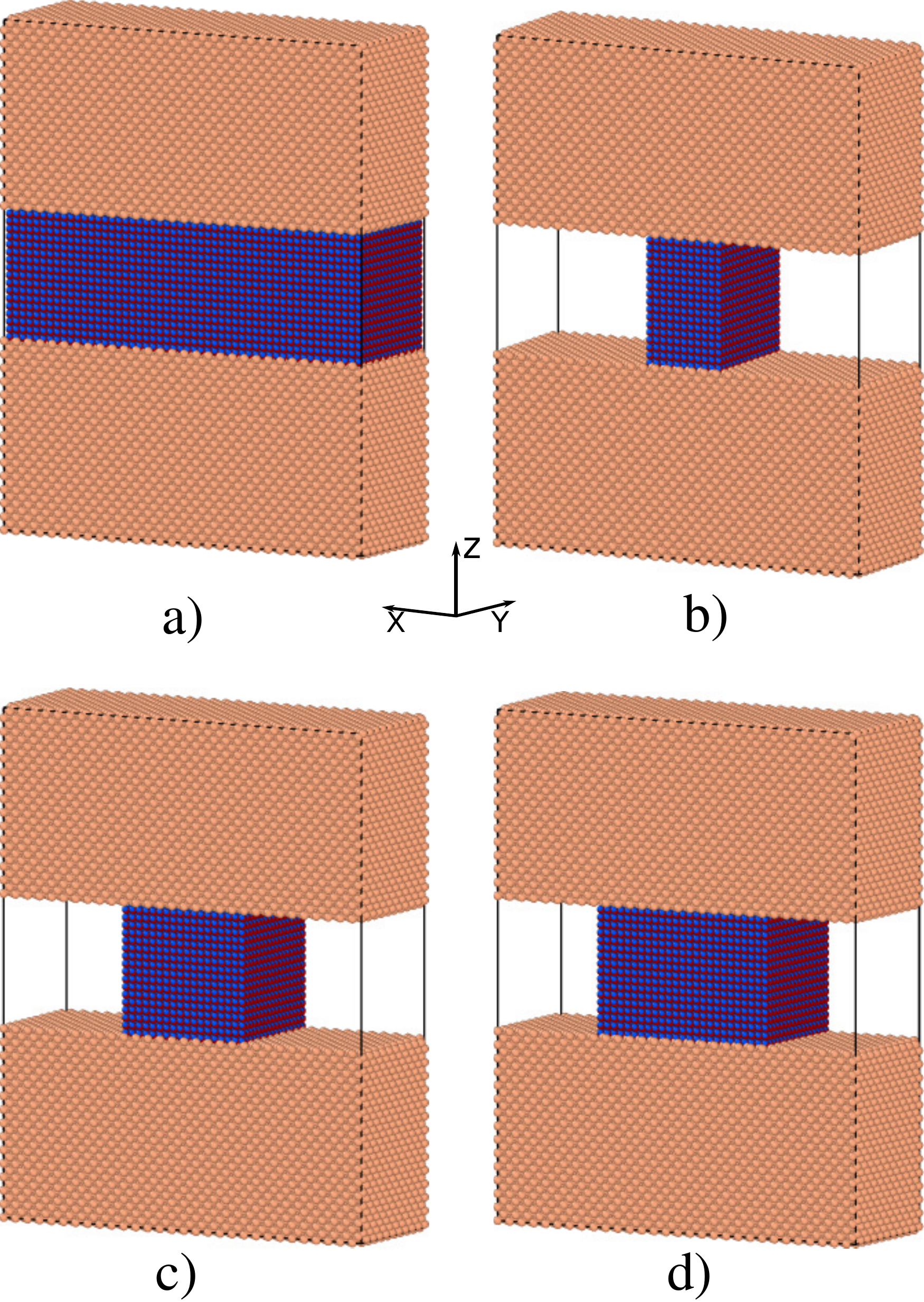}
 	\caption{\label{fig:5} The initial configuration for: a) confined water (case i), b) meniscus $6\text{a}_0$ (case ii), c) meniscus $10\text{a}_0$ (case ii), and d) meniscus $14\text{a}_0$ (case ii). Periodic boundary conditions were applied in all three dimensions.}
 \end{figure*}

To investigate heat transfer in solid/liquid systems, we performed non-equilibrium molecular dynamics (NEMD) simulations with LAMMPS~\cite{PLIMPTON19951,LAMMPS}. Distinct systems were considered. Specifically, the following systems were used: i) \emph{‘‘water layer confined between two Si layers''} (or confined water), where a slab of water totally fills the gap between two solid walls, and ii) \emph{a ‘‘water meniscus between two Si layers}'', where the water domain fills only the middle part of the gap between the two Si layers, (see Fig.~\ref{fig:5}). For the latter we performed simulations for different water drop sizes: $6\text{a}_0$, $10\text{a}_0$, and $14\text{a}_0$, where $\text{a}_0=0.543$ nm is the lattice parameter for Si. The initial systems were equilibrated at $T=300$~K, then a temperature gradient set between $T=270$~K to $T=330$~K was applied in Si regions. More details about the models and simulations are provided in (SI).

The interaction between silicon atoms of the solid subsrate and oxygen atoms of the water was simulated with a Lenard-Jones potential, given as:

\begin{equation}
V(r) = 4 \varepsilon \left(\left(\frac{\sigma}{r}\right)^{12}-\left(\frac{\sigma}{r}\right)^6\right),
\end{equation}
where $r$ is the distance between atoms, $\varepsilon$ is the depth of the potential, $\sigma$ is the distance at which potential is equal to zero. In all our simulations, we take the distance from the Matthiessen's mixing rule: 

\begin{equation}
\sigma = \sigma_{Si-O}=\frac{\sigma_{Si-Si}+\sigma_{O-O}}{2}.
\end{equation}

\begin{figure*}
 	\centering
 	\includegraphics[width = 6in]{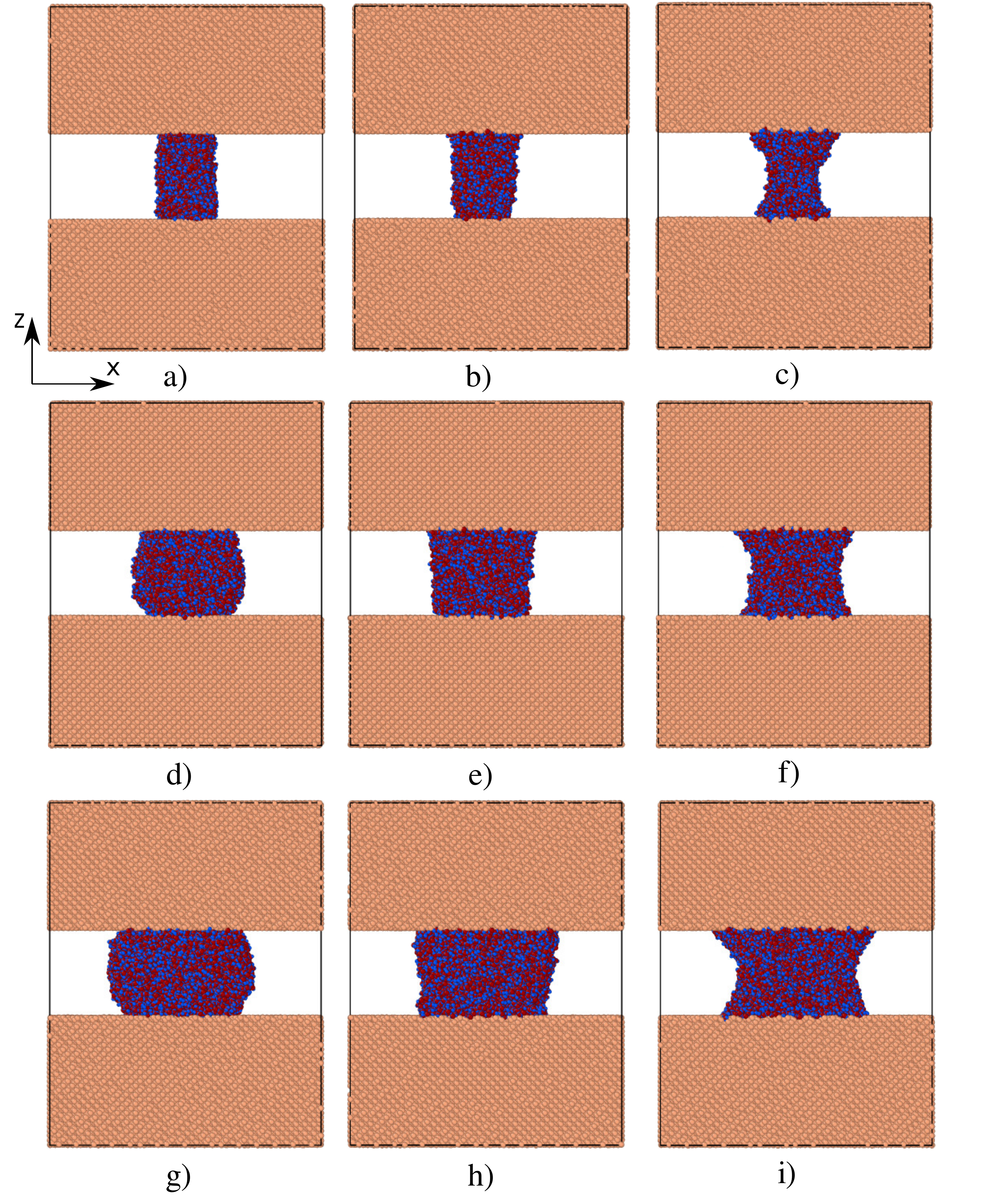}
 	\caption{\label{snaps} The thermalized configuration of water menisci  ($6\text{a}_0$, $10\text{a}_0$, and $14\text{a}_0$ from top to bottom) for different parametrizations: a), d), g) $\varepsilon=10$~meV;  b), e), h) $\varepsilon=16$~meV;  c), f), i)  $\varepsilon=21$~meV. Periodic boundary conditions were applied in all three dimensions.}
 \end{figure*}

The value of the $\varepsilon$ parameter was varied to obtain the different wetting state. The variation range of $\varepsilon$ was chosen between 10 to 21 meV, with step of 1 meV, that corresponds to a continuous variation of the nanoscale wetting angle between 126 to 49$^{\circ}$~\cite{Isaiev2015}. The snapshots of water atoms in the meniscus  for all the studied values of potential depth after procedure of thermalisation are represented in Fig.~\ref{snaps}.  

This range of $\varepsilon$ values allows us to mimic water/substrate behaviour going from hydrophobic to hydrophilic cases respectively. For each considered value of the $\varepsilon$ parameter, we performed five independent NEMD simulations.

\subsection{Heat flux evaluation}

In order to determine the heat flux flowing through the water junction, two independent methods were used. The first method, named \textit{energy balance method} (\textbf{EBM})~\cite{Ikeshoji1994} is based on the added/extracted energy, $\left<E\right>$,  in thermostats (source/sink), and the heat flux was defined as the amount of energy passing through a unit area, $A=L_x\times L_y$, per unit time $t$: 
\begin{equation}
\label{eq:1}
J = \frac{\left<E\right>}{t\times A},
\end{equation}

where $L_x$, $L_y$ are the system sizes of the simulated box. 

The second method called all \textit{atom contributed method} (\textbf{ACM}) for the evaluation of the heat flux makes possible to decompose different contributions of heat transport~\cite{Surblys2019,Boone2019} and to consider more precisely the spatial distributions of heat fluxes:
\begin{equation}
\label{eq:2}
\boldsymbol{J} = \frac{1}{V}\left[\sum_ie_i\boldsymbol{v}_i-\sum_i\hat{S}_i\boldsymbol{v}_i\right],
\end{equation}
where $e_i$, $v_i$, and $\hat{S}_i$ are the per-atom energy (sum of potential and kinetic), velocity and stress tensor respectively, and $V$ is the volume of a bin where the calculations were performed. The first term of Eq.~\ref{eq:2} is called convective term and relies to heat transfer due to matter displacement. The second term, called virial, corresponds to heat transfer due to stress mechanisms.
Knowing the heat flux as well as temperature variations is mandatory to evaluate ITR as it will shown hereafter.

\section{Results and discussions}

\subsection{Meniscus morphology}

\begin{figure*}[t!]
	\centering
	\includegraphics[width =7.7in]{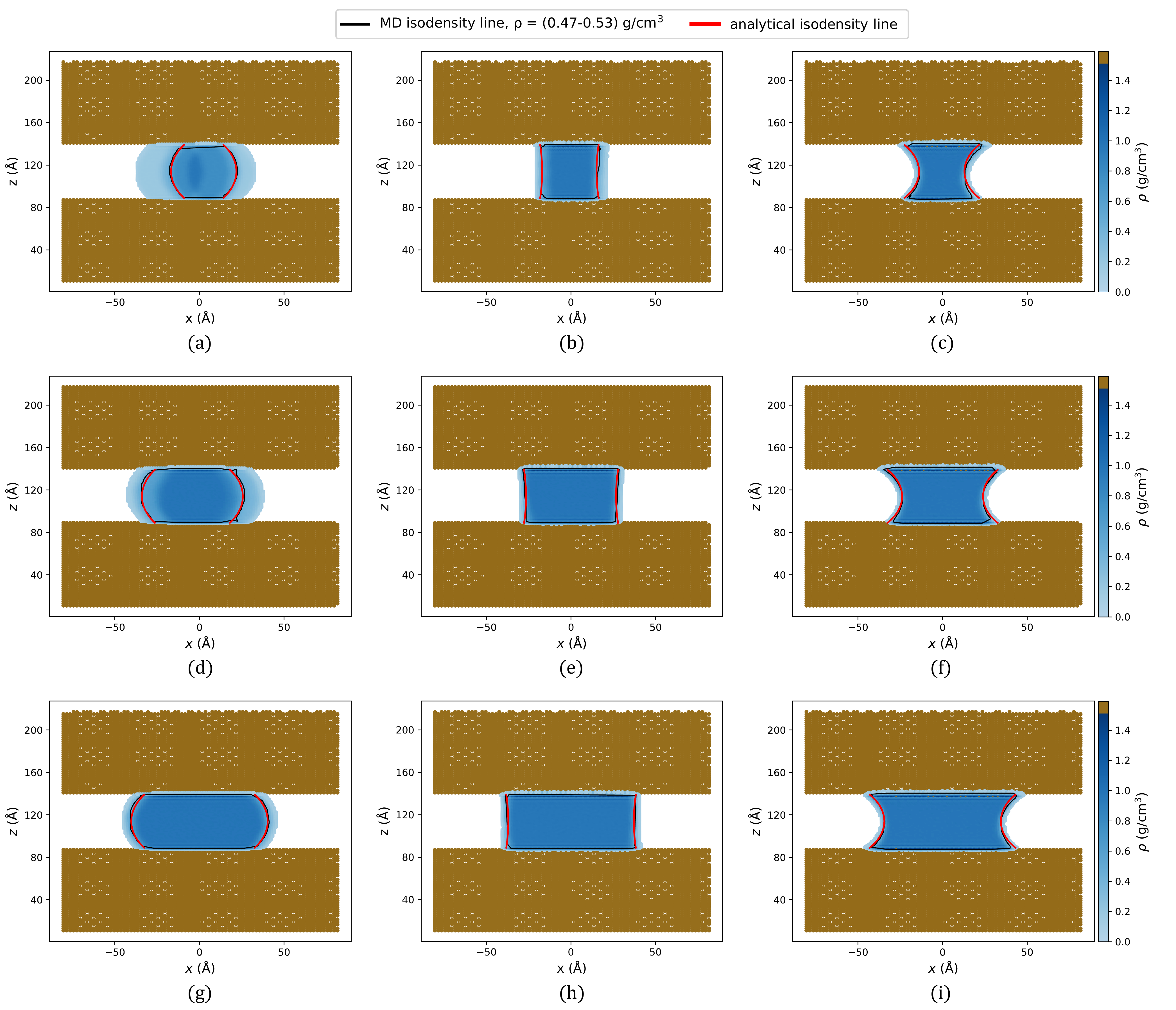} 
	\caption{\label{fig:1} MD density profiles of water menisci ($6\text{a}_0$, $10\text{a}_0$, and $14\text{a}_0$ from top to bottom) for different parametrizations: a), d), g) $\varepsilon=10$~meV;  b), e), h) $\varepsilon=16$~meV;  c), f), i)  $\varepsilon=21$~meV. The black solid line connects the points where the density of meniscus is equal to the half of the density of confined water $\rho=(0.47 - 0.53)$~g/cm$^{3}$. Red line corresponds to the analytic evaluation of the meniscus shape based on the assumption of its circular geometry (see (SI)).}
\end{figure*}

Mass density maps of water menisci with different sizes (case ii) are presented in Fig.~\ref{fig:1}. The distribution of the density inside the menisci is also shown for three different wettability. For small $\varepsilon$ value,10meV, (i.e. large wetting angle, see Fig.~\ref{fig:1}a, d, g) the expected hydrophobic case is recovered. With the increase of the $\varepsilon$ parameter the hydrophilicity increases as expected. In addition, we can observe that the density of water at the edge of the meniscus (see Fig.~\ref{fig:1}a, d, g)) is lower for small $\varepsilon$ values (light blue area). Only the central part of the domain is filled with water with a density close to the bulk one. Intuitively, this situation will not be favorable to heat transfer. As $\varepsilon$ increases,  the wetting angle decreases as well as the amount of ``low density water'', present on the different edges, for all sizes of meniscus (Fig.~\ref{fig:1}). In the latter cases, energy coupling through the water meniscus is more efficient as it will be demonstrated hereafter.

The shape of the water menisci was deduced from the isodensity line corresponding to $\rho=0.5 \rho_{bulk}$ ($\rho_{bulk}$ is the density of confined water). The isodensity lines are depicted in the Fig.~\ref{fig:1} by black solid lines. Such lines were compared with the analytical predictions shown by red lines. For the analytic description of the menisci shapes, we use the assumptions of: ``circular shape'' of the water profile, a constrained volume of the meniscus, and assumes that the wetting angle can be taken the same as for a  droplet with the same $\varepsilon$ value~\cite{Isaiev2015} (see details in (SI)). Such theoretical profiles are also presented in Fig.~\ref{fig:1} with the red lines. As one can see, the excellent agreement between the results of simulations and analytic modeling can be stated. The latter model will be used to define system boundaries in FEM simulations. Density maps for all considered values of wetting angle together with density profiles can be founded in (SI).

\subsection{Interfacial thermal boundary resistance between the liquid meniscus and the solid substrate}

Knowing the thermal energy going through across the interface is important to estimate efficacy of heat transport performance. The related power can be calculated as the energy required for maintaining the temperature difference in thermostats per unit of time in EBM, $P = \left<dE\right>/dt$.

In the frames of ACM approach the power was calculated from the heat fluxes as follows:

\begin{equation}
\label{eq:powerACM}
P = \int_\Sigma \boldsymbol{J} \cdot \boldsymbol{ds},
\end{equation}
where $\boldsymbol{J}$ is calculated with the Eq.~\ref{eq:2}, $\boldsymbol{ds}$ is the unit area taken as $\boldsymbol{ds} = dxdy\cdot\boldsymbol{e_z}$ in our calculations, and $\Sigma$ is the cross-sectional area of the meniscus normal to $z$ . In order to decrease numerical errors caused by atoms fluctuations, we averaged the power over different slices in $z$ direction.

\begin{figure*}[h!]
	\centering
	\includegraphics[width = 6.7in]{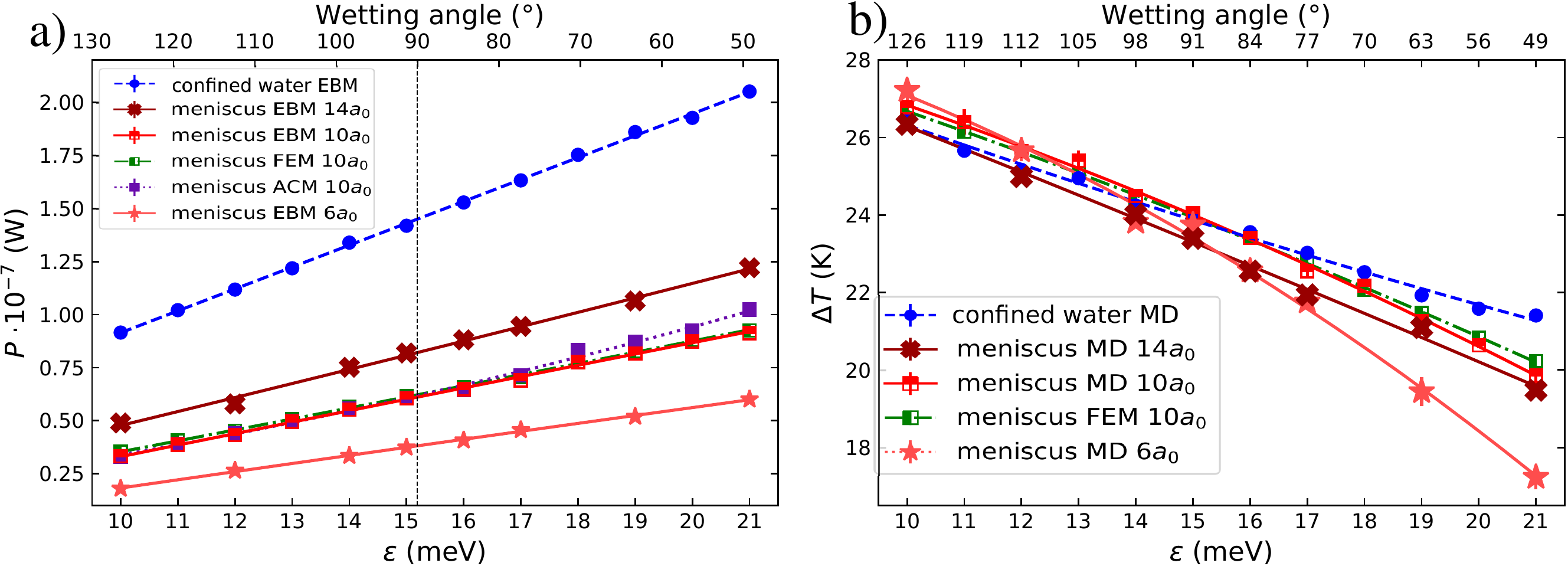}
	\caption{\label{fig:3} (a) Power balance and (b) temperature jump  as a function of $\varepsilon$ for case i) (blue line) and case ii) obtained by MD, FEM, and ACM approaches.}
\end{figure*}

The Fig.~\ref{fig:3}a presents dependence of the exchanged power with the interaction potential $\varepsilon$ for different initial meniscus sizes as well as for confined water layer. One can see that the exchanged power shows a linear trend with the increase of $\varepsilon$. In addition, the rate of the change is increasing with the meniscus size. Eventually, as it can be seen in Fig.~\ref{fig:3}a for the confined water, both EBM and ACM methods gives very similar results (relative error lower than 3\%). Beyond the fact that both approaches can be used, it also confirms the correctness of the computed heat flux values.

Temperature jump as a function of $\varepsilon$ at the interface is shown in Fig.~\ref{fig:3}b. The decrease of the temperature difference can be stated with increasing of the strength of interaction.  This is coherent with a better water adsorption at Si interface and the occurrence of a higher water density layer that improves the thermal contact. It is interesting to note, that for the smaller $\varepsilon$, the temperature difference is almost the same for all meniscus sizes as well as for confined water layer. In the same time, there is a visible difference in the value of $\Delta T$ for higher interactions. This may caused by the capillary nature of the edges of the meniscus, which allow transverse modes propagate~\cite{Landau1987Fluid}, that may significantly improve transverse phonon interactions with the liquid. 

The evaluated exchanged power and temperature jump at the interface give us the possibility to calculate the absolute thermal resistance ($R_{th}$) at the interface by:
\begin{equation}
\label{eq:powerACM}
R_{th} = \frac{\Delta T}{P}.
\end{equation}

\begin{figure*}[h!]
	\centering
	\includegraphics[width =6.7in]{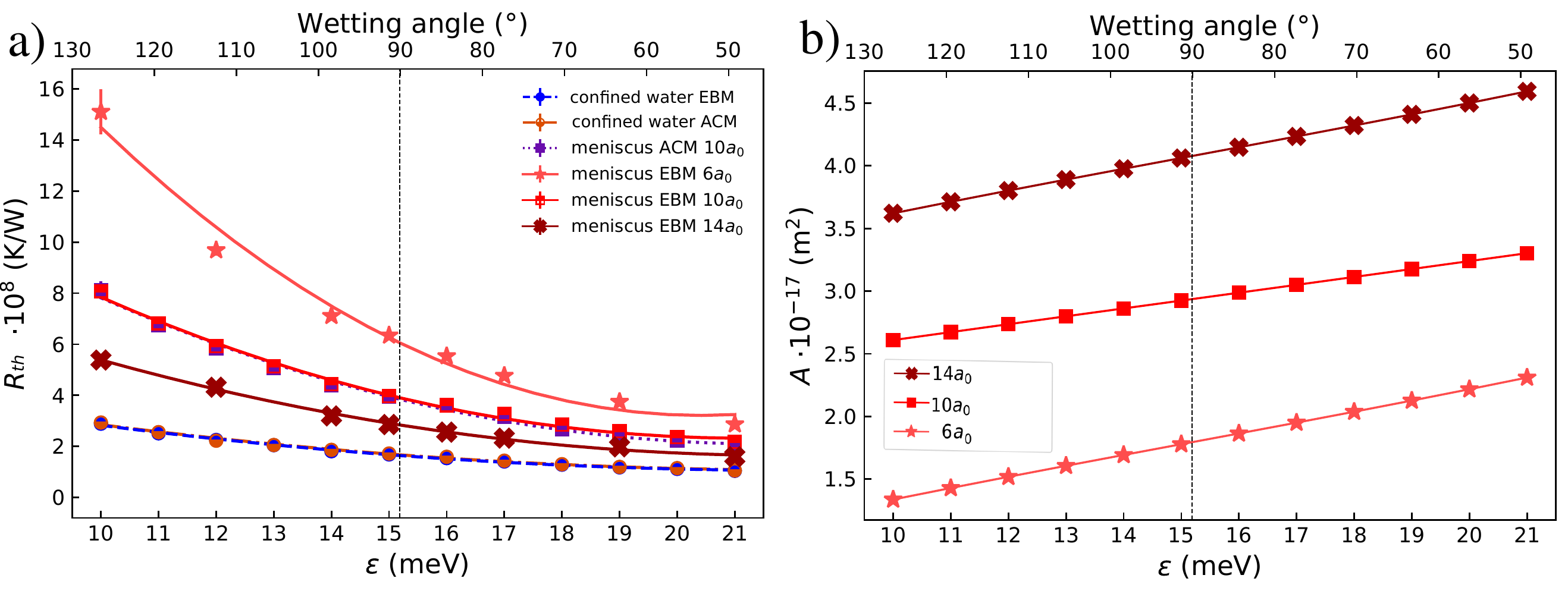}
	\caption{\label{fig:Rth}Left graph (a) shows the dependence of the thermal resistance normalized by the silicon/water contact area on different values of $\varepsilon$  for: system with confined water layer (case i) and water menisci with different sizes, $6\text{a}_0$, $10\text{a}_0$, and $14\text{a}_0$  (case ii). The contact area is shown in the right graph (b). Dependence of   silicon/water contact area with $\varepsilon$ for water meniscus with different sizes, $6\text{a}_0$, $10\text{a}_0$, and $14\text{a}_0$  (case ii) is shown in the left graph.}
\end{figure*}

As can be seen from Fig.~\ref{fig:Rth}a, the value of  absolute thermal resistance is decreasing with increasing $\varepsilon$, showing rather similar behavior for all meniscus sizes. Nevertheless, the difference of the $R_{th}$ is significant for the considered sizes in the case of the weak interaction, and this difference gets smaller for the high values of $\varepsilon$. Additionally, it can be mentioned that while increasing the $R_{\text{th}}$ meniscus size the curve cowerage forward the curve corresponding to the confined water layer.

The silicon/water contacted area ($A$) as a function of $\varepsilon$ shows linear variation with increasing interaction parameter $\varepsilon$, and obvious slope dependence with the size of meniscus. Those results are  represented in Fig.~\ref{fig:Rth}b. This information allows us to evaluate the interfacial thermal resistance (ITR) with the use of the following relation:

\begin{equation}
\label{eq:ITR}
R = R_{th} \cdot A.
\end{equation}

\begin{figure*}[h!]
	\centering
	\includegraphics[width =3.7in]{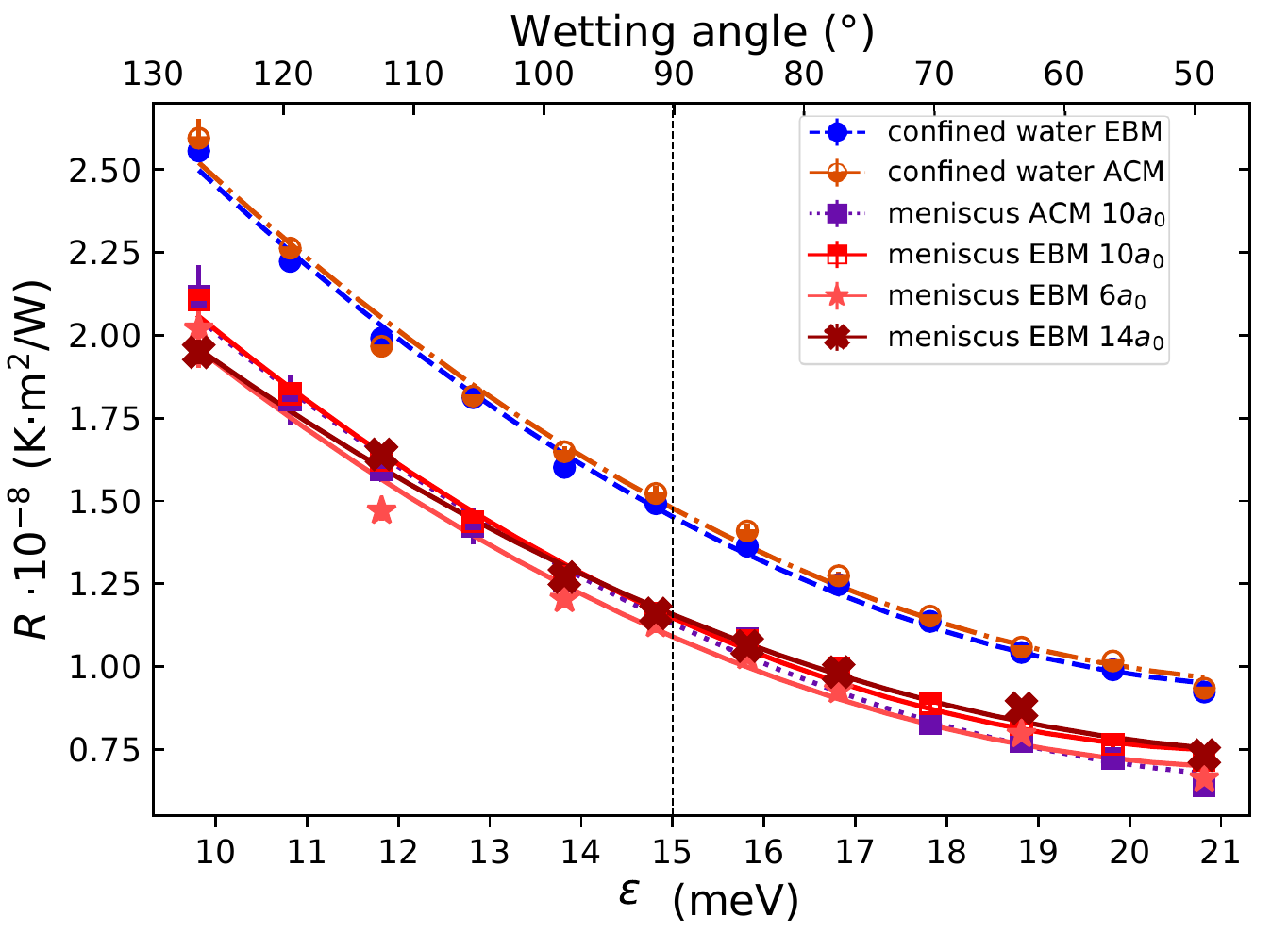}
	\caption{\label{fig:2} Dependence of interfacial thermal resistance (ITR) with $\varepsilon$ for: system with confined water layer (case i) and water meniscus with different sizes, $6\text{a}_0$, $10\text{a}_0$, and $14\text{a}_0$  (case ii).}
\end{figure*}

Comparisons of the ITR values of $R$ vs. $\varepsilon$, both respectively determined by EBM and ACM methods, are shown in Fig.~\ref{fig:2}. The computed ITR values for confined water layer are similar to the literature results which are typically in the range of $0.4 \times 10^{-8} <R< 2.3 \times 10^{-8}$ m$^2$ K/W, see Barisik et al~\cite{Barisik2014}.

For both situations of confined water layer and menisci, $R$ decreases with the increase of $\varepsilon$. This is naturally due to the growth of interaction strength, as the adhesion behavior started to be more and more hydrophylic. The latter ($\varepsilon=21$meV) creates a better coupling between the water and the silicon molecules. ITR values for the meniscus (case ii) are smaller than for confined water (case i). Mechanisms that play an important role are thus the weak bonding of water molecules at the interface and edge effects where water with ``low density'' surrounds the meniscus. Occurrence of both lead to a degradation of the energy transfer. The difference between the corresponding $R$ values for the meniscus and confined water slightly decreases with increasing $\varepsilon$. The wetting angle was determined based on already estimated values~\cite{Isaiev2015}. It should be noted that the value of $R$ is almost the same for all meniscus sizes. Thus, the decreasing of the ITR between cases ii) and i) can be connected with heat transfer channels related to the presence of liquid/vapor interface. 

\subsection{MD insights for upper scale approaches}

MD simulations give interesting insights about ITR variations with the contact angle at the microscale. However, engineering applications cannot be currently described by MD tools. To tackle this issue, we propose to use parameters resulting from the atomic scale modelling as inputs for FEM modelling. To perform FEM calculations, we used the COMSOL\textsuperscript{\tiny\textregistered} software (see subsection~\ref{subsec:FEM_procedure} in SI).

For such simulations, we approximate the shape of meniscus edges with isodensity lines (see (SI)) as it was mentioned above, which were presented in Fig.~\ref{fig:1} (red lines) and which match the isodensity lines computed in MD (black lines).

In order to calculate temperature profile with the FEM, we assume that there is no exchange on the lateral edge of the water domain. The wetting angle~\cite{Isaiev2015} for chosen $\varepsilon$ at the contact points with solid was set as a boundary condition. To find the position of the edges, we also constrained the volume of the meniscus. More details about the procedure and resulted figures are provided in (SI). 
In COMSOL\textsuperscript{\tiny\textregistered}, silicon slabs were modelled by two rectangles with lengths and spacing coinciding with the corresponding dimensions used in MD. The temperatures of the lower and upper boundaries were set at 330 and 270~K respectively. Periodic boundary conditions were applied. As input data, we used ITR values obtained from MD. The thermal conductivity of silicon and water were chosen equal to 220 and 0.86~W/(m K)~\cite{Isaiev2020}.

\begin{figure*}[h!]
	\centering
	\includegraphics[width = 7.9in]{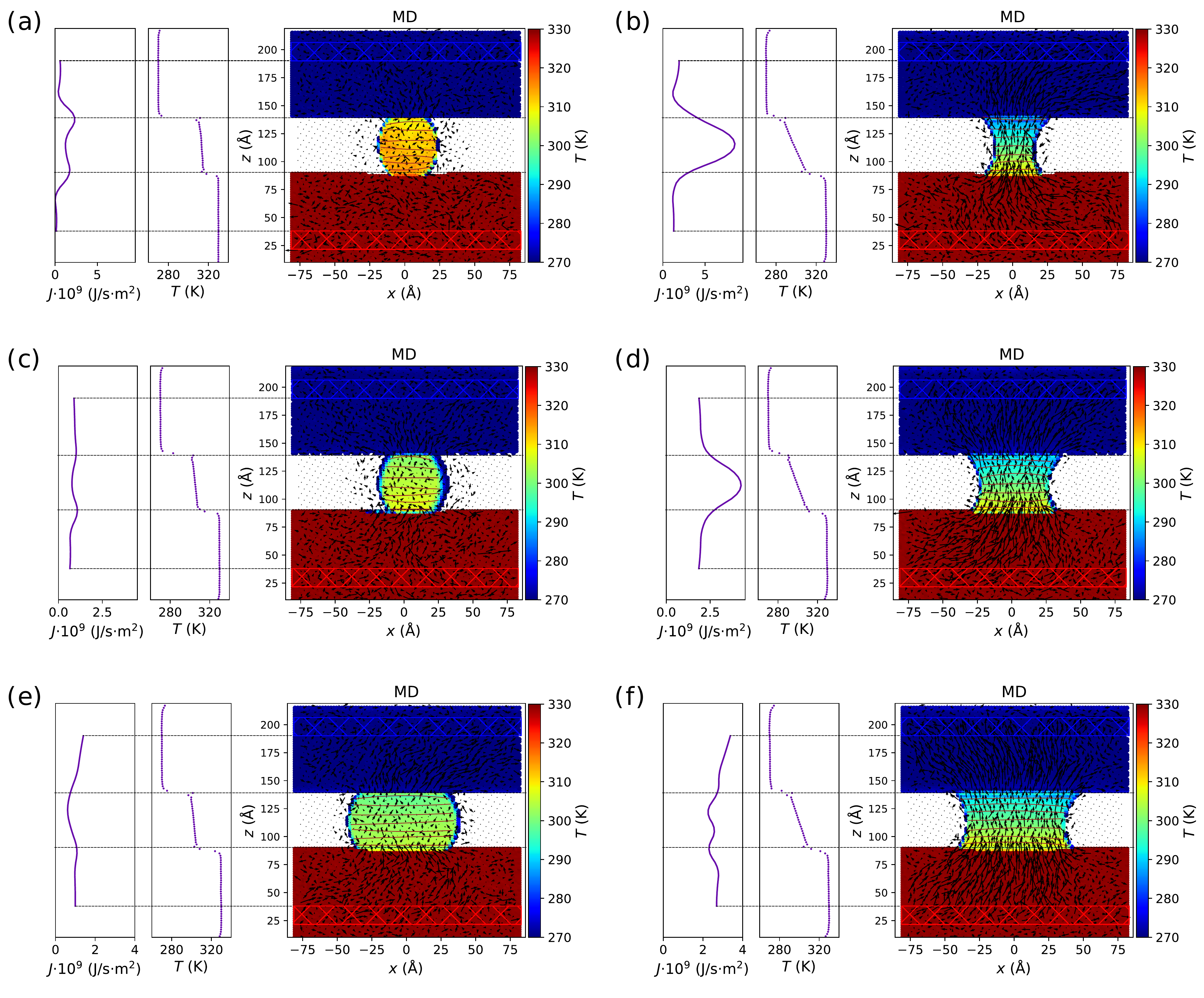}
	\caption{\label{fig:4} Temperature profiles of the menisci ($6\text{a}_0$, $10\text{a}_0$, and $14\text{a}_0$ from top to bottom) with heat flux distributions (black arrows) for different hydrophobicities: a), c), e) $\varepsilon=$10~meV,  b), d), f) $\varepsilon=$21~meV. In each central plot is shown the one-dimensional distribution of heat fluxes (left side) and temperatures (right side). The regions where the thermostats with $T=270$ K and $T=330$ K were applied are shown with blue and red areas respectively. Brown lines the 2D distributions of isotemperature are shown.} 
\end{figure*}

To take into account the non-constant cross-sectional area of the drop, it was decided to characterize the energy transfer as the total heat flux transferred from hot to cold sides. The results are shown in Fig.~\ref{fig:4}. One can see from Fig.~\ref{fig:4} that, at small $\varepsilon$ values, the heat flux distributions obtained with FEM and MD are a bit different, although at $\varepsilon=$21 meV they almost converge. This trend could be related to ITR results given in Fig.~\ref{fig:2}. While $\varepsilon$ is increases, $R$ gets smaller as well as channels for heat transport are better defined (dense water slab). 

\begin{figure*}
	\centering
	\subfloat{\includegraphics[width=3.7in]{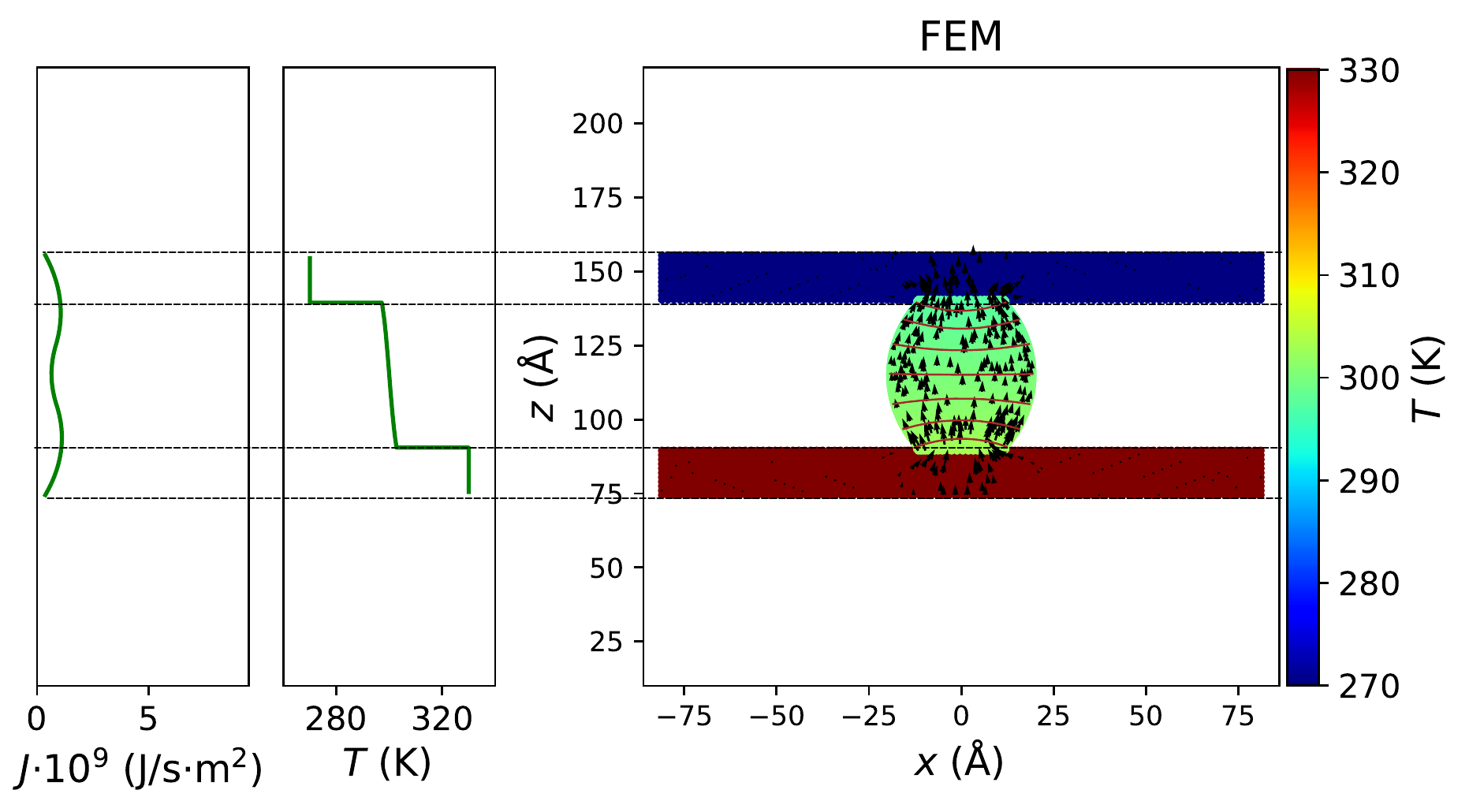}} 
	\subfloat{\includegraphics[width = 3.7in]{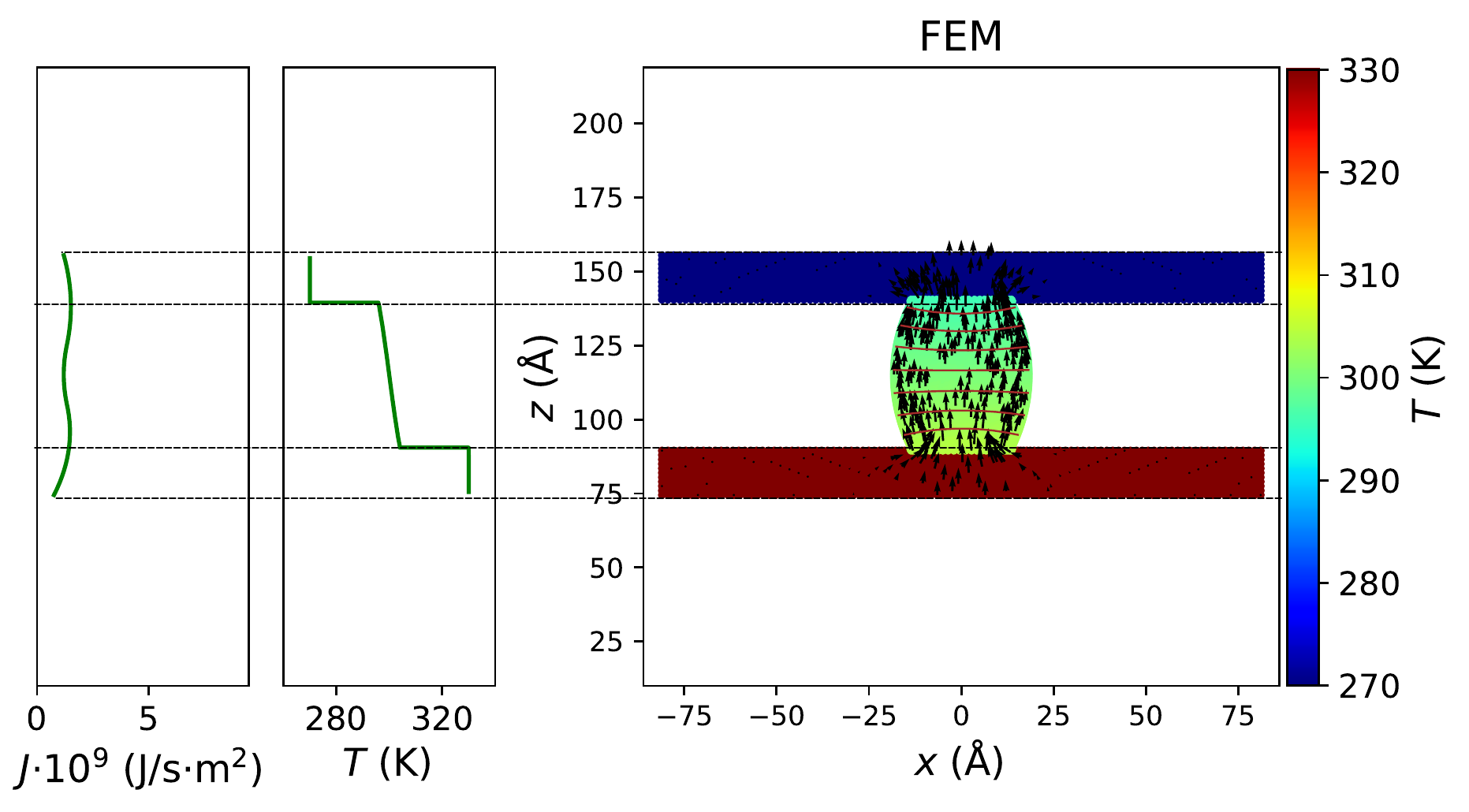}}\\
	\subfloat{\includegraphics[width = 3.7in]{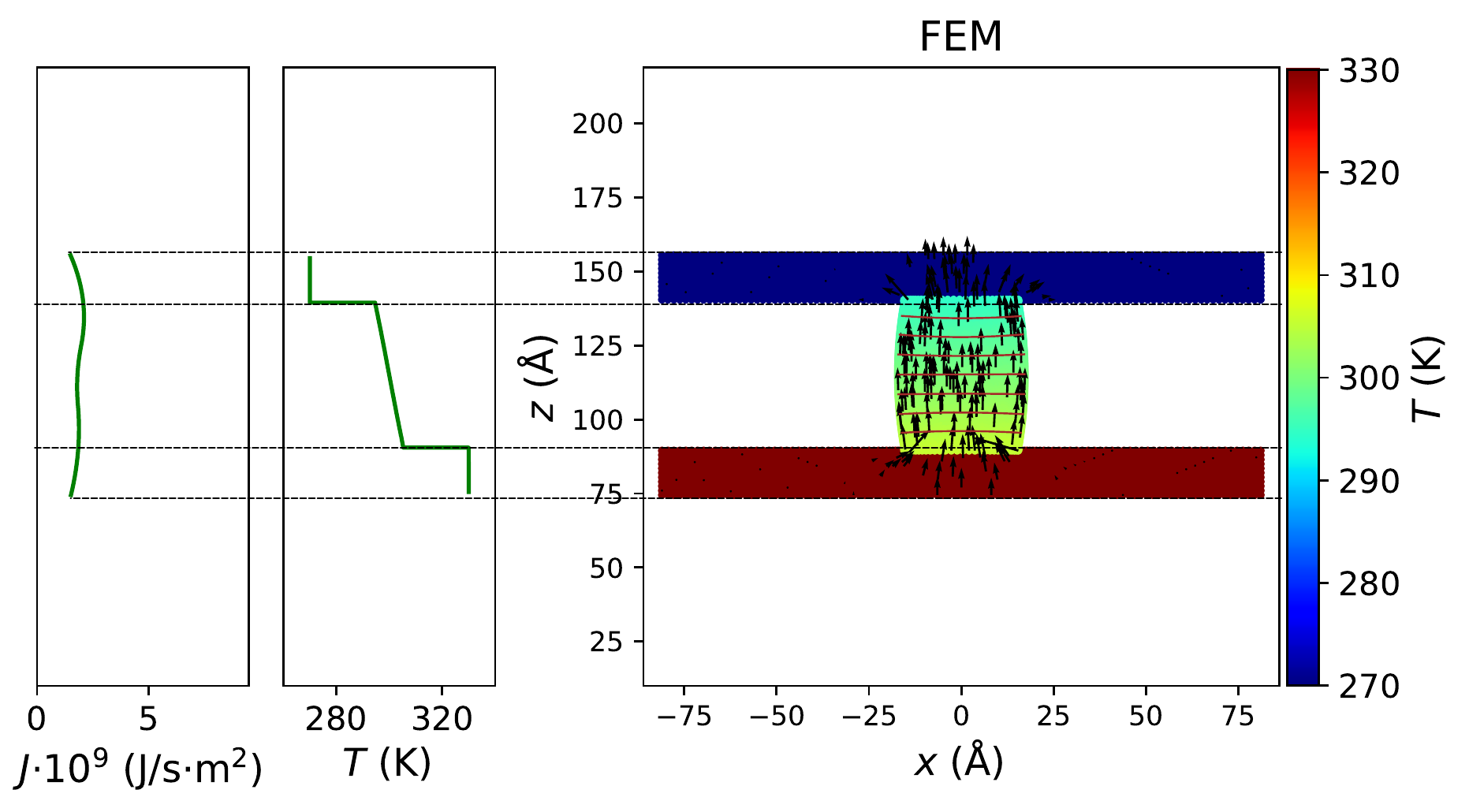}} 
	\subfloat{\includegraphics[width = 3.7in]{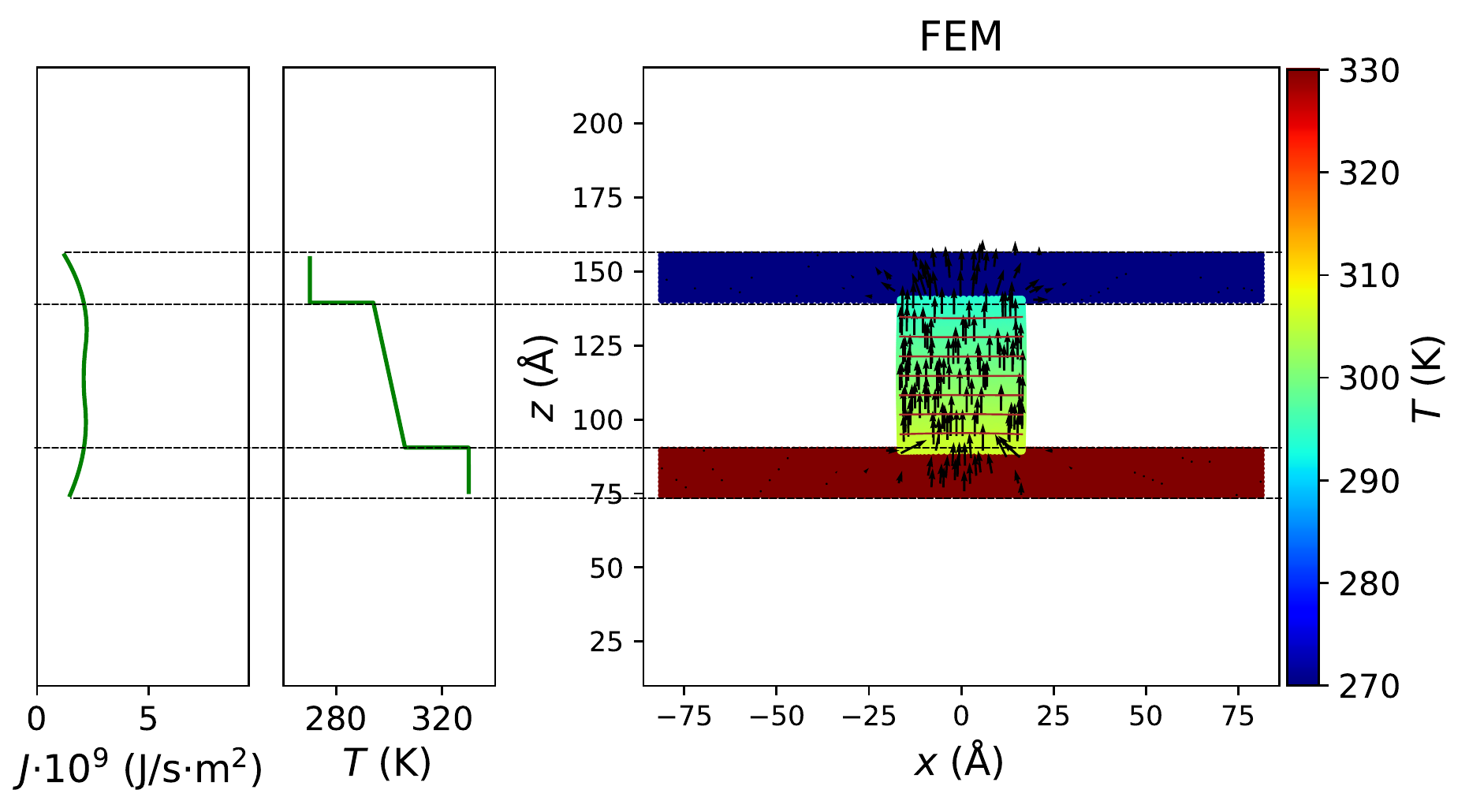}}\\
	\subfloat{\includegraphics[width = 3.7in]{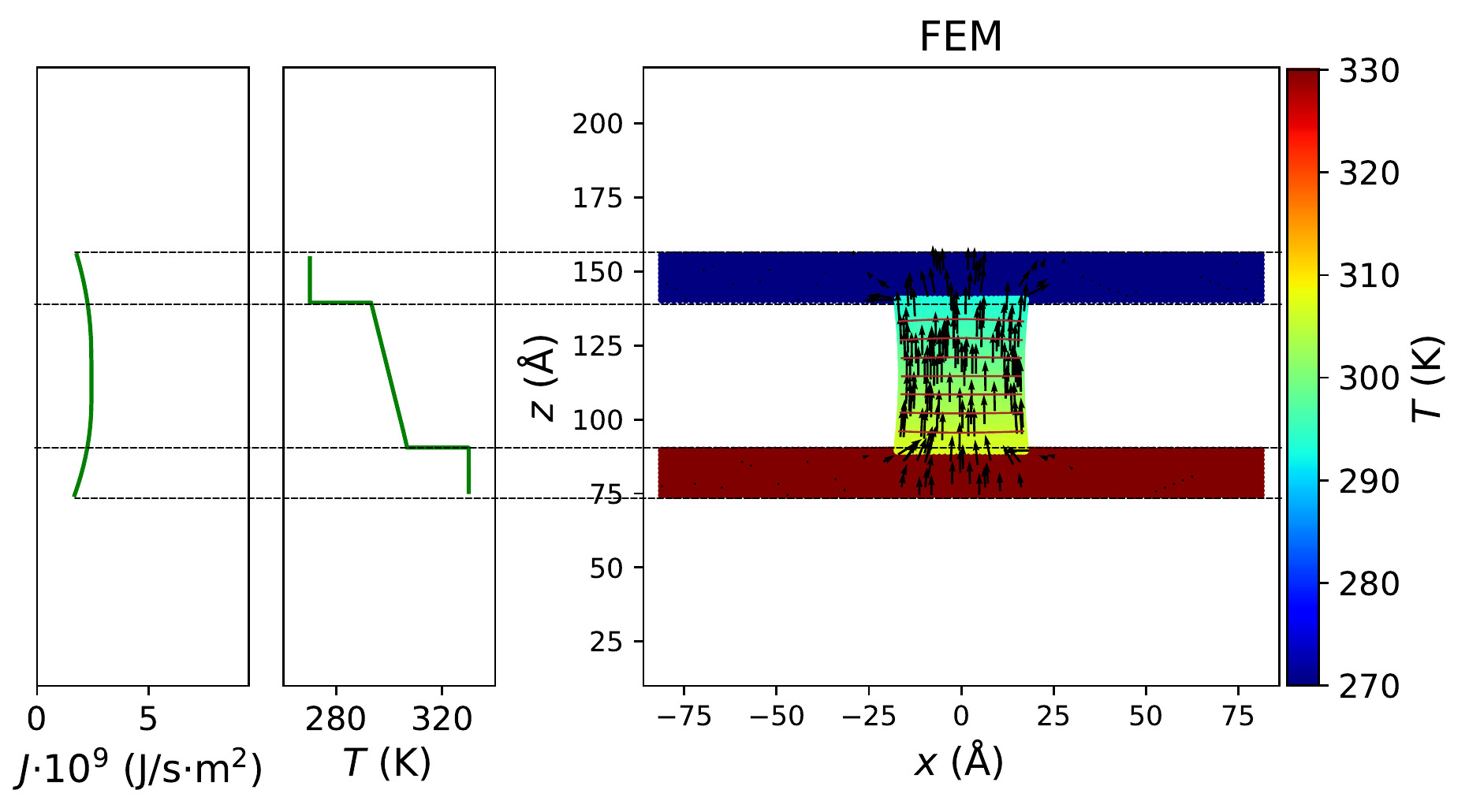}}
	\subfloat{\includegraphics[width = 3.7in]{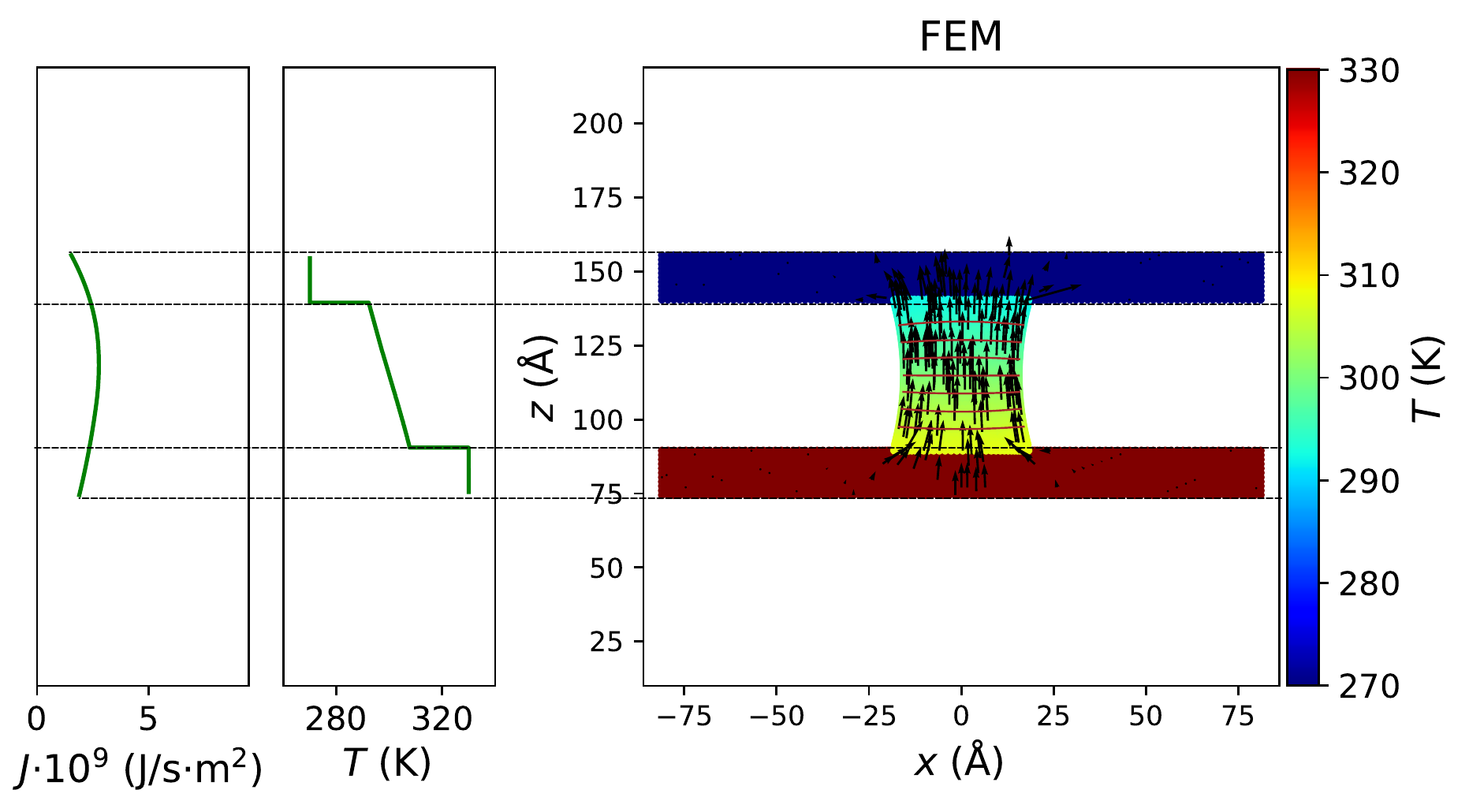}} \\
	\subfloat{\includegraphics[width = 3.7in]{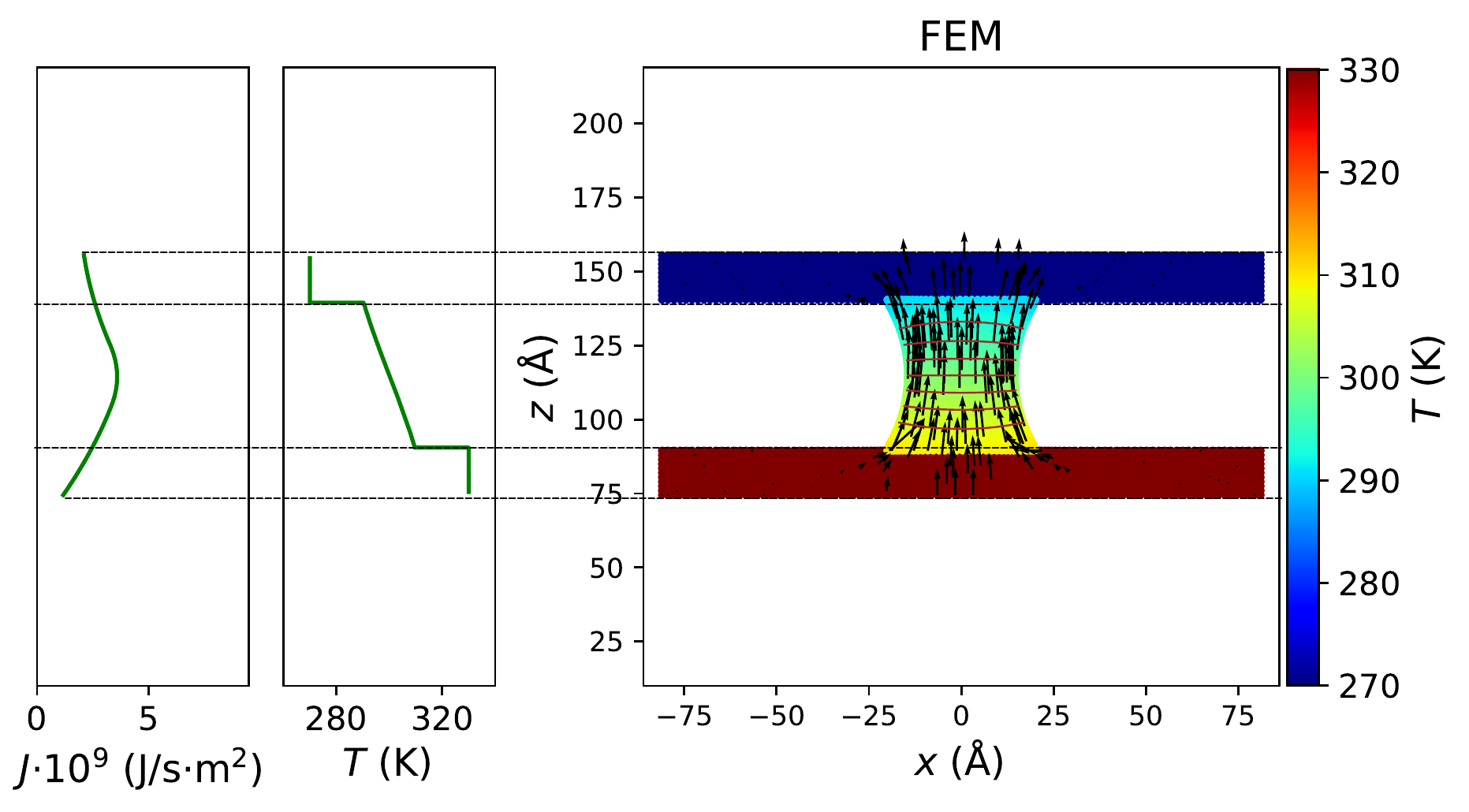}}
	\subfloat{\includegraphics[width = 3.7in]{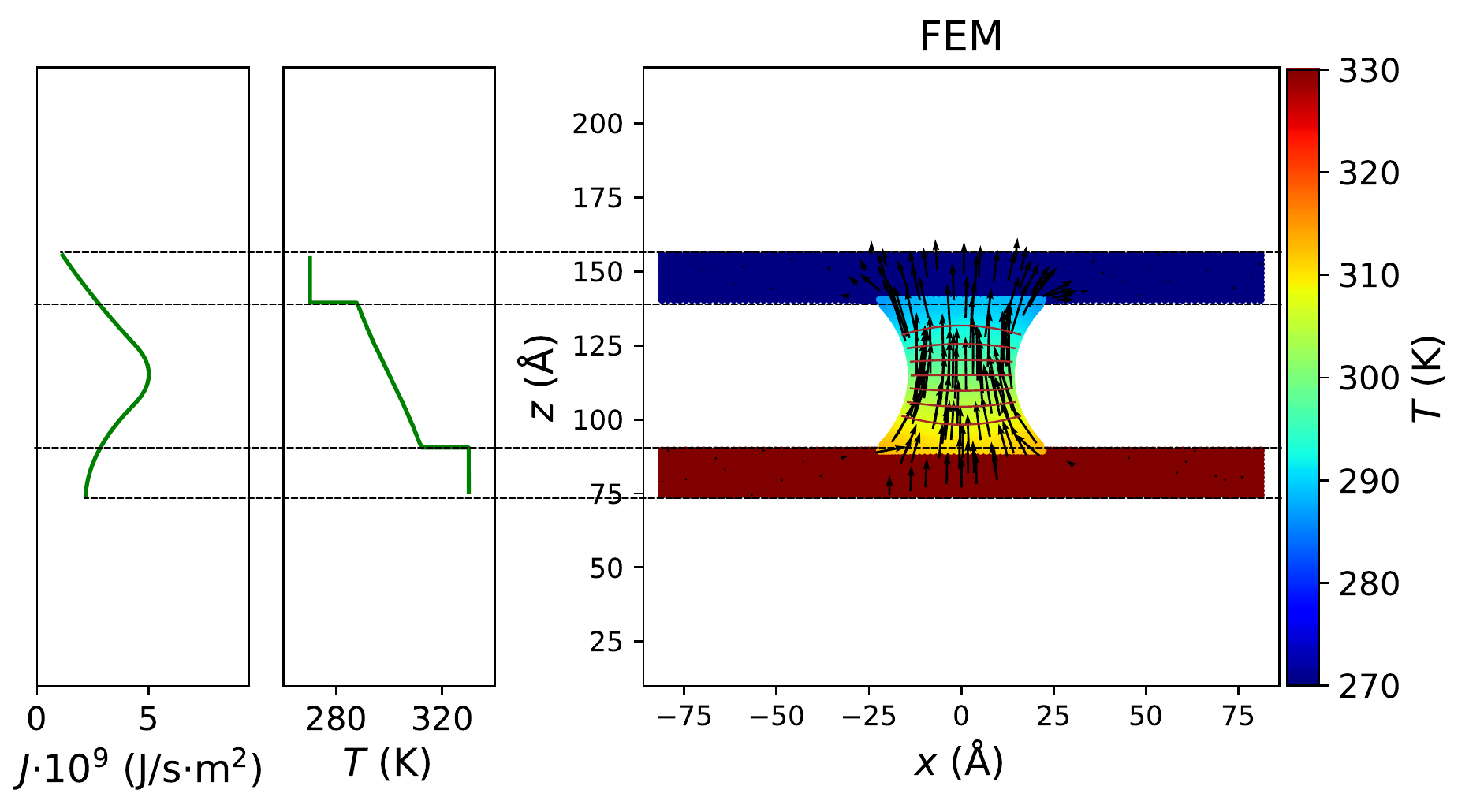}} 
	\caption{\label{fig:FEM_SM_6} Temperature profiles of the meniscus with size $6\text{a}_0$, ($\varepsilon$ = 10 - 21 meV) with heat fluxes distributions obtained from FEM simulations.}
	
\end{figure*}

\begin{figure*}
	\centering
	\subfloat{\includegraphics[width=3.7in]{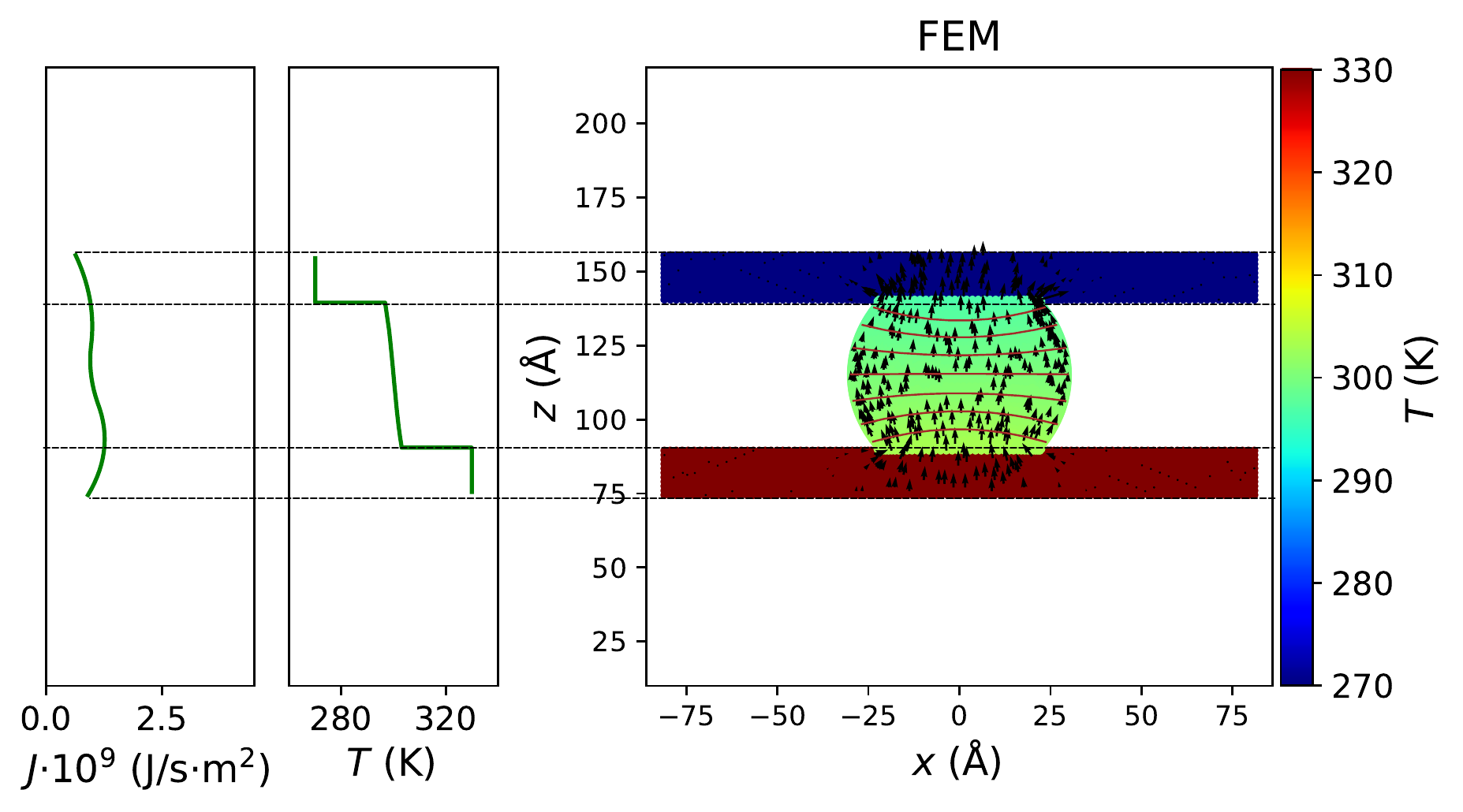}} 
	\subfloat{\includegraphics[width = 3.7in]{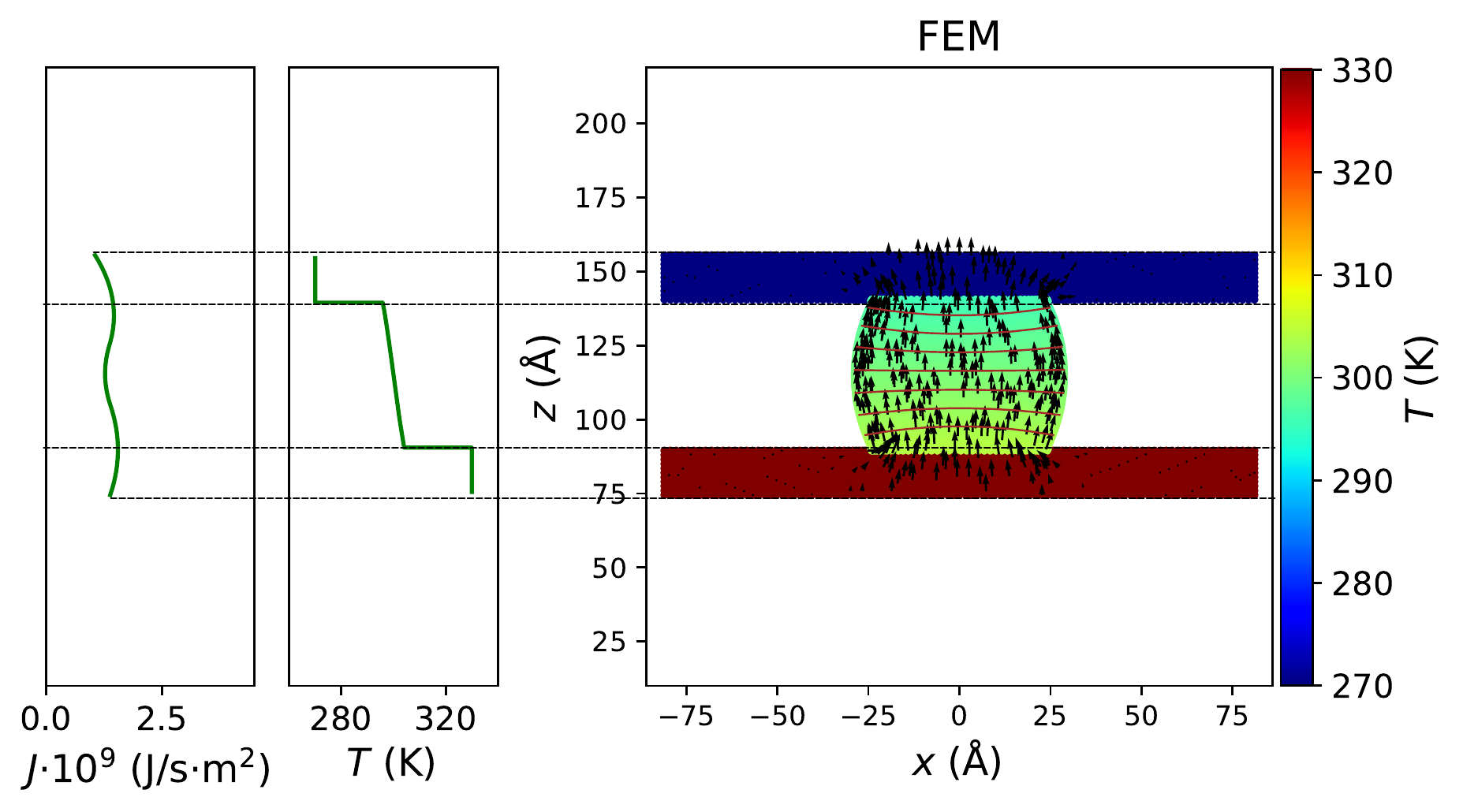}} \\
	\subfloat{\includegraphics[width = 3.7in]{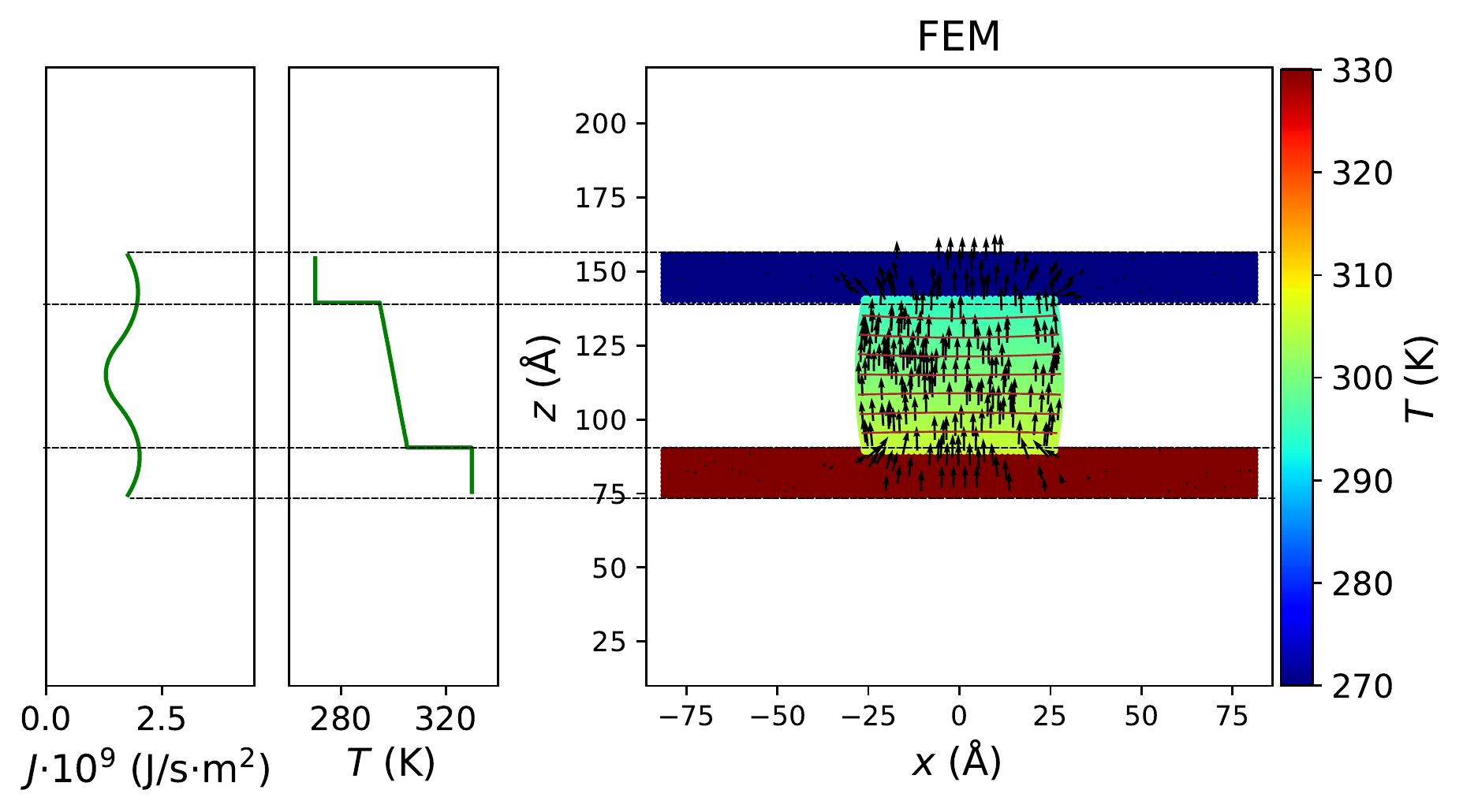}}
	\subfloat{\includegraphics[width = 3.7in]{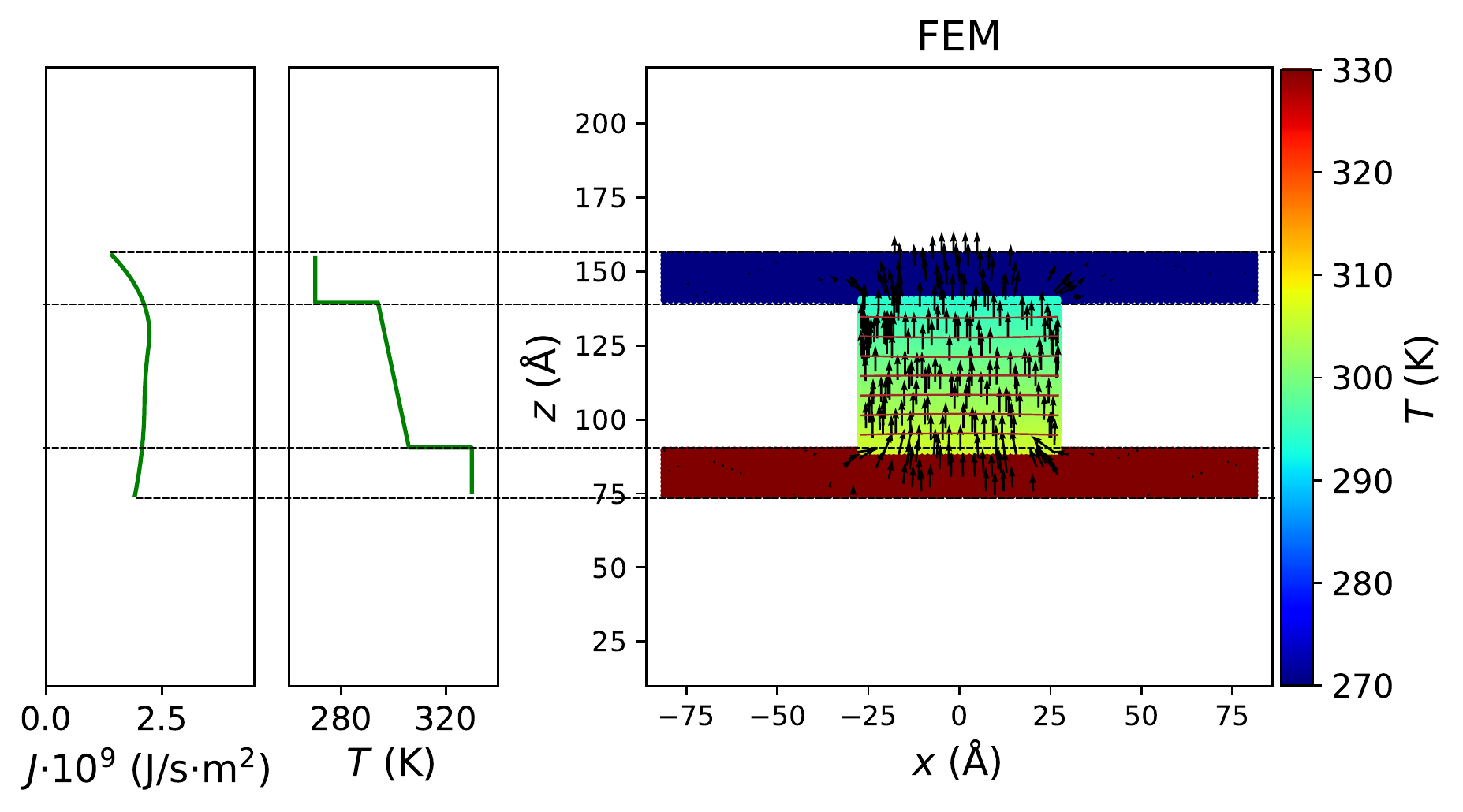}} \\
	\subfloat{\includegraphics[width = 3.7in]{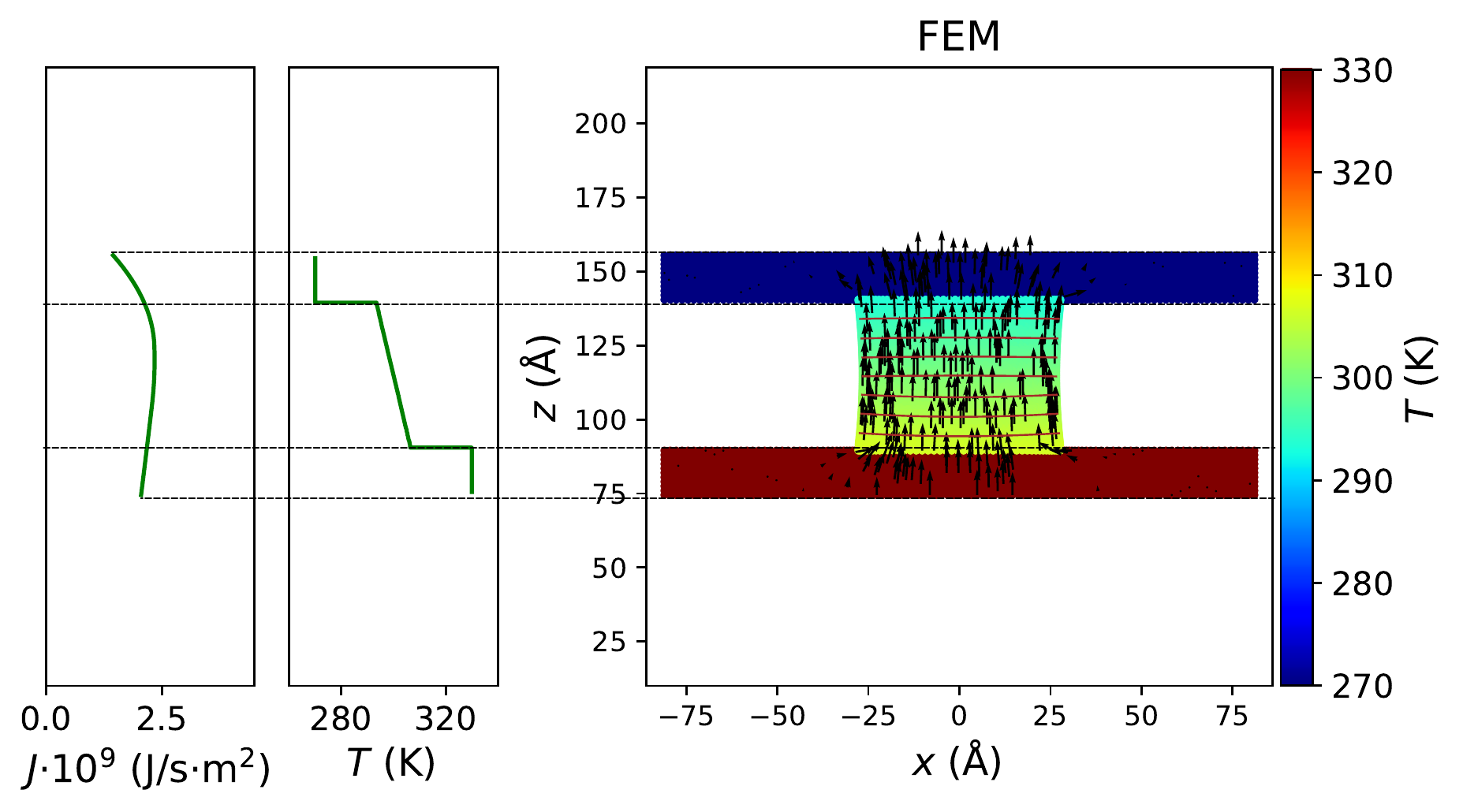}}
	\subfloat{\includegraphics[width = 3.7in]{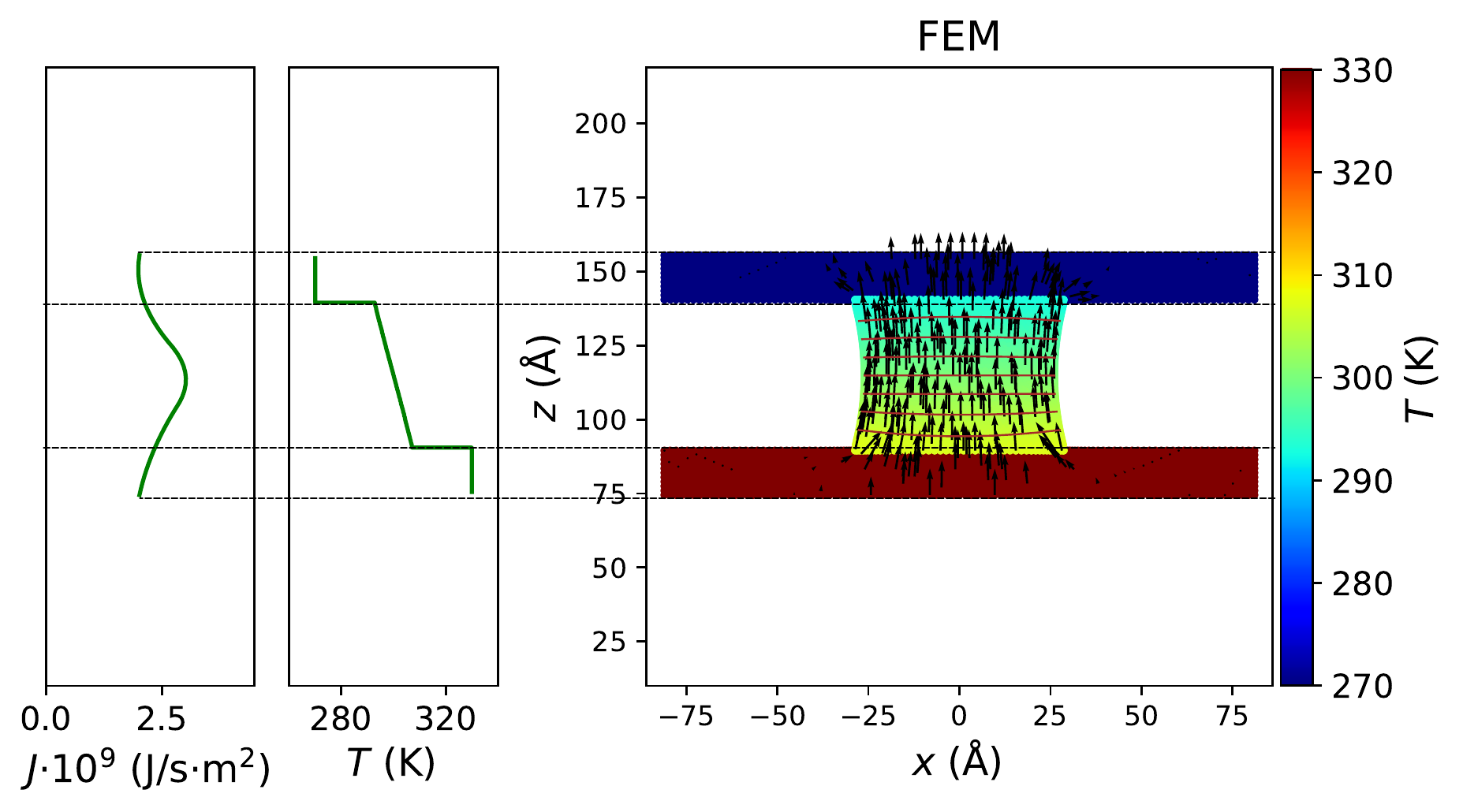}}\\
	\subfloat{\includegraphics[width = 3.7in]{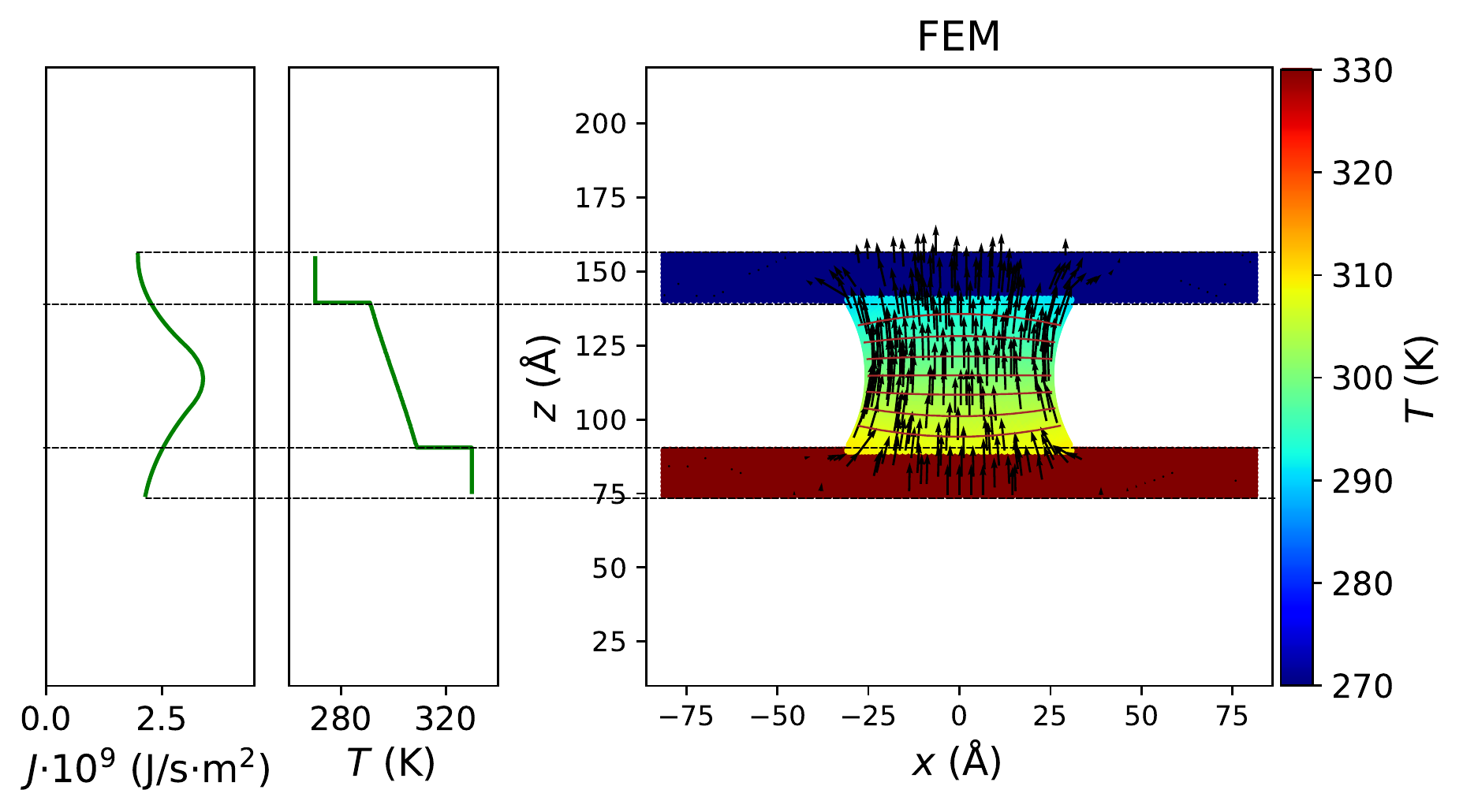}}	
	\subfloat{\includegraphics[width = 3.7in]{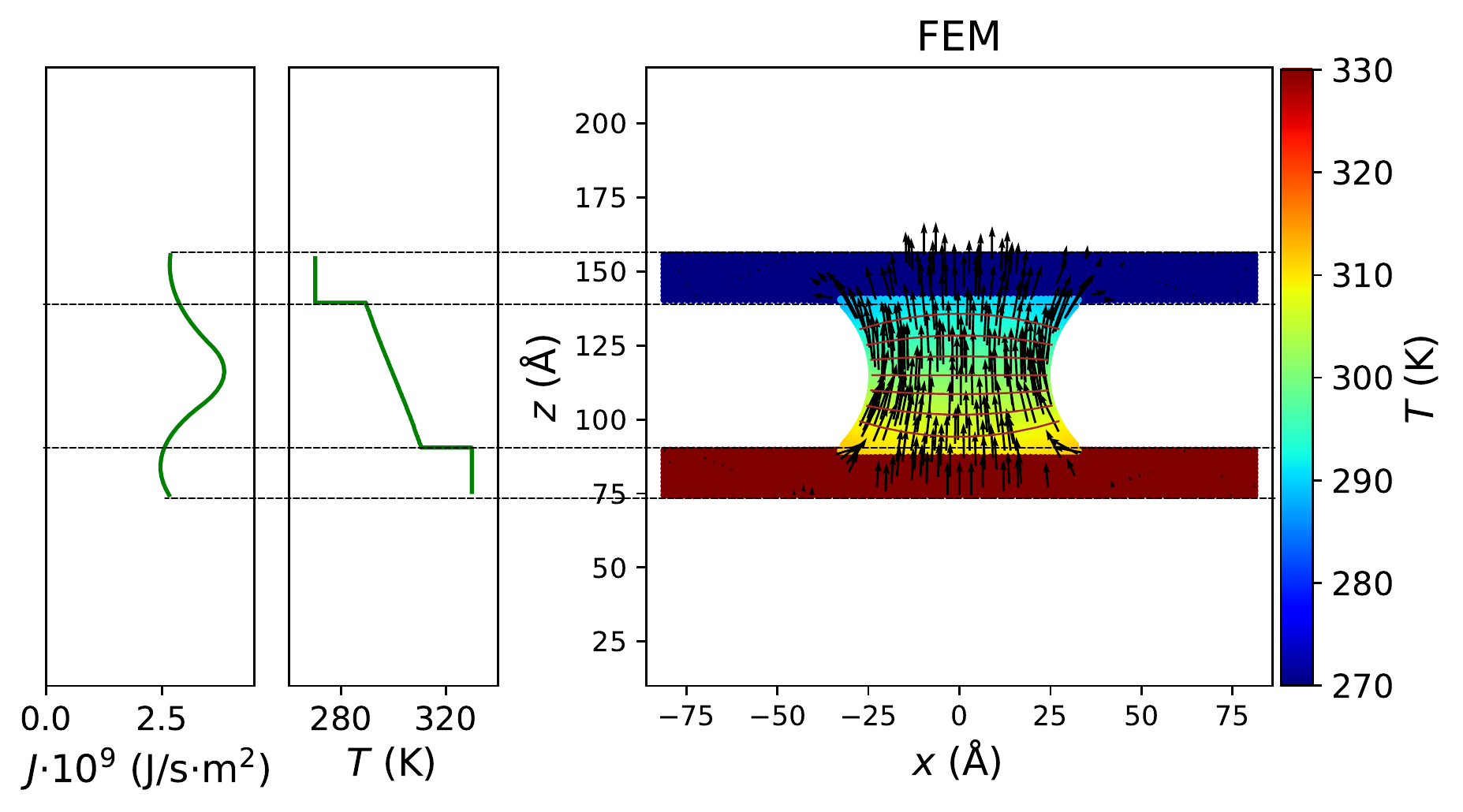}} 
	\caption{\label{fig:FEM_SM_10} Temperature profiles of the meniscus with size $10\text{a}_0$, ($\varepsilon$ = 10 - 21 meV) with heat fluxes distributions obtained from FEM simulations.}
	
\end{figure*}

\begin{figure*}
	\centering
	\subfloat{\includegraphics[width=3.7in]{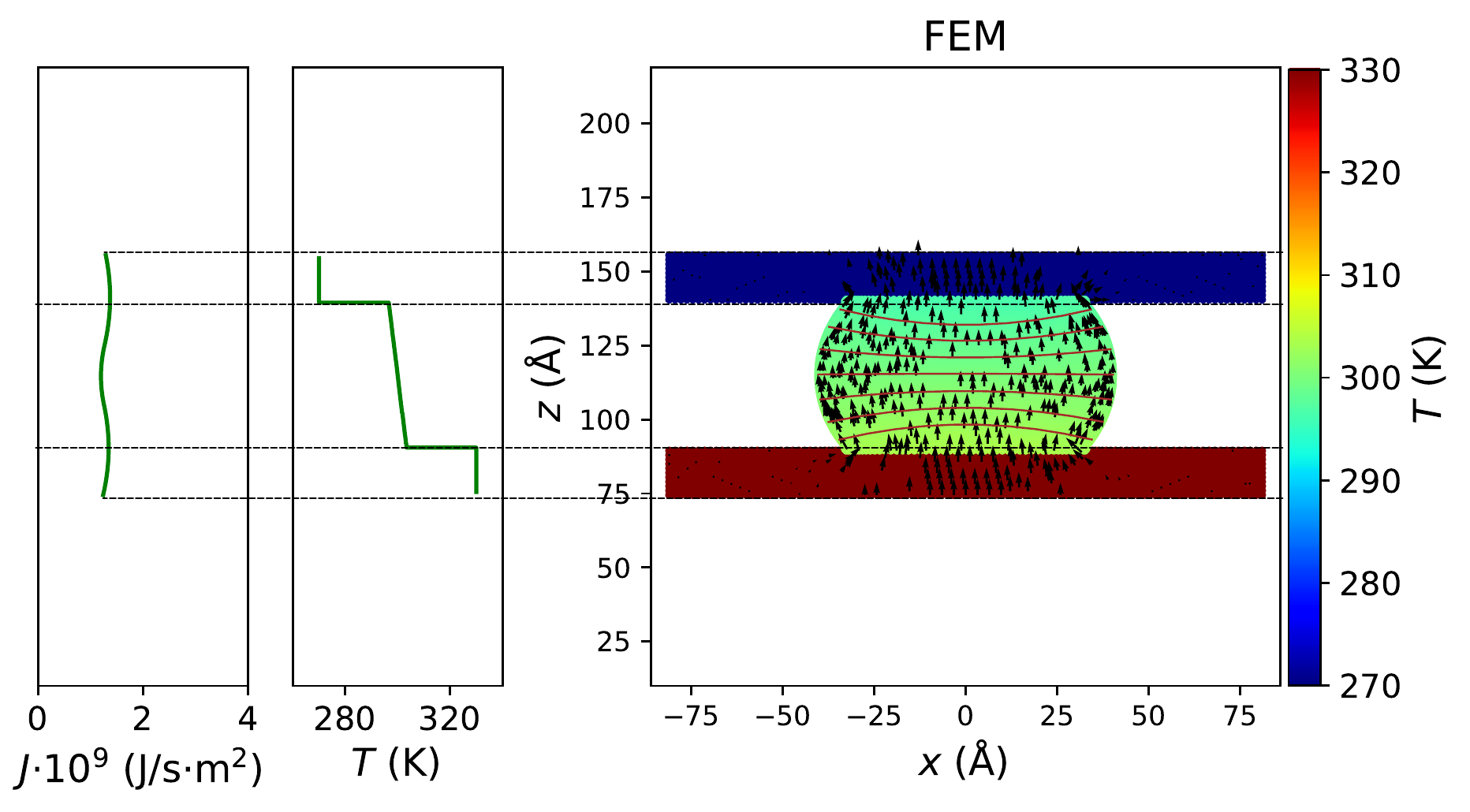}} 
	\subfloat{\includegraphics[width = 3.7in]{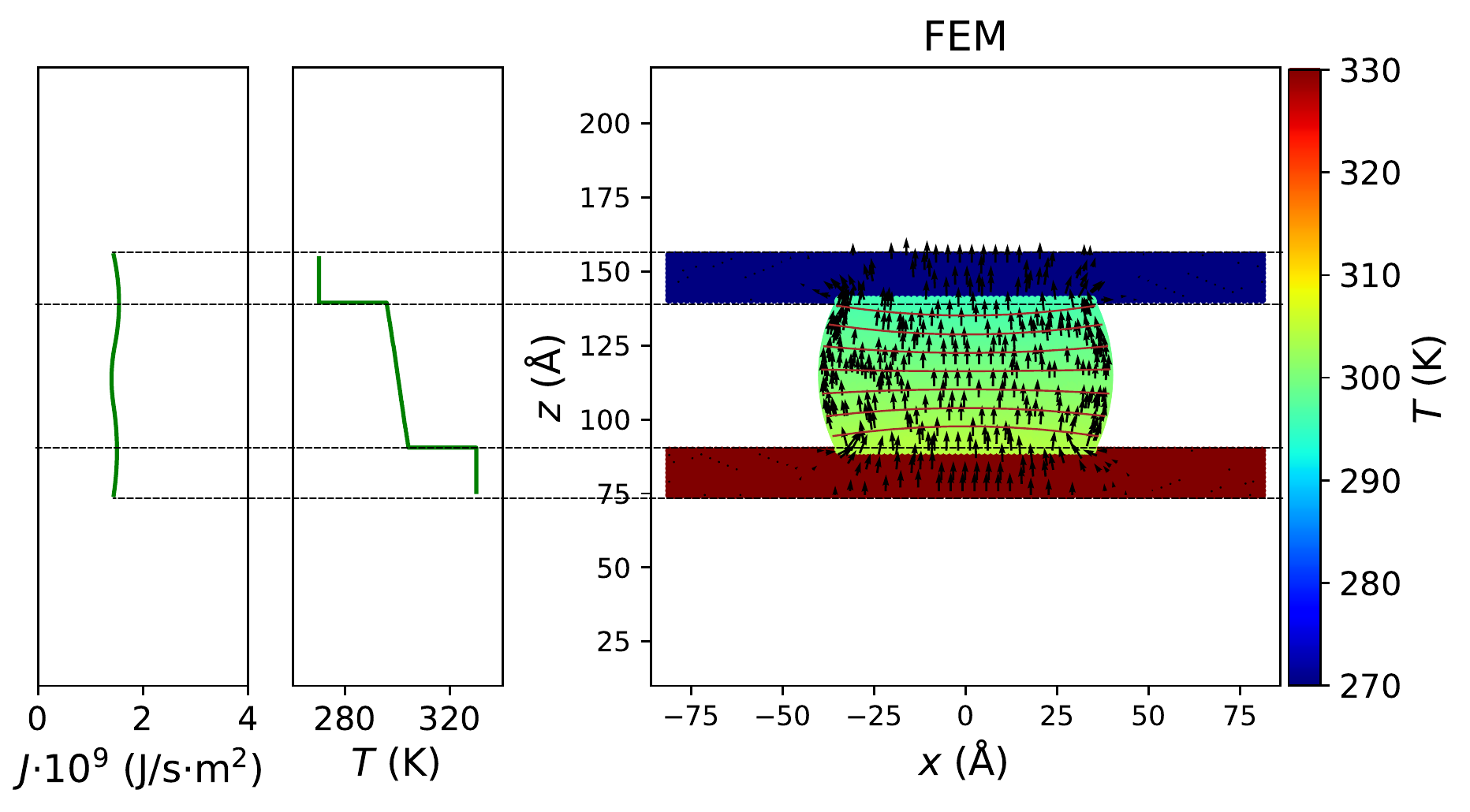}}\\
	\subfloat{\includegraphics[width = 3.7in]{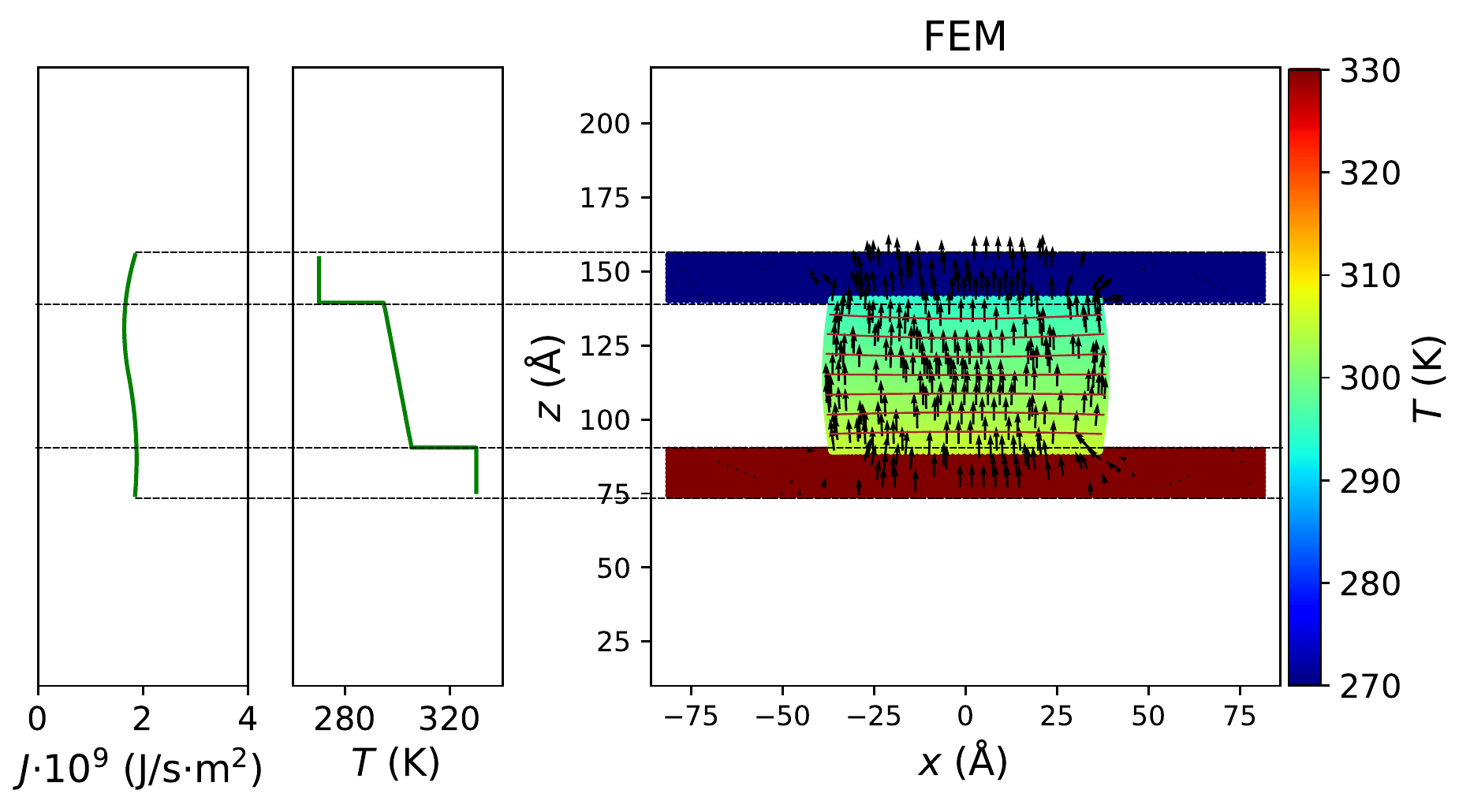}} 
	\subfloat{\includegraphics[width = 3.7in]{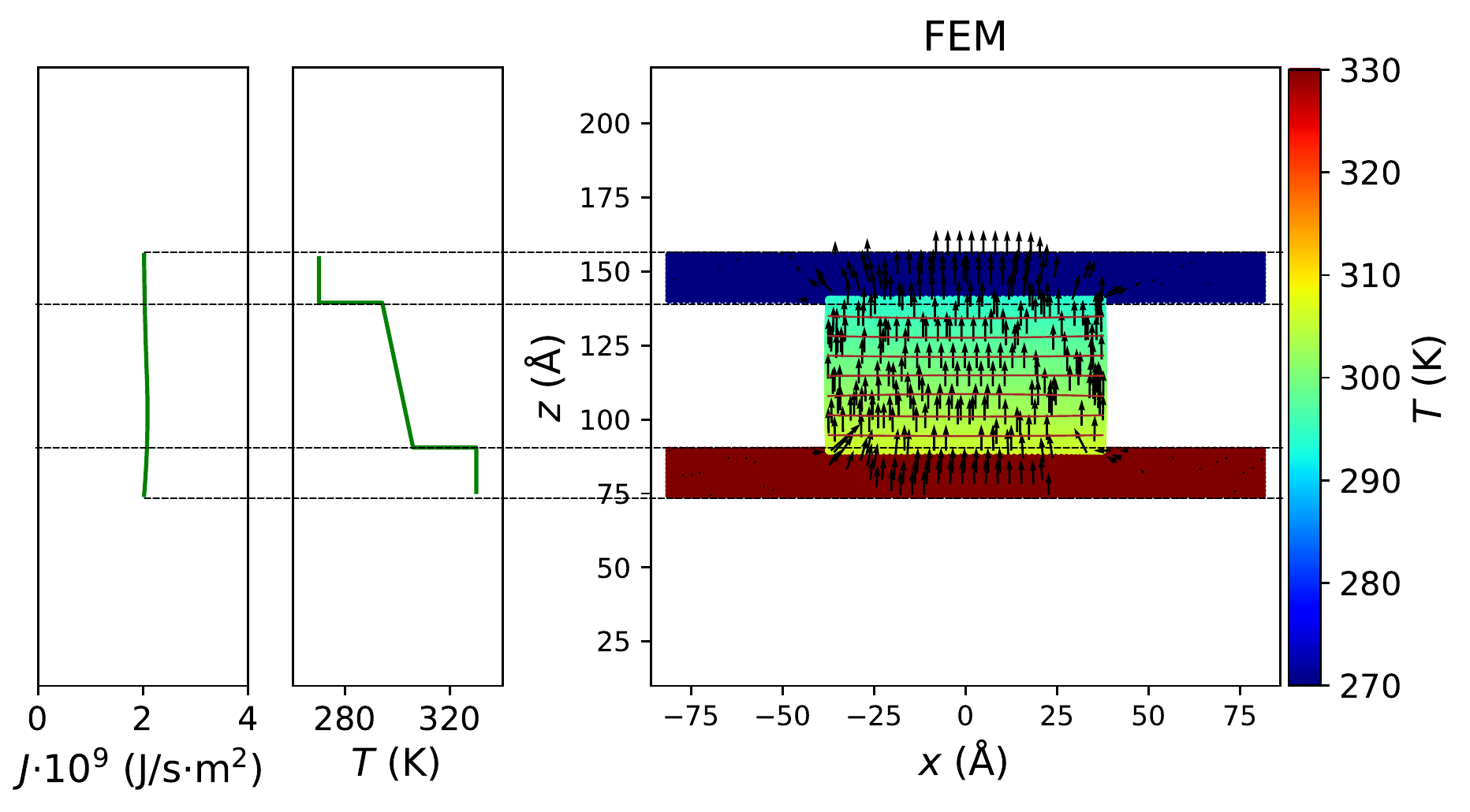}}\\
	\subfloat{\includegraphics[width = 3.7in]{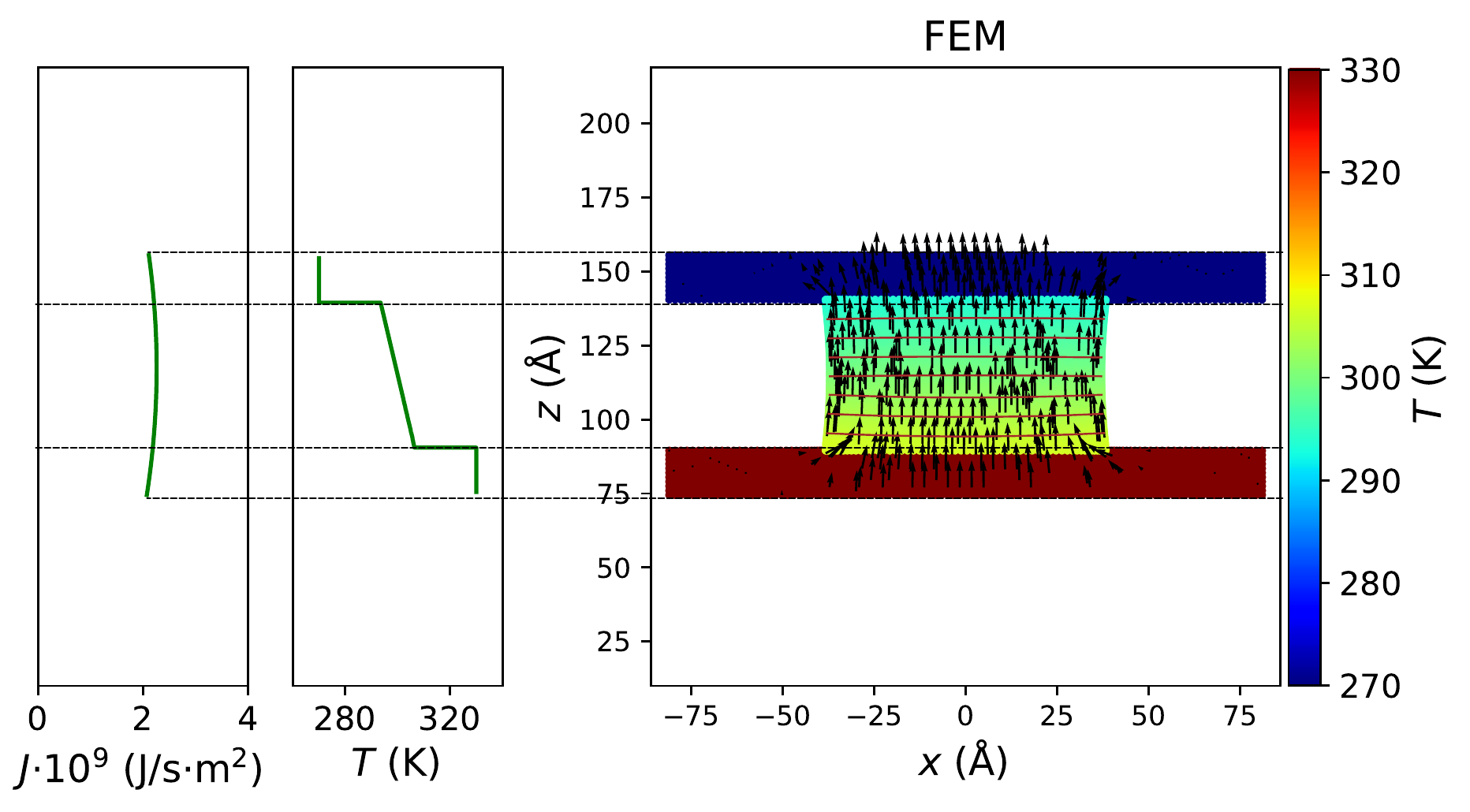}}
	\subfloat{\includegraphics[width = 3.7in]{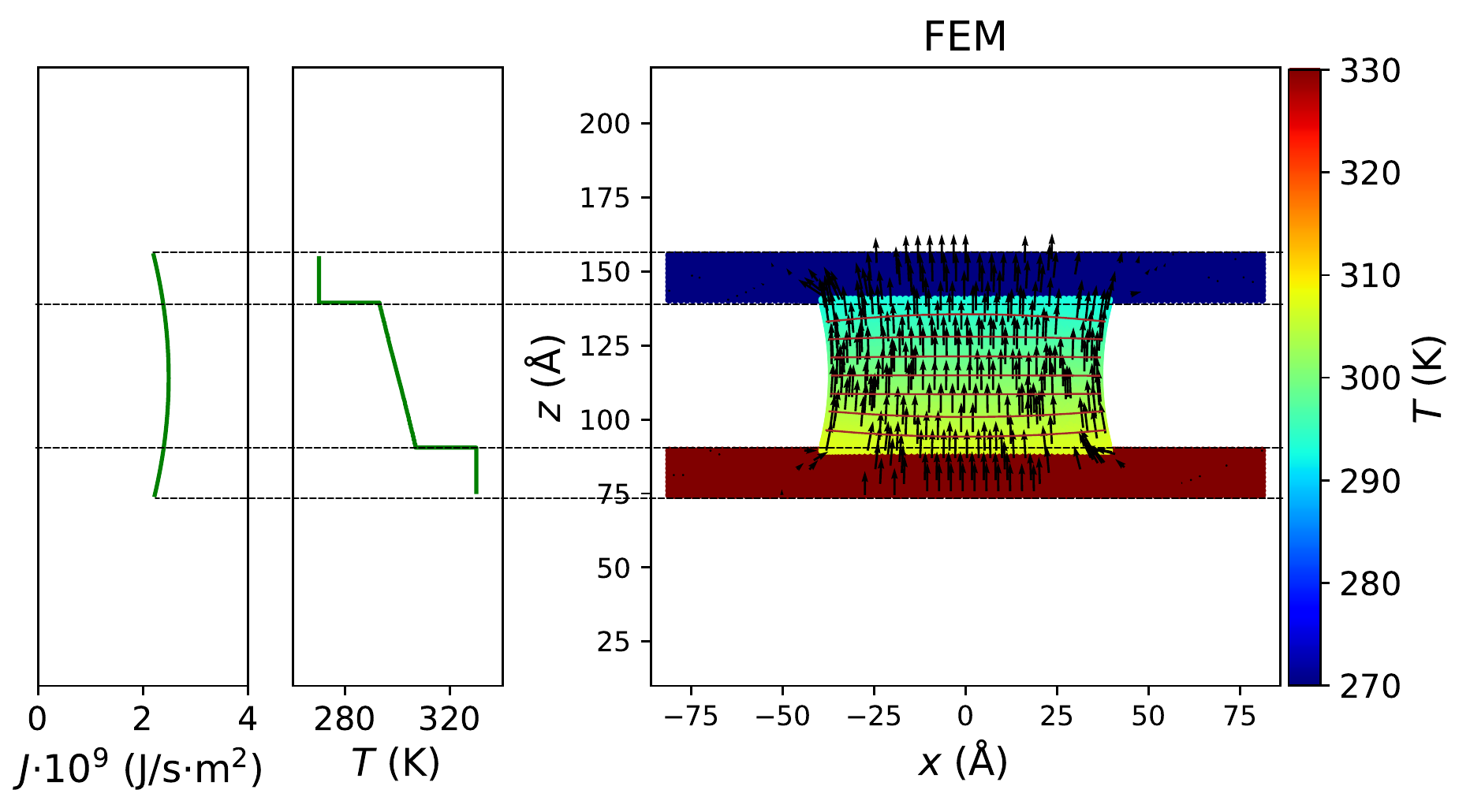}} \\
	\subfloat{\includegraphics[width = 3.7in]{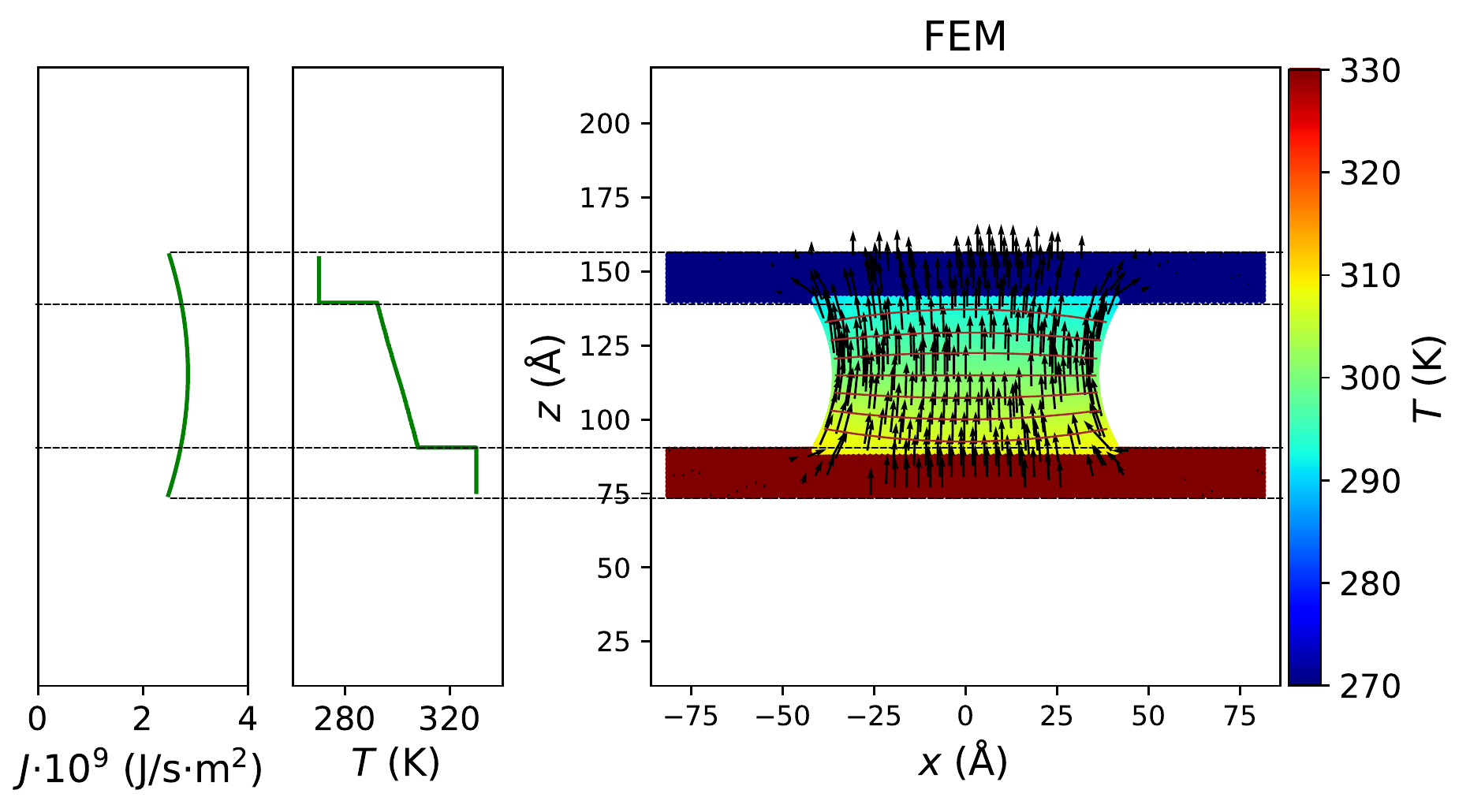}}
	\subfloat{\includegraphics[width = 3.7in]{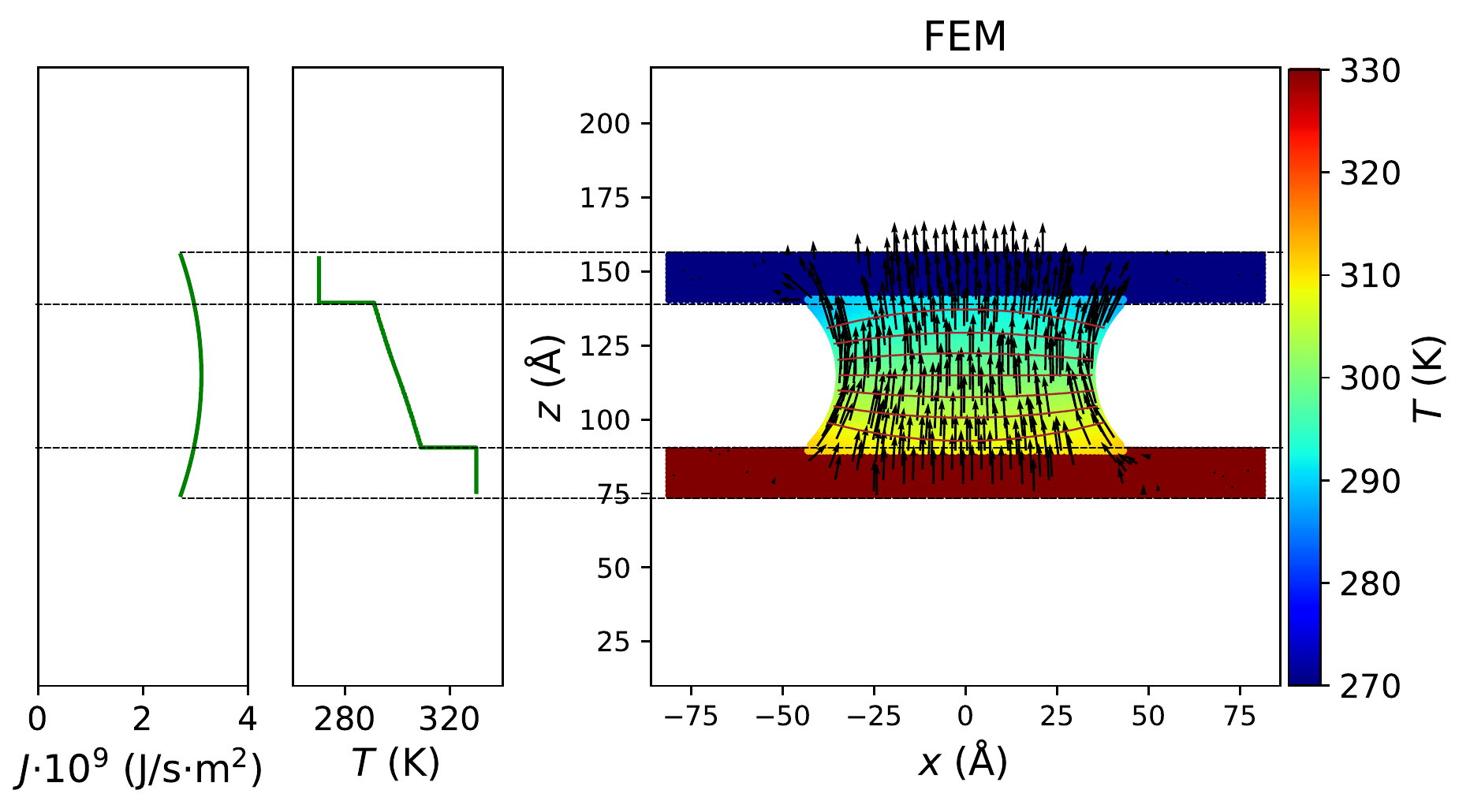}} 
	\caption{\label{fig:FEM_SM_14} Temperature profiles of the meniscus with size $14\text{a}_0$, ($\varepsilon$ = 10 - 21 meV) with heat fluxes distributions obtained from FEM simulations.}
	
\end{figure*}

The right part of Fig.~\ref{fig:4} shows the temperature profile in the water meniscus at $\varepsilon =10,16,21$ meV with heat flux distribution represented by black arrows, and the colorbar displays the temperature range. Those arrows are given along $x$ and $z$ axes of the system. 
Results obtained with FEM and MD show similar trends (see Figs.~\ref{fig:FEM_SM_6}--Figs.~\ref{fig:FEM_SM_14} in SI). While increasing $\varepsilon$, exchanged heat flux rises and becomes more structured. Fluxes obtained with MD exhibit more fluctuations than the one obtained with FEM due to the inherent variations of MD method. Averaging over more than one simulation reduces those fluctuations, improving the contact between hot and cold baths by reducing the ITR, and also gives more clear patterns. As can be seen from the 1D distributions, the temperature inside the meniscus varies linearly, and the temperature difference increases with  $\varepsilon$. For all cases, the iso-temperature lines (red lines) are mostly parallel to the interface.

The temperature jump at the interface as well as total exchanged power as function of $\varepsilon$ calculated using FEM are presented in Fig.~\ref{fig:3} for the meniscus with the initial size equals to $10a_0$. As one can see, there is an excellent agreement between the results obtained with two different approaches.


\section{Conclusions}

To conclude, inspired by the impact of the three-phase contact line between solid, liquid and gas on the thermal transport at the nanoscale, we investigated features of thermal transport across the liquid meniscus constrained between two solid walls.
Firstly, we considered morphology of the capillary interface for different interaction strengths between atoms of solid and liquid and for different meniscus sizes. An analytic model allows us to describe the shape of the meniscus with respect of wetting angle and volume of the liquid.

Further, linear trend of the exchanegd power across the meniscus to maintain temperature difference between the walls as function of interactional potential depth ($\varepsilon$) was observed for a considered range of wetting angles varrying between hydrophobic and hydrophilic states. We found that the temperature jump at the solid/liquid interface is decreasing with increasing of $\varepsilon$. 

Additionally, it was stated that a significant difference exists between absolute thermal resistance of the smallest meniscus and confined water layer for the hydrophobic case. The latter becomes small while the system is reaching the hydrophilic state. With increasing of the meniscus size the system behavior goes to the same as for the confined water layer.

Regarding the interfacial thermal resistance (ITR) between solid/liquid interface, it was found that for the meniscus, the value of ITR shows  the same behavior as for the confined water layer -- it decreases with increasing of $\varepsilon$. However, the ITR for the meniscus is systematically lower then one for the confined water layer. Furthermore, the size dependence of ITR for all meniscus sizes is almost negligible. It may appear because of arising of additional possible channel for heat transfer due to capillary nature of liquid/vapour interface.

Finally, the ITR and wetting angle values evaluated with MD approach were used to combine atomistic simulation with FEM. It was shown that with such parametrization the results obtained with both approaches are consistent. This gives us the possibility  to link simulations at different scales.

\vspace{10.00mm}

\textbf{Acknowledgements} \par 
\vspace{5.00mm} 
This paper contains the results obtained in the frames of the project ``Hotline'' ANR-19-CE09-0003. This work was performed using HPC resources from GENCI-TGCC and GENCI-IDRIS (Grant 2020-A0080907186), in addition HPC resources were partially provided by the EXPLOR centre hosted by the Universitt\'e de Lorraine. The authors thanks the ERASMUS+ for providing mobility grant for V. Mandrolko.

\medskip
\bibliography{small_mi_vertion.bbl}

\clearpage
\newpage

\medskip
\textbf{Supporting Information} \par 
\setcounter{section}{0}
\section{The model and simulation approaches }
\label{app:A}
\subsection{Molecular dynamics calculation}
\label{subsec:MD_procedure}
Simulation domain has size $30\text{a}_0 \times 10\text{a}_0 \times 42 \text{a}_0$ in $x$,$y$, and $z$ directions respectively, where $\text{a}_0=0.543$ nm is the lattice parameter for Si. Both systems consist of two silicon slabs with a slab of water. For the case of confined water (case i), the slab of water covers all silicon surface, and for the meniscus (case ii) it covers only the central part. All the systems consist of $67800$ atoms of Si in total, but with variety of water molecules: $15028$ for confined water, $3179$ for meniscus with size $6\text{a}_0$, $5201$ for meniscus with size $10\text{a}_0$, and $7225$ for meniscus with size $14\text{a}_0$. The distance between two silicon slabs is set as $10a_0$.

\begin{figure*}[h!]
	\centering
	\includegraphics[width = 3.5in]{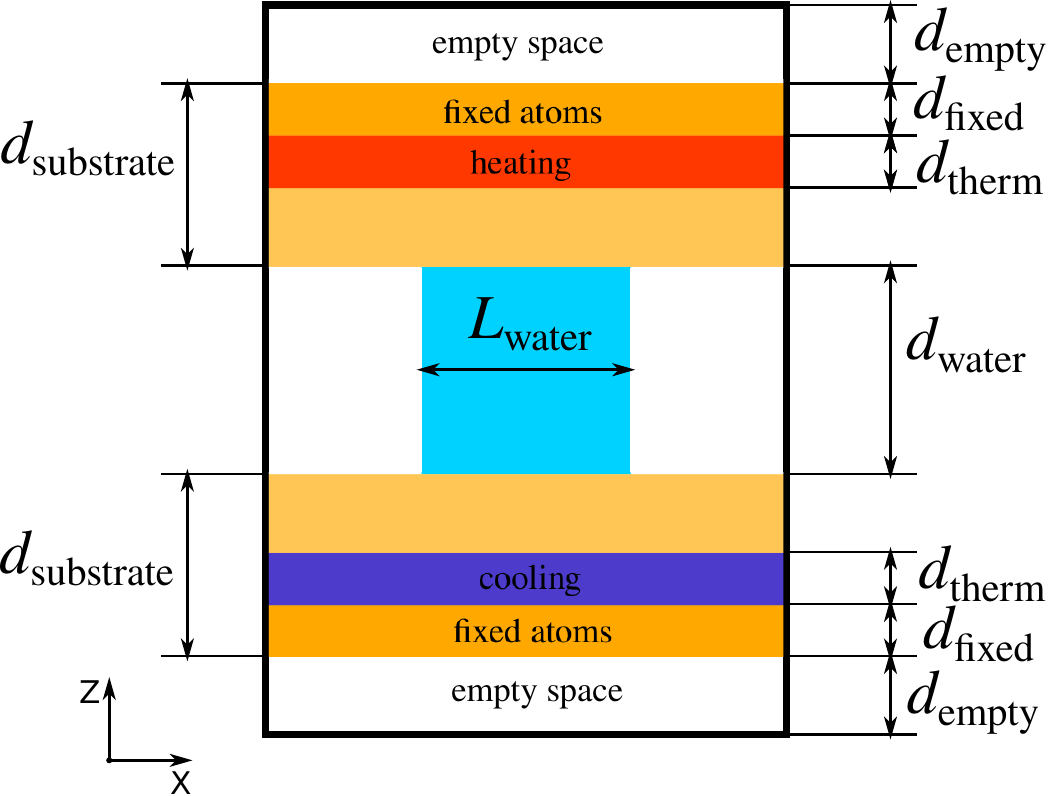}
	\caption{\label{fig:sketch} Representation of the initial systems. Here $L_{\text{water}}$ is the size of the initial systems cases i) $30a_0$, and ii) $6a_0$, $10a_0$,$14a_0$, $d_{\text{water}}=10a_0$, $d_{\text{empty}}=2a_0$,  $d_{\text{fixed}}=2a_0$, $d_{\text{substrate}}=14a_0$, $d_{\text{therm}}=3a_0$.}
\end{figure*}

Silicon was simulated as diamond lattice with lattice parameters of $0.543$ nm and its surface is in contact with water with $(0,0,1)$ crystal plane. The interaction between silicon atoms were modeled via the Stillinger-Weber potential~\cite{Stillinger1985}. Atoms in the bottom and top of silicon slabs ($1.086$ $<$ $z_1<$ $2.172$ nm, $20.634$  $<z_2<21.72$ nm) were fixed with string forces to maintain their position close to the initial. The simple sketch that shows how input parameters are arranged compared to the system is showm in Fig.~\ref{fig:sketch}.

Water was simulated in frame of extended simple point charge model, $SPC/E$~\cite{Orsi2013}. The interaction potential was determined by a combination of the Lennard-Jones, and the Coulombic potentials with cutoff distances $r_c=1.0$ nm for both of them. For long-range Coulomb forces, each atom is assigned a charge such as $q(H)=+0.4238 e$, and $q(O)=-0.8476 e$. Usage of selected parameters give us a neutral charge for water molecules.
To keep the O-H bond length, $l=0.096$ nm, and H-O-H angles, $109.46^{\circ}$, sustainable the SHAKE algorithm was applied. For O-O interaction the values of $\varepsilon = 6.736$ meV, and $\sigma$ = 0.3166 nm  were chosen. In case of  O-Si interaction, epsilon varied within 10 – 21 meV, and $\sigma$ = 0.26305 nm. 

Firstly, the initial system was tempered at temperature $T=1$ K by using  Gaussian distribution for velocity field. Then the system evolved in the $NVE$ ensemble with a gradual temperature rescaling up to 300 K for 150 ps. (Temperature rescaling to 10 K was applied for 50 ps, to 50 K for 50 ps and finally to 300 K for another 50 ps).
Further the systems were equilibrated in $NVT$ ensemble for 150 ps. 

After equilibration, temperature rescales to 270 K and 330 K were applied to the regions 2.172 < $z$ < 3.801 nm  and 19.005 < $z$ < 20.634 nm  respectively.  
Then the simulation was performed for 1 ns. The values of temperature, density, potential and kinetic energy of particles and stress tensors were obtained. The obtained values are averaged from all values calculated every 0.01 ps. The binning of the system was 0.2 nm. In the region with water, calculations were performed with binning 0.1 nm for more precise calculations of water density. 

\subsection{Usage of FEM procedure}
\label{subsec:FEM_procedure}

In order to analyze system by using finite element method (FEM) we performed simulation in  COMSOL\textsuperscript{\tiny\textregistered} software in steady-state mode with convergence criteria = 0.001 (or relative tolerance). All the boundaries that confined the droplet and  silicon slabs were thermally insulated (except those to which a temperature gradient from 270 to 330 K was applied). Within simulations a triangular grid with maximum resolution (or ``extremely fine'' resolution) was constructed. The estimated time that we spend for one seed is about $t=2\sim5$ sec.

\section{The droplet shape derivation}
\label{app:B}

\begin{figure*}[h!]
	\centering
	\includegraphics[width = 3.3in]{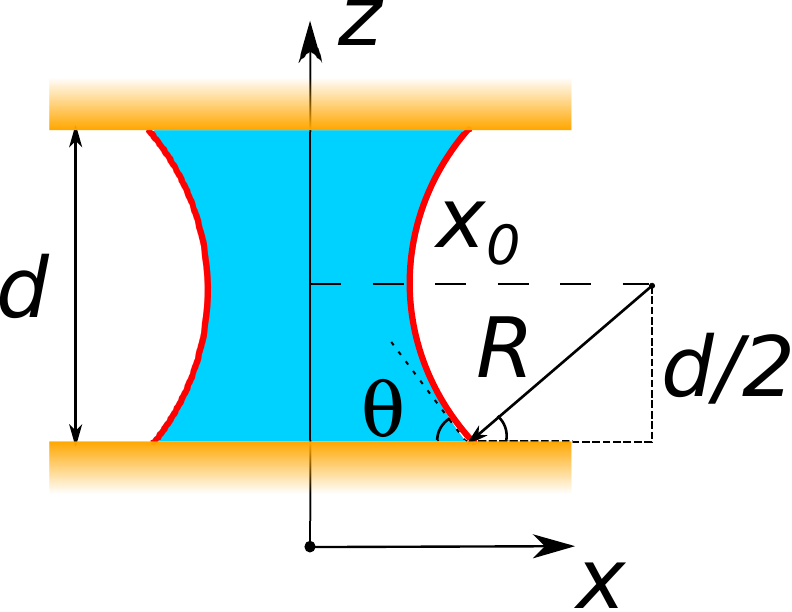}
	\caption{\label{fig:sc} A simple sketch that shows connection between curves corresponding to the boundaries of the droplets and the system, that we used to obtain curves in Figs.~\ref{fig:1},~\ref{fig:8}-~\ref{fig:8_2}. The input parameters that are shown here are well described in equation above.}
\end{figure*}

Curves corresponding to the boundaries of the droplets were obtained based on the circle approximation (see Fig.~\ref{fig:sc}):
\begin{equation*}
(x-x_0 )^2+(z-z_0 )^2=R^{2}_\text{circle},
\end{equation*}

where $x_0$ and $z_0$ – coordinates of the center of the circle.
If $d$ is the distance between silicon slabs, and $\phi$ is the wetting angle:

\begin{equation*}
\begin{aligned}
&d/2=R_\text{circle} \cos\left(\phi\right), \\
&(x-x_0 )^2+(z-z_0 )^2 = R^{2}_\text{circle},\\
&(x-x_0 )^2+(z-z_0 )^2 = d^2/(4 \cos^2\left(\phi\right)), \\
&x = x_0 \pm ( d^2/(4 cos^2\left(\phi\right)-(z-z_0 )^2 )^{1/2}, \\
\end{aligned}    
\end{equation*}

where the roots with ``$\pm$'' and sign corresponds to a hydrophobic (+) and hydrophilic (-) cases respectively.
$x_0$ was calculated based on the fact that the volume $v$ of the droplet for all cases remains constant.

We calculated the volume of the droplet based on the number of water molecules and its density.

\begin{equation*}
\frac{v}{2b} = \int_0^{x_0 \pm ( d^2/(4 \cos^2\left(\phi\right) )-(z-z_0 )^2 )^{1/2}} \int_{-d/2}^{d/2}dxdz, 
\end{equation*}

$b$ is the thickness of droplet which is not used for curve fitting because we are considering a 2D system.
After calculating the integral, we obtain:
\begin{equation*}
\begin {aligned}
\frac{v}{2b} = x_0d &\pm \frac{1}{8}\Big[\frac{d^2}{\cos^{2}\left(\phi\right)}\cdot\arcsin\left(\frac{\left(
	2z_0+d\right)\cos\left(\phi\right)}{d}\right)-\\
&-\frac{d^2}{\cos^{2}\left(\phi\right)}\cdot\arcsin\left(\frac{\left(
	2z_0-d\right)\cos\left(\phi\right)}{d}\right)+\\
&+ \left(d-2z_0\right)\sqrt{4z_0\left(d-z_0\right)+d^2\tan^{2}\left(\phi\right)} + \\
&+ \left(d+2z_0\right)\sqrt{-4z_0\left(d+z_0\right)+d^2\tan^{2}\left(\phi\right)}\Big]
\end{aligned}   
\end{equation*}

Thus, we obtained an analytical equation of the curve bounding the drop based on the distance between the silicon slabs and the wetting angle.

\section{Temperature profiles of the meniscus with heat fluxes distributions}
\label{app:C}

\begin{figure*}
	\centering
	\subfloat{\includegraphics[width=3.7in]{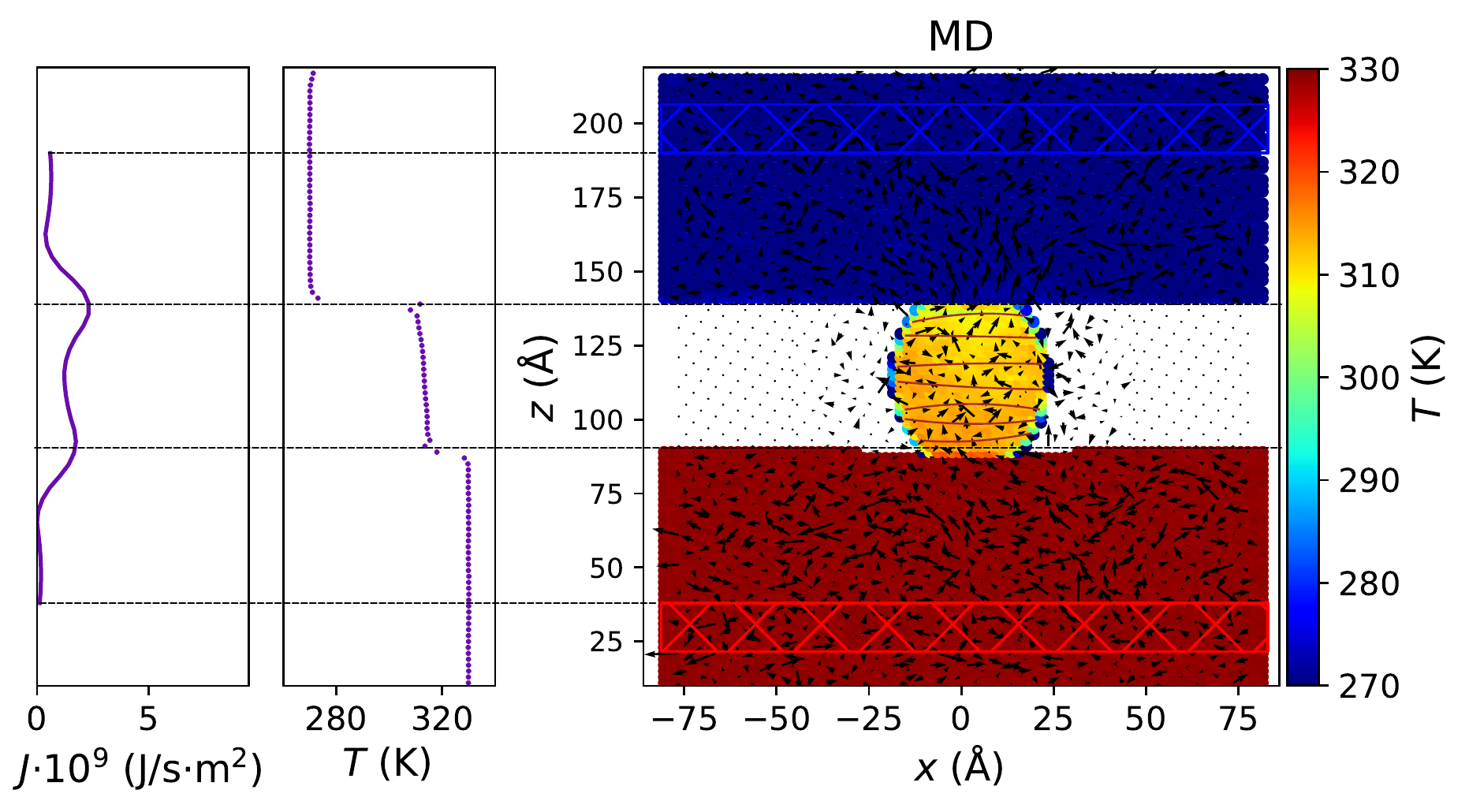}} 
	\subfloat{\includegraphics[width = 3.7in]{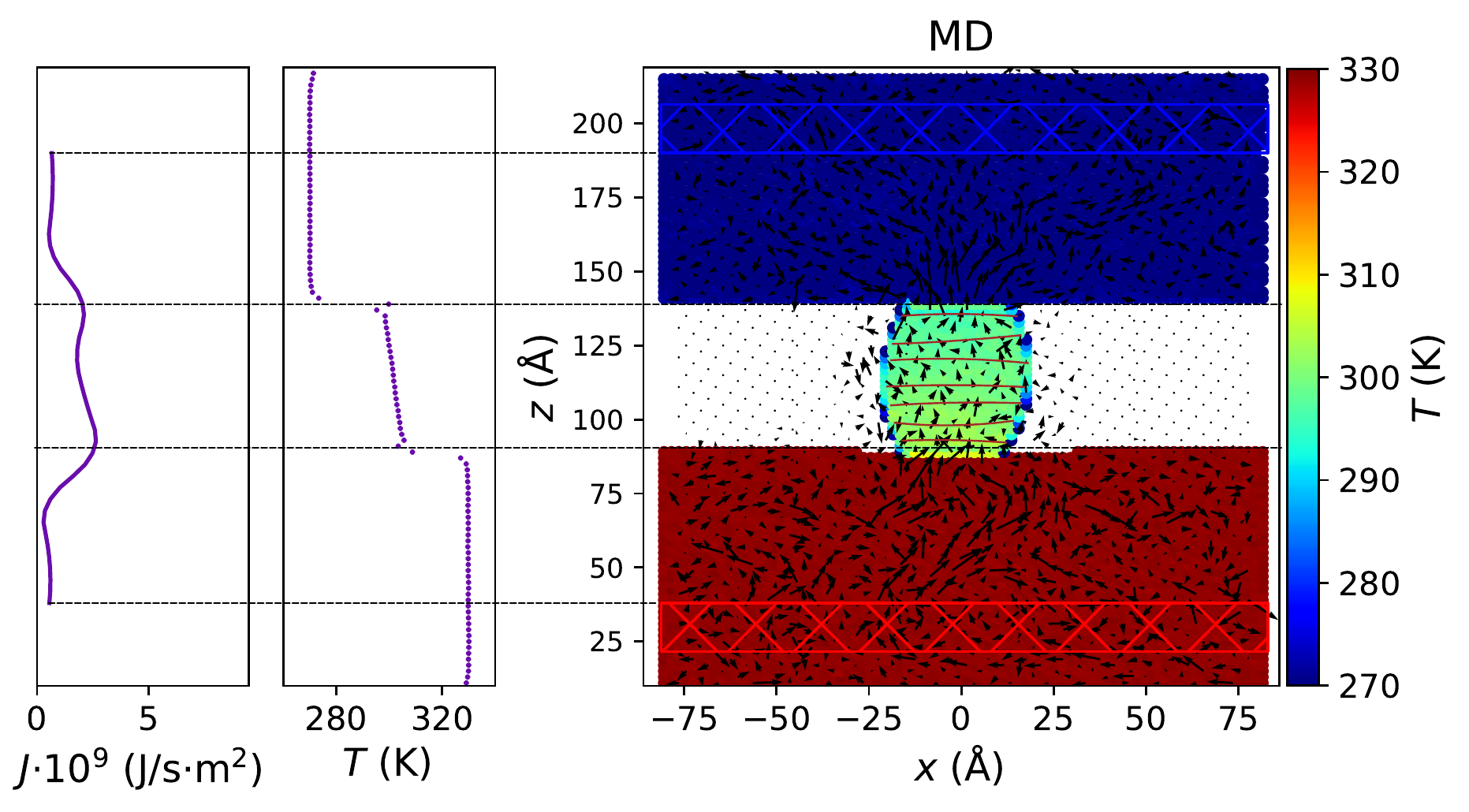}}\\
	\subfloat{\includegraphics[width = 3.7in]{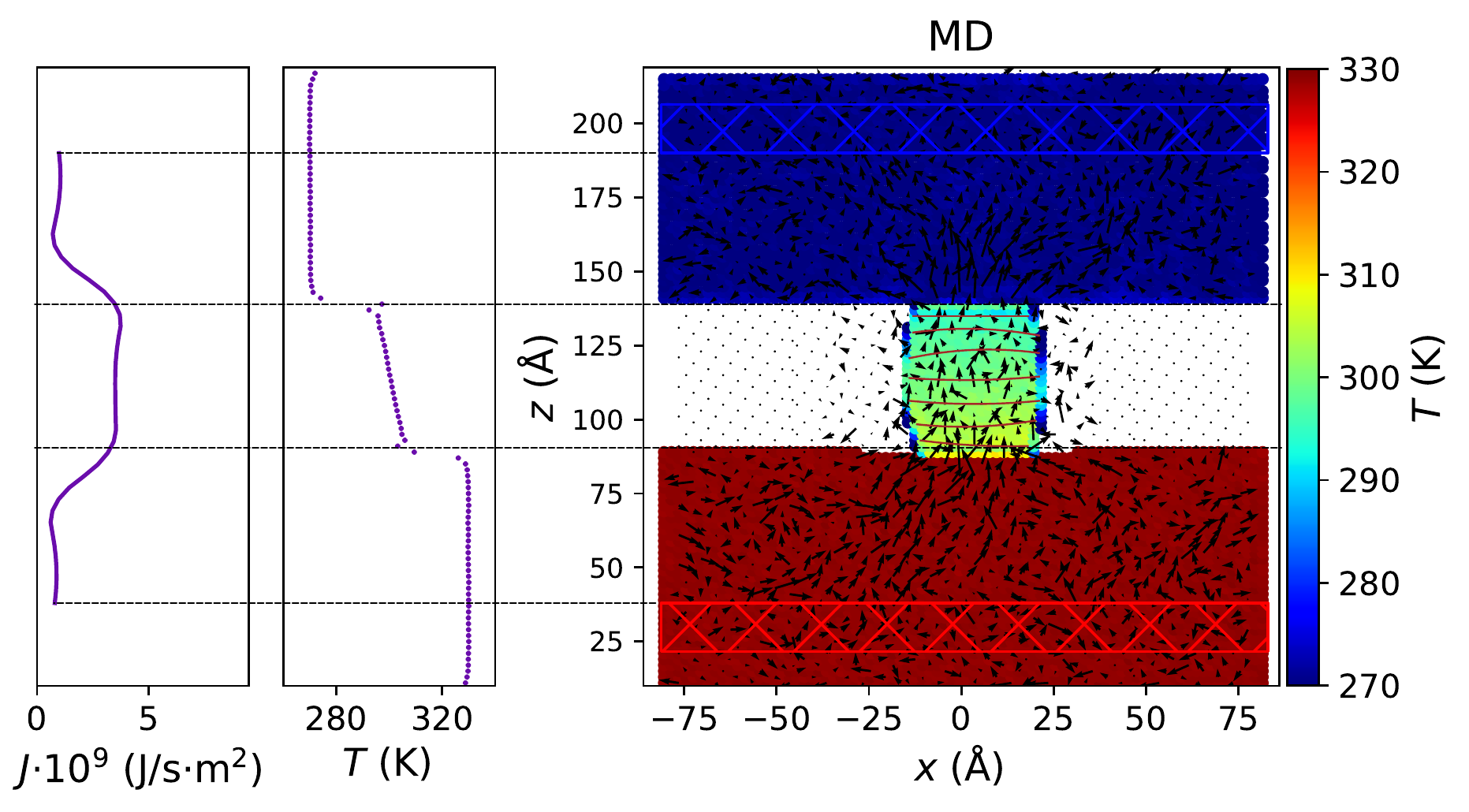}} 
	\subfloat{\includegraphics[width = 3.7in]{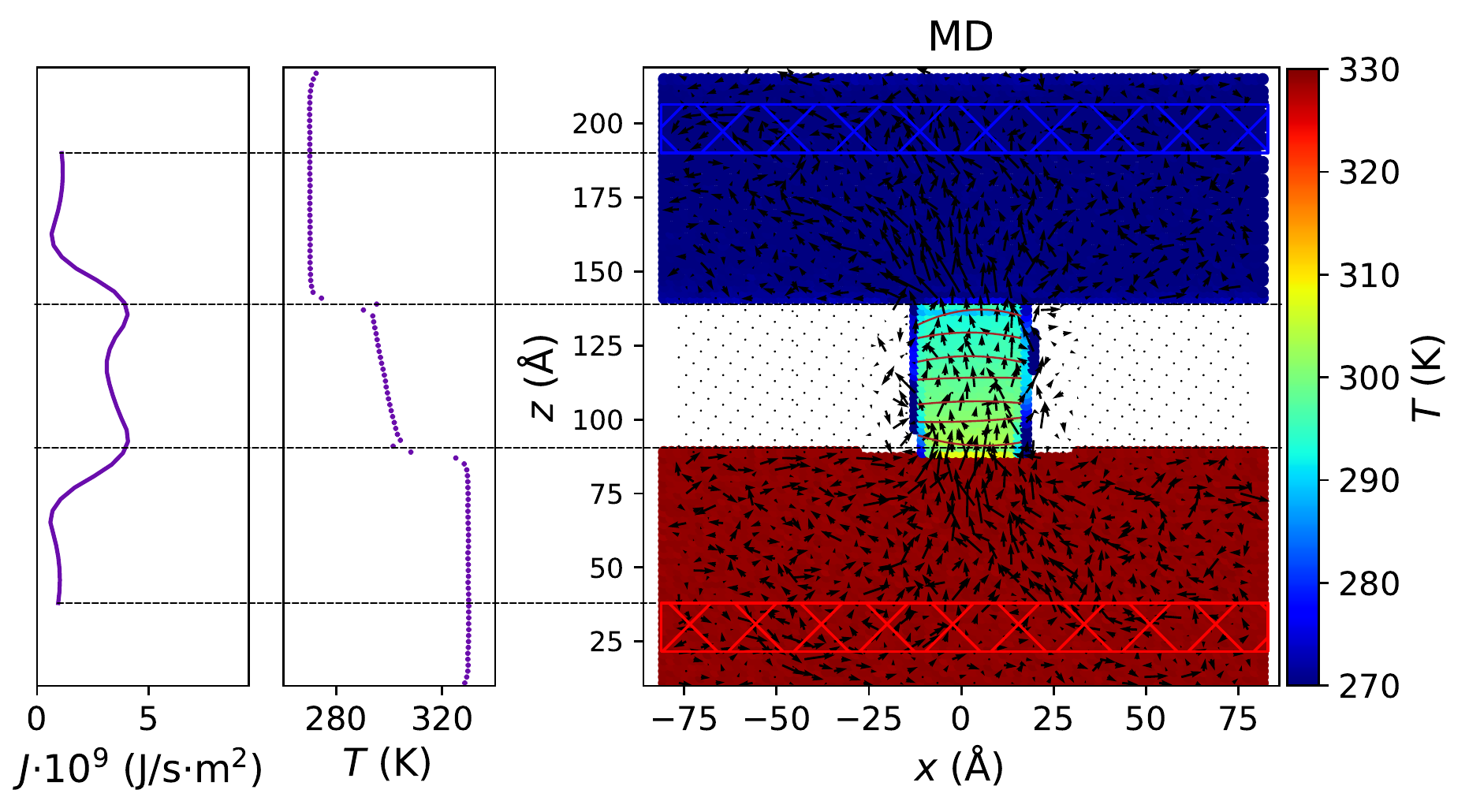}}\\
	\subfloat{\includegraphics[width = 3.7in]{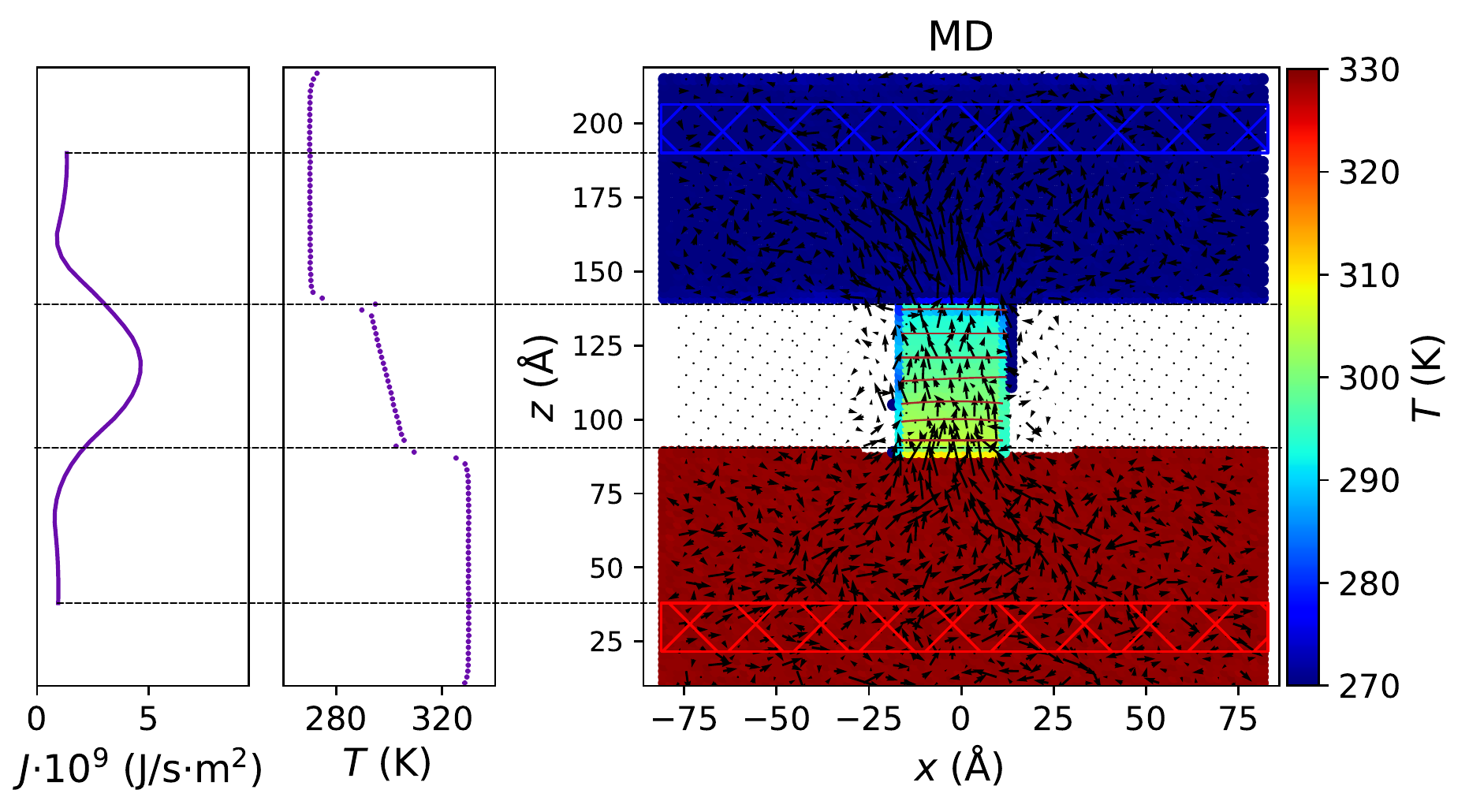}}
	\subfloat{\includegraphics[width = 3.7in]{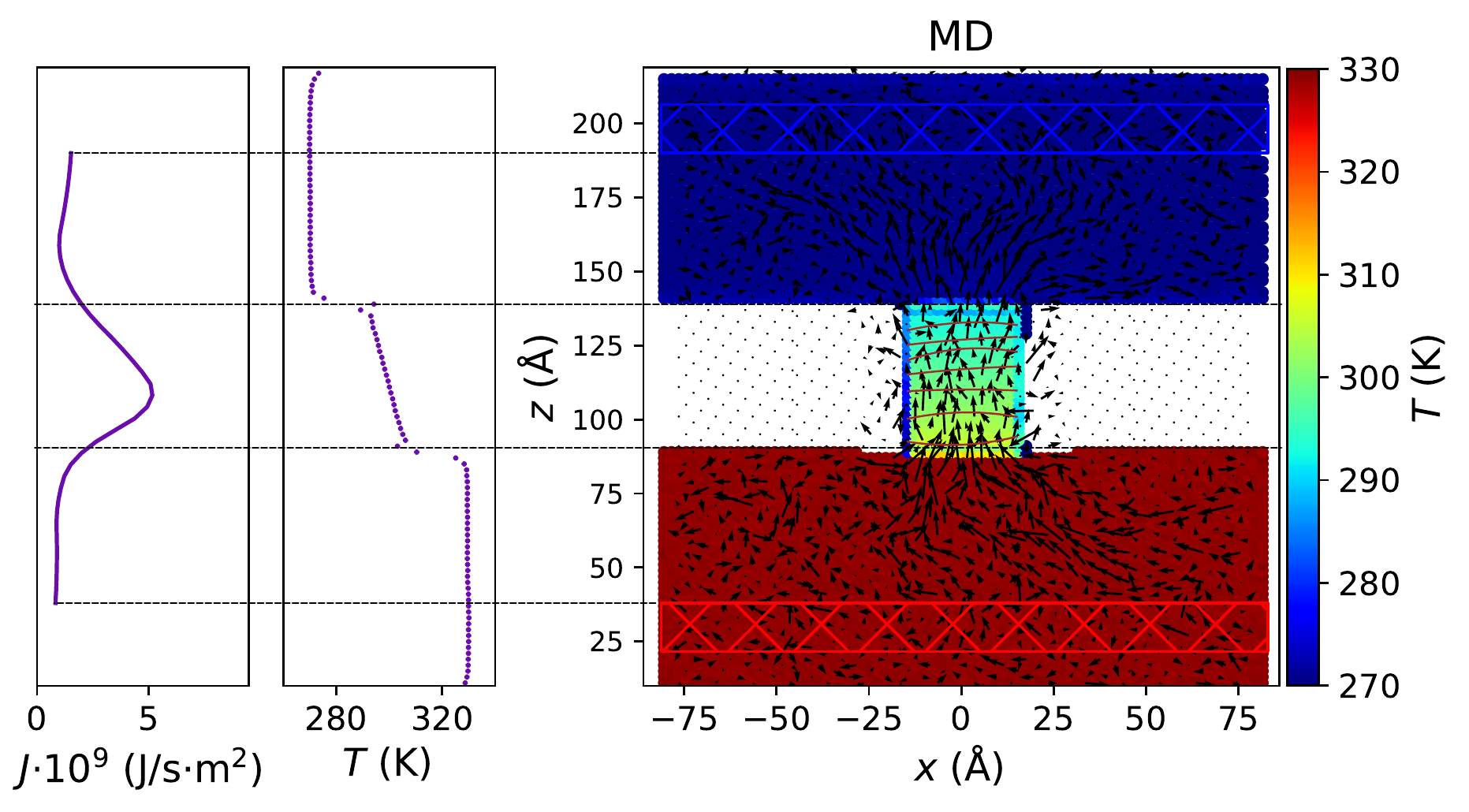}} \\
	\subfloat{\includegraphics[width = 3.7in]{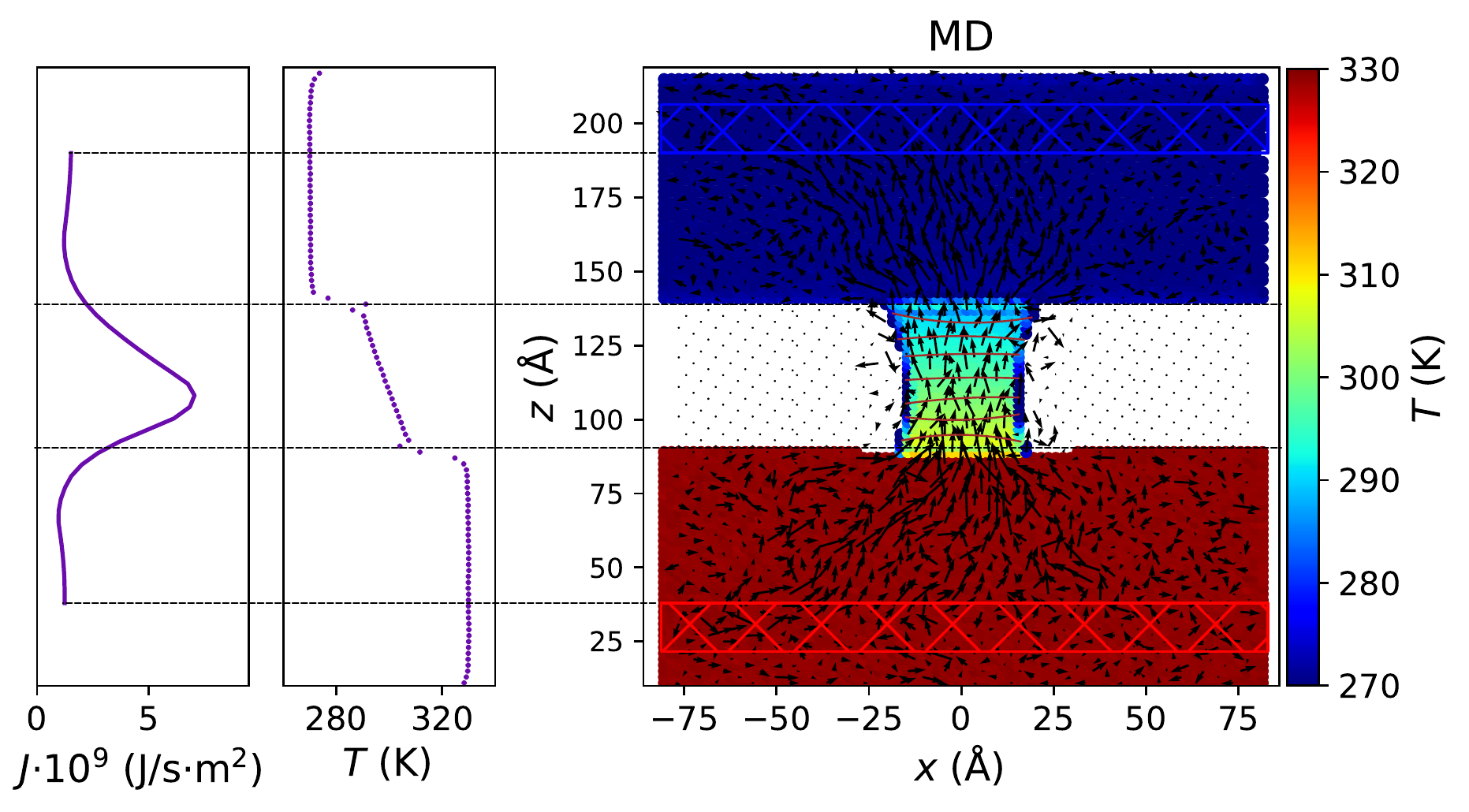}}
	\subfloat{\includegraphics[width = 3.7in]{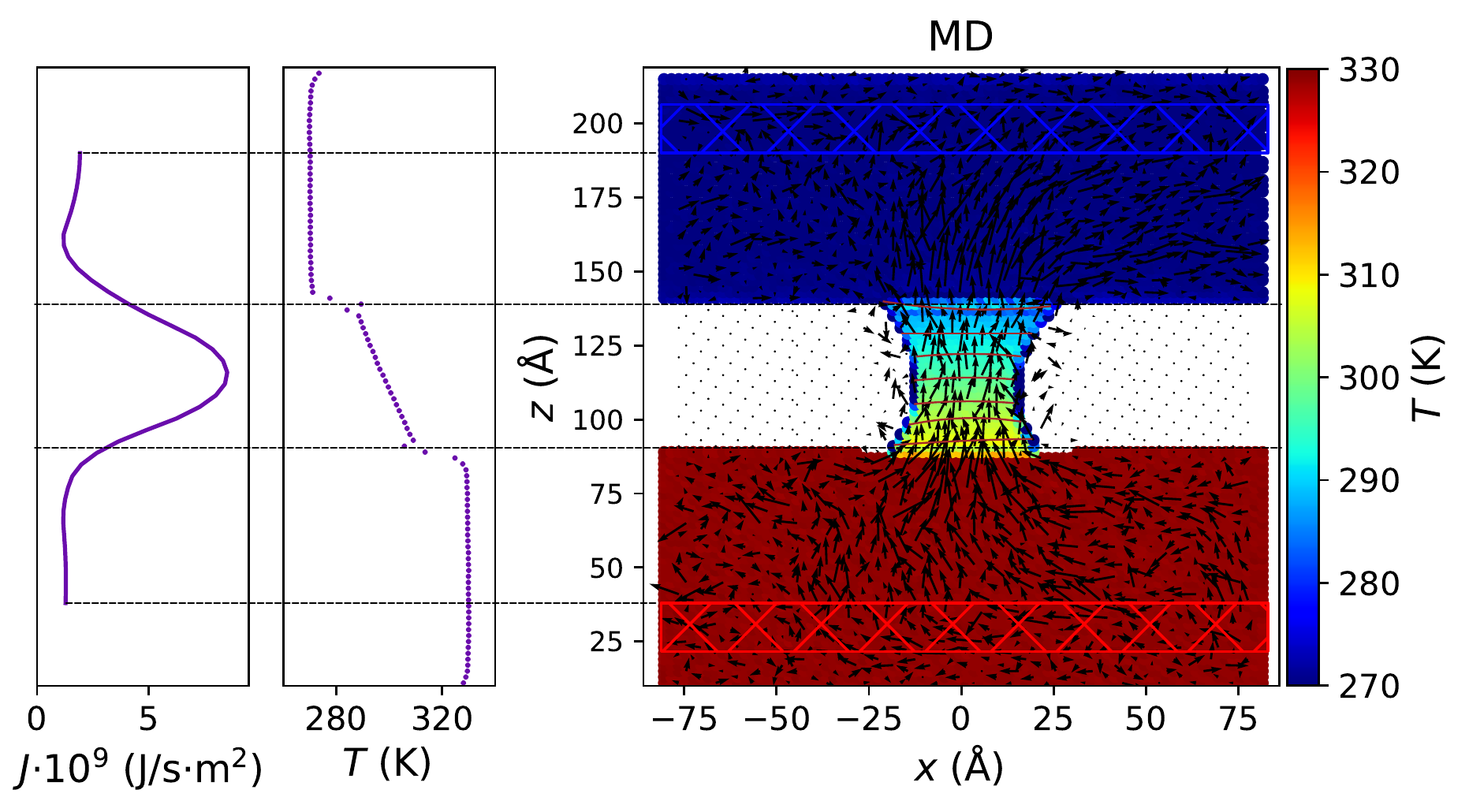}} 
	\caption{\label{fig:6} Temperature profiles of the meniscus with size $6\text{a}_0$, ($\varepsilon$ = 10 - 21 meV) with heat fluxes distributions obtained from MD simulations.}
	
\end{figure*}

The same temperature difference between two silicon slabs (60 K) was applied.  Fig.~\ref{fig:6} shows the temperature profiles with heat fluxes obtained using FEM (left) and MD (right) methods. The central plot shows one-dimensional temperature distributions in the middle of the system obtained from both of these approaches (solid green line represents FEM values and purple dots represents MD values). As it can be seen, the fluxes on the graph for MD are less structured than for FEM. This is expected due to fluctuations of physical quantities that are present in the MD simulation. Distribution for MD was obtained by averaging five simulations, each of them separately gives much more disordered fluxes. Averaging over more simulations would give the more directional fluxes with an increased computational costs.

\begin{figure*}
	\centering
	\includegraphics[width = 7in]{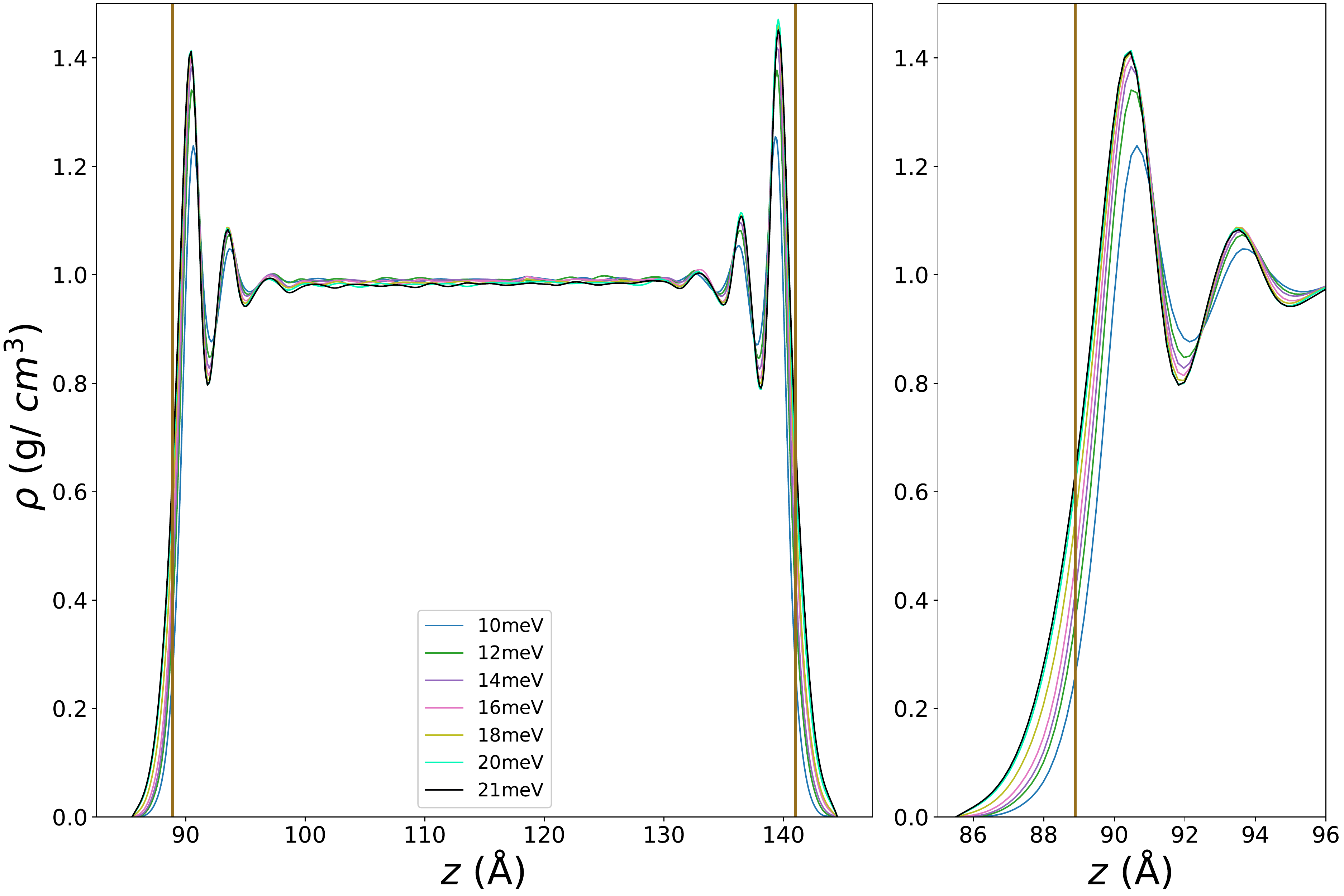}
	\caption{\label{fig:7}Density profile of confined water obtained at different values of $\varepsilon$. }
\end{figure*}

The increase of the density peak  near the surface with increasing $\varepsilon$ is represented in Fig.~\ref{fig:7} as well. Moreover, from Fig.~\ref{fig:7}, it can be seen how water penetrates deeper into silicon for high $\varepsilon$. Increasing the $\varepsilon$ creates a stronger coupling between water and silicon, which increases the depth of water penetration into the silicon lattice, and the density of the surface layer of water, which in turn reduces the mismatch in the density of vibration states and provides better heat transfer.

\begin{figure*}
	\centering
	\subfloat{\includegraphics[width=3.7in]{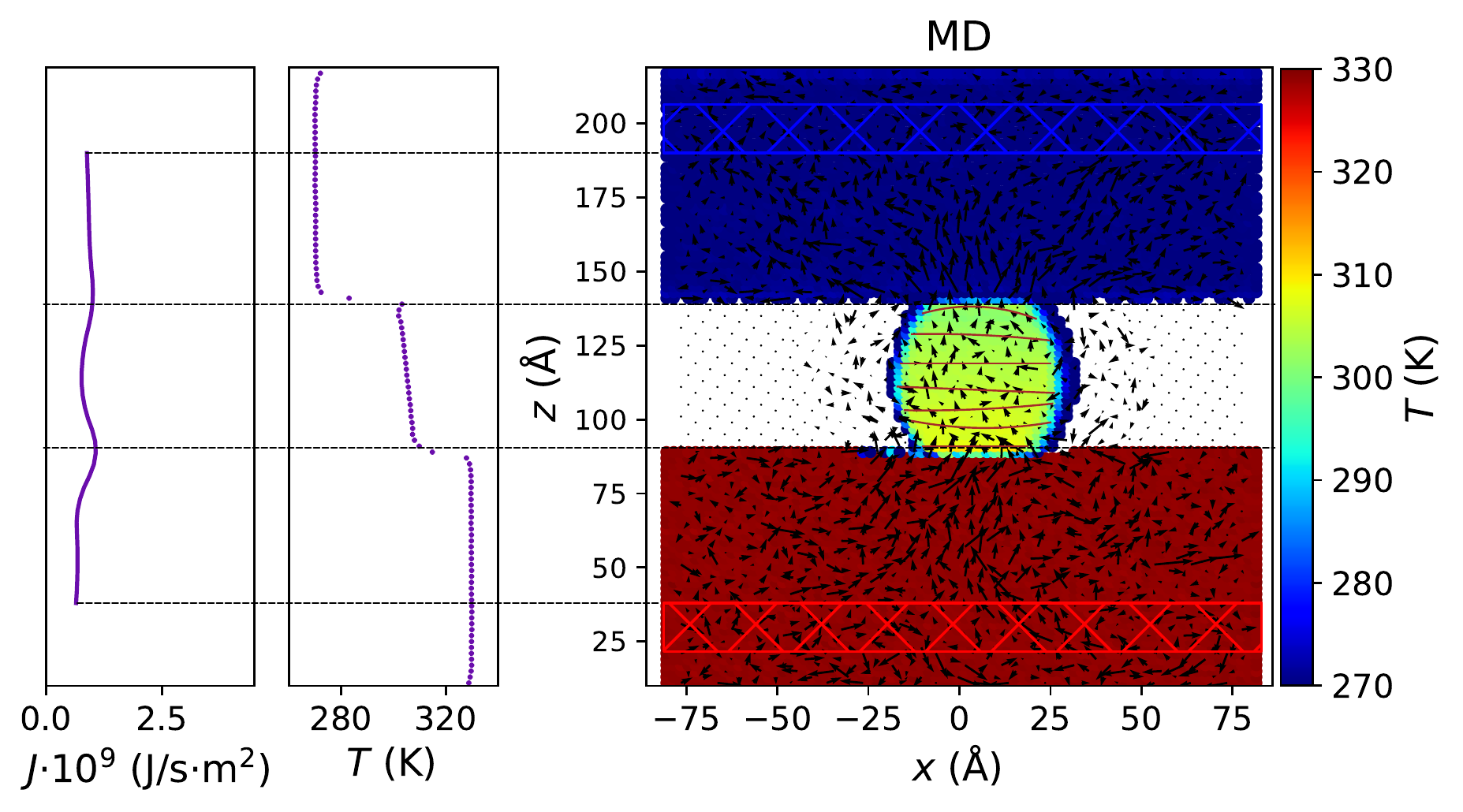}} 
	\subfloat{\includegraphics[width = 3.7in]{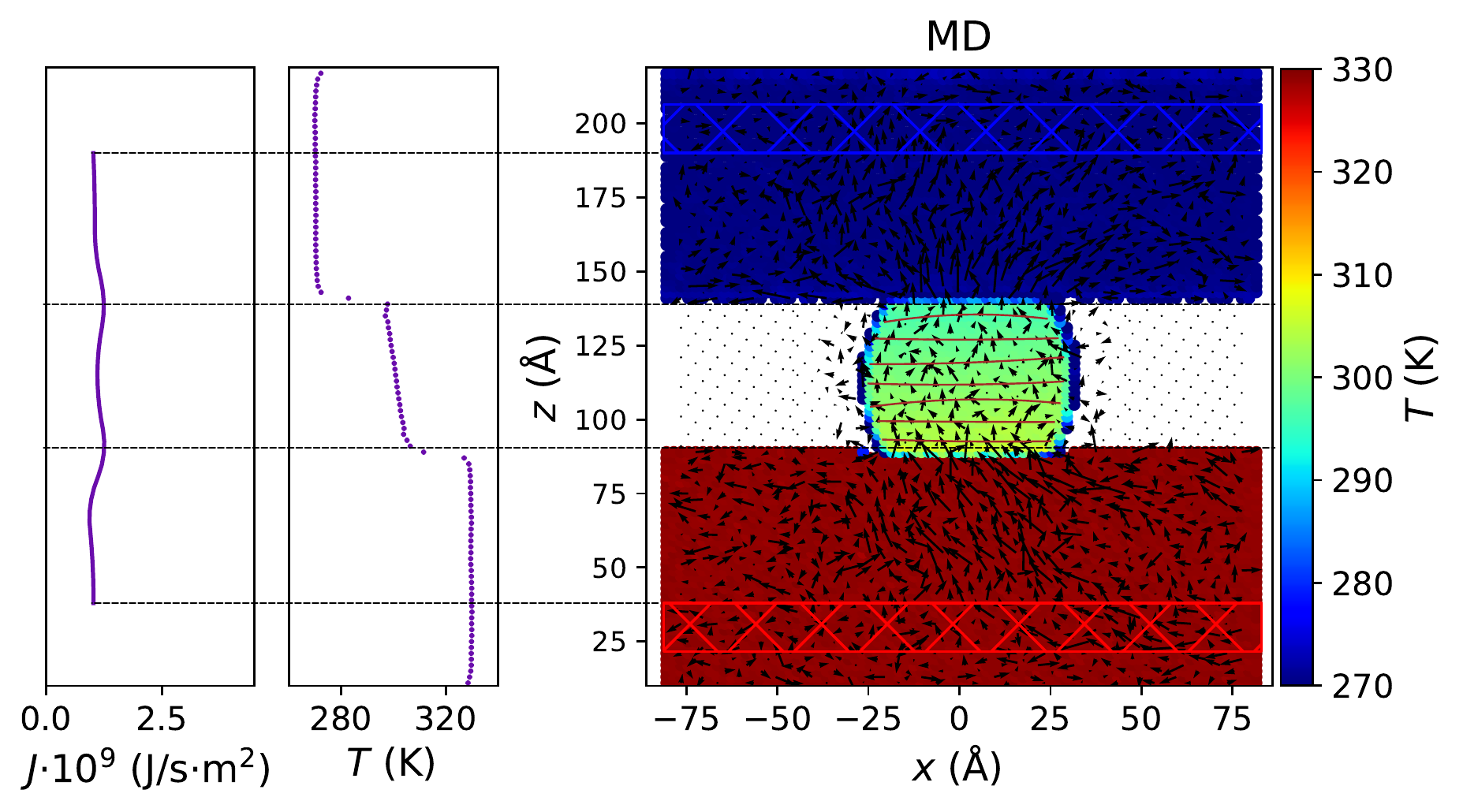}} \\
	\subfloat{\includegraphics[width = 3.7in]{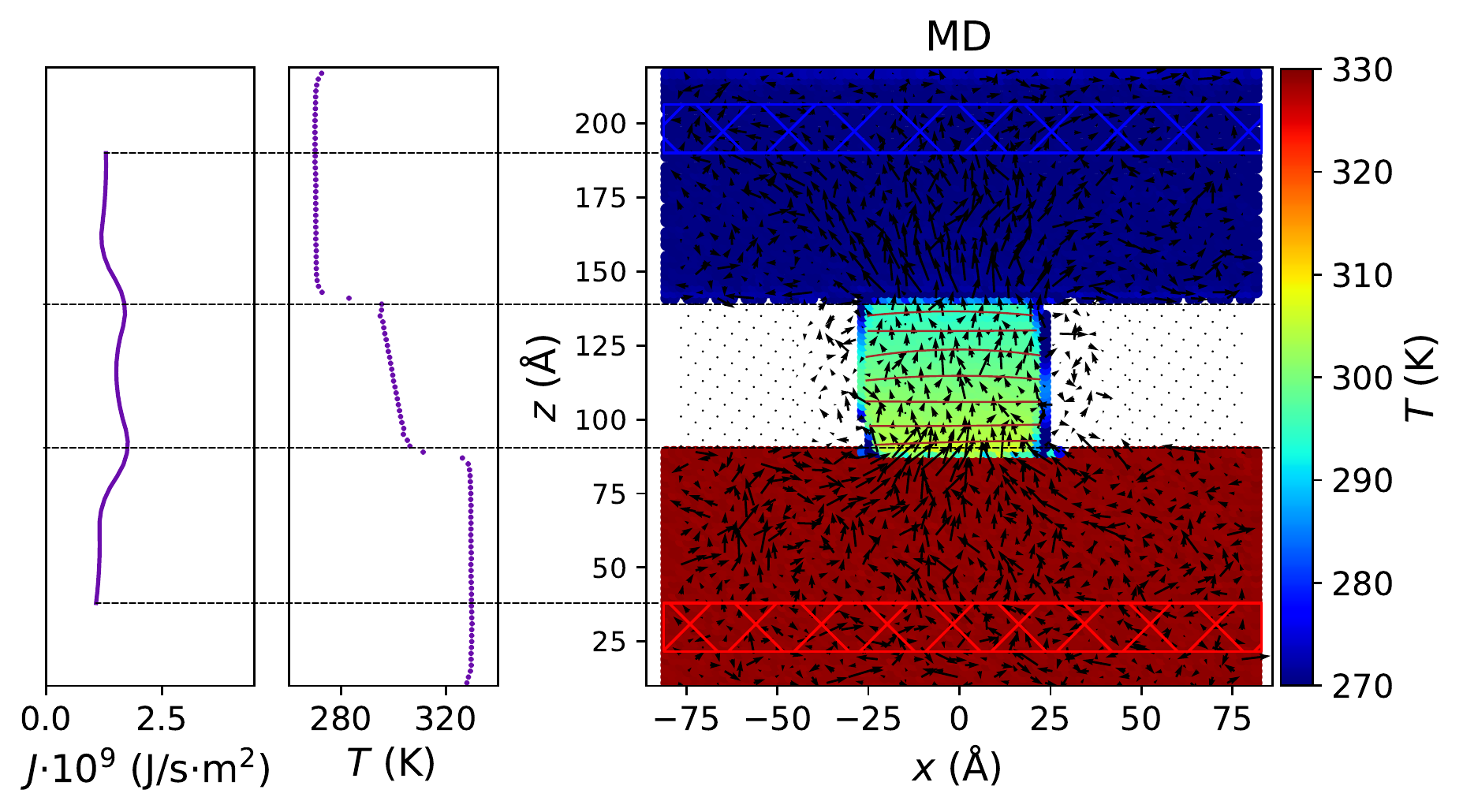}}
	\subfloat{\includegraphics[width = 3.7in]{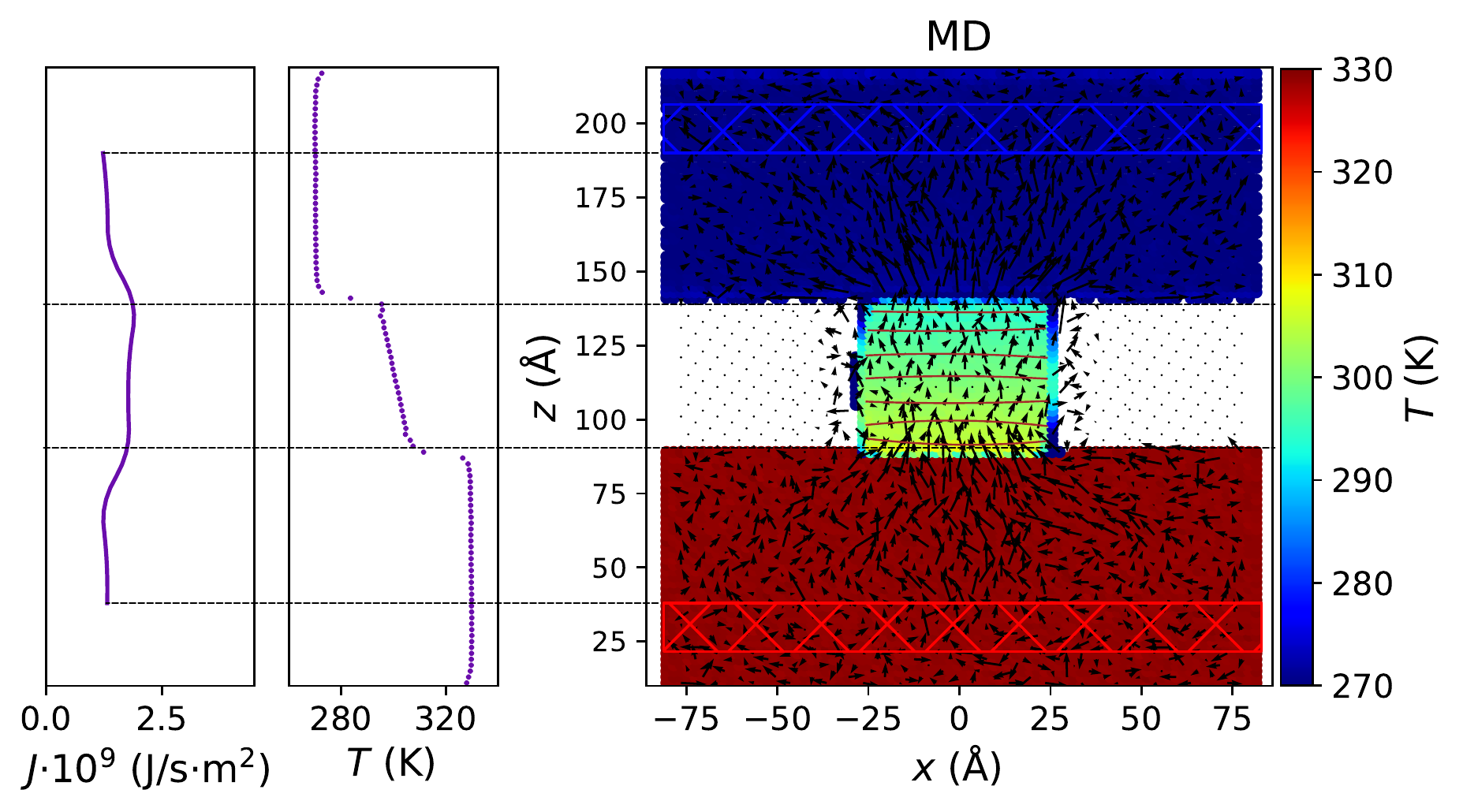}} \\
	\subfloat{\includegraphics[width = 3.7in]{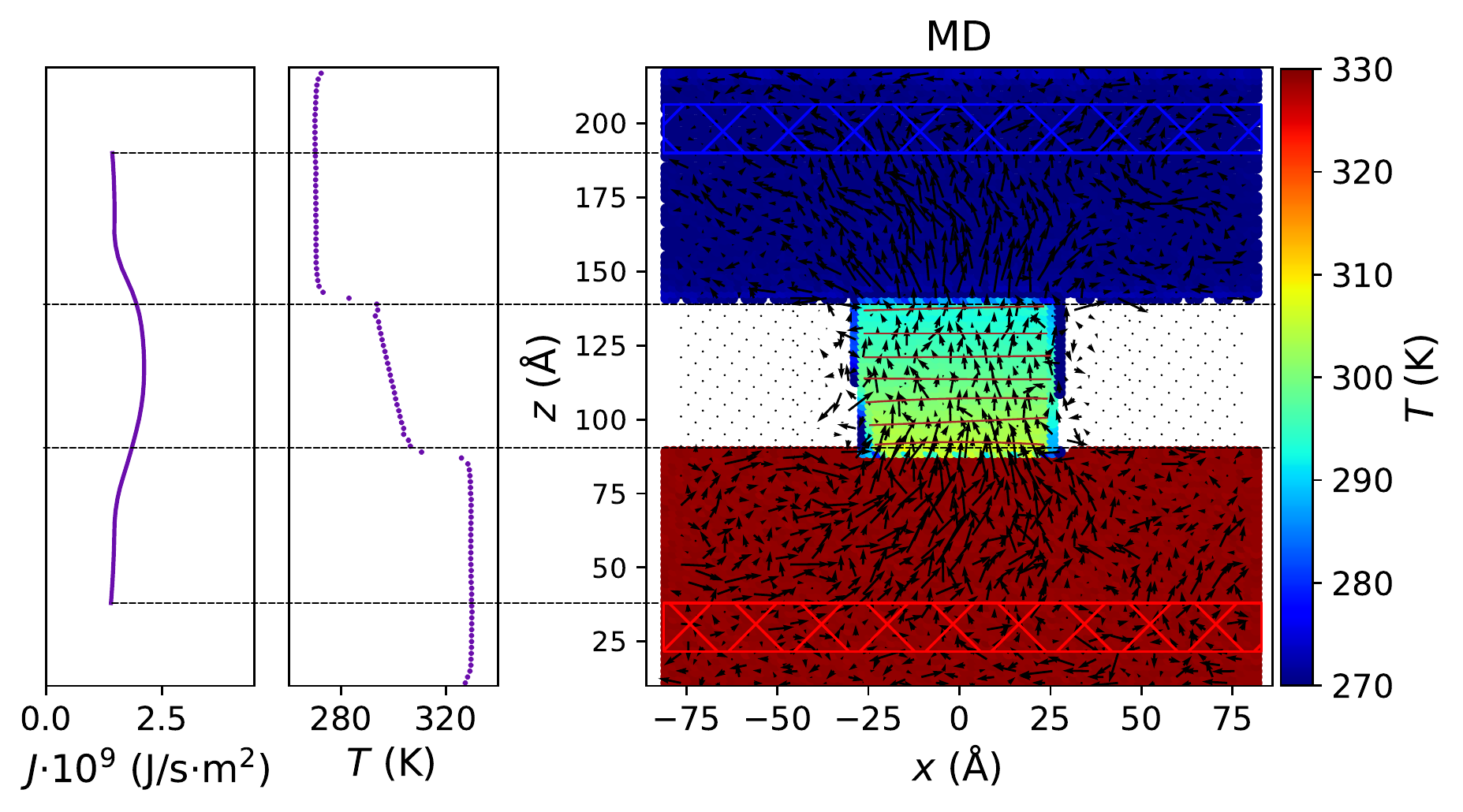}}
	\subfloat{\includegraphics[width = 3.7in]{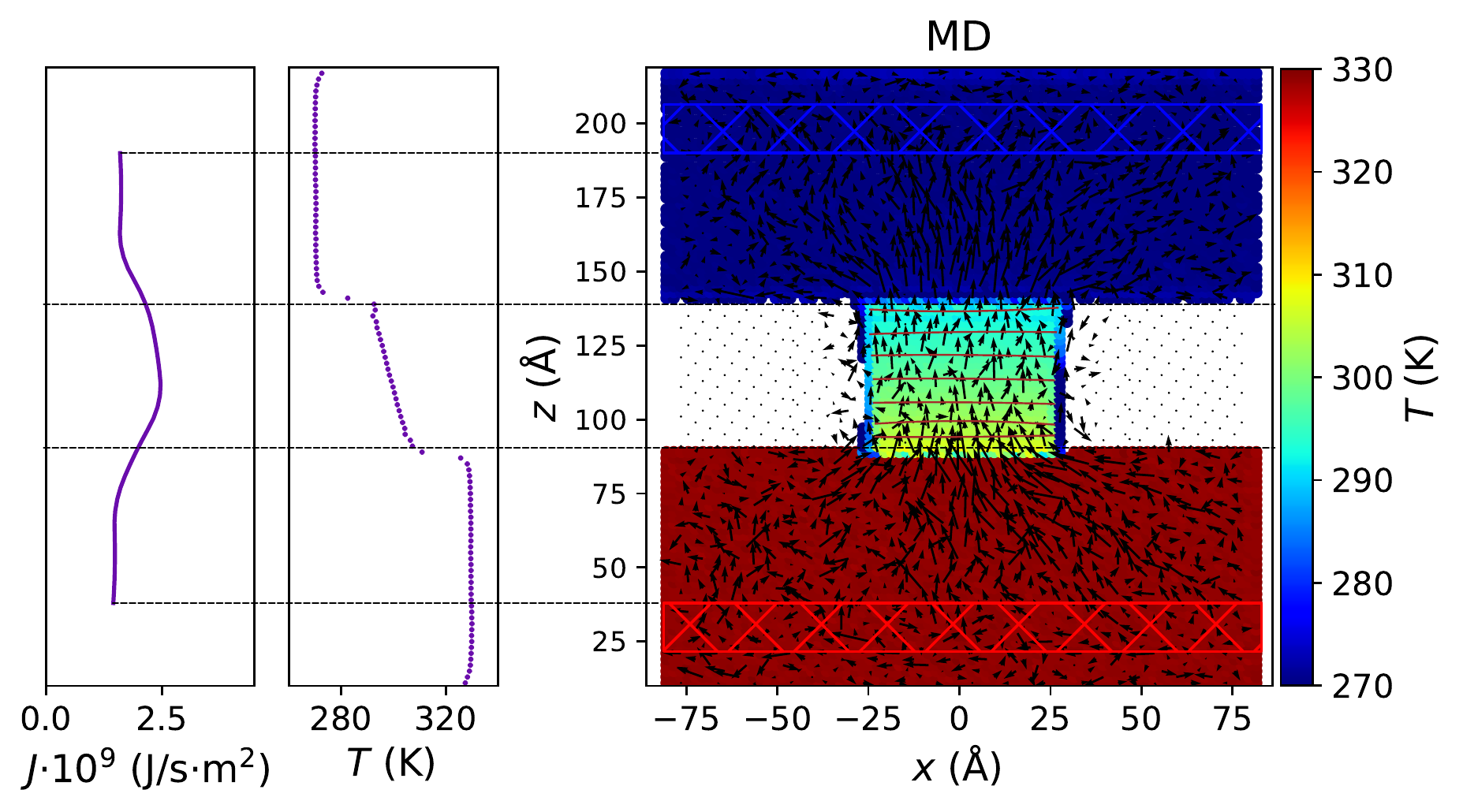}}\\
	\subfloat{\includegraphics[width = 3.7in]{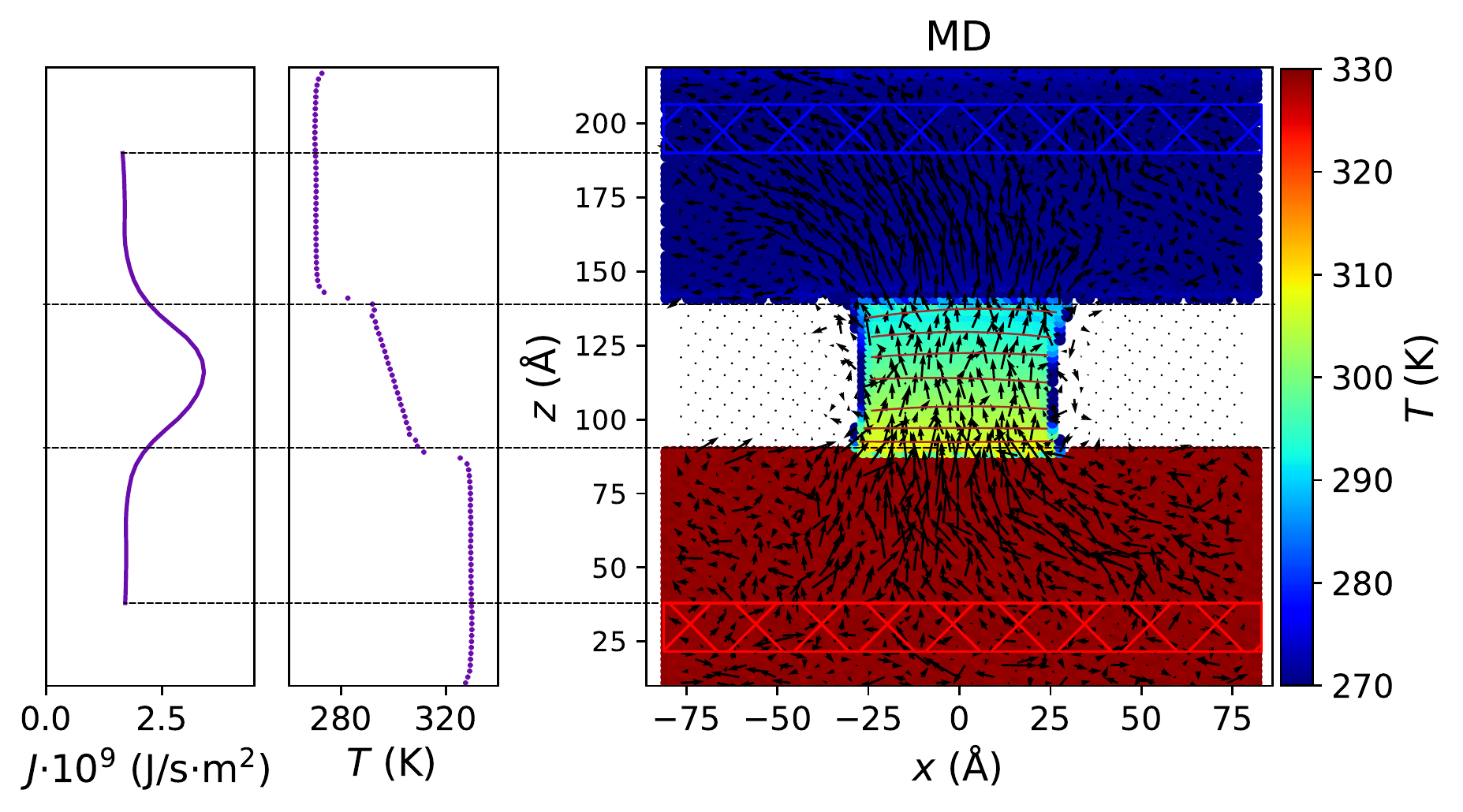}}	
	\subfloat{\includegraphics[width = 3.7in]{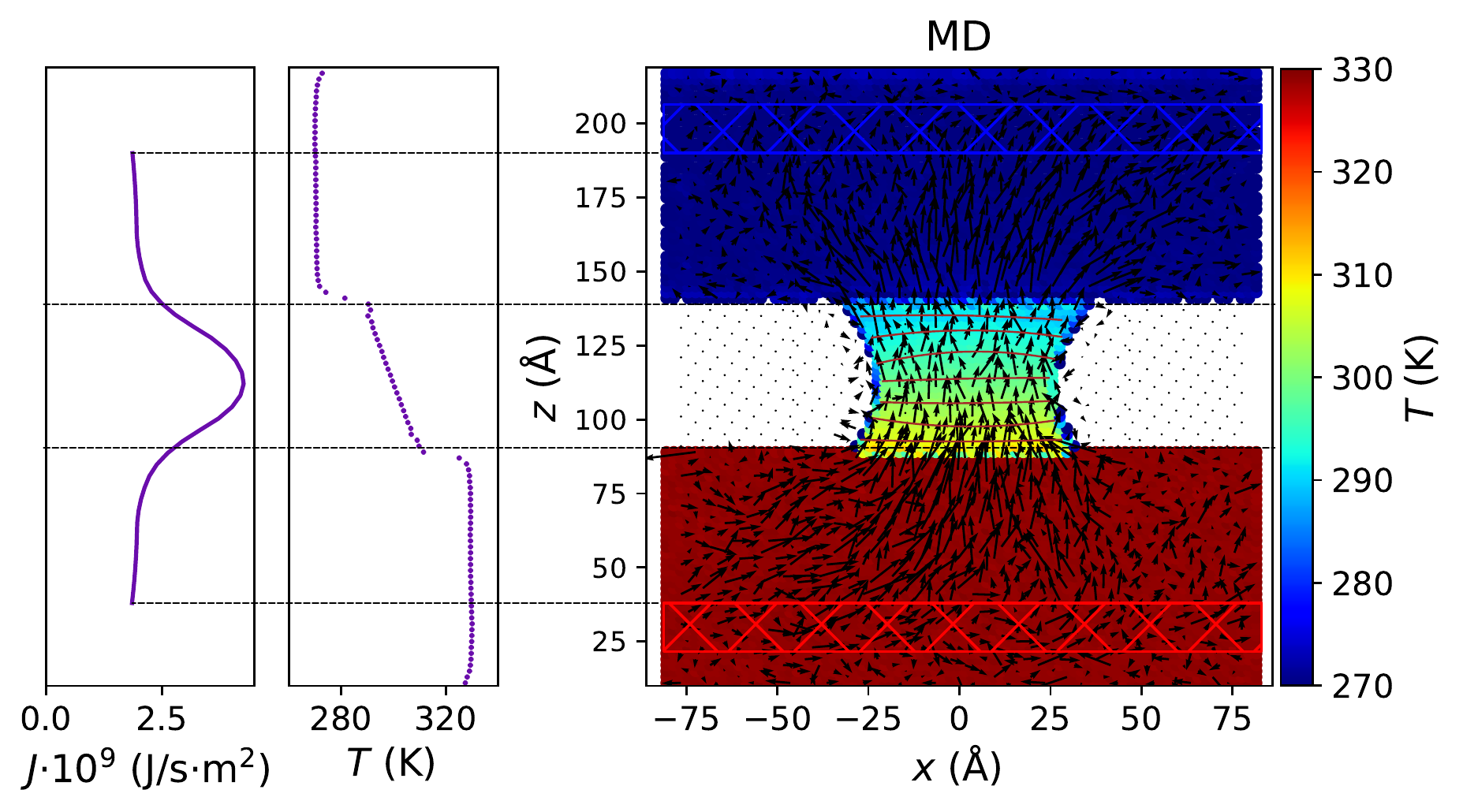}} 
	\caption{\label{fig:6_1} Temperature profiles of the meniscus with size $10\text{a}_0$, ($\varepsilon$ = 10 - 21 meV) with heat fluxes distributions obtained from MD simulations.}
	
\end{figure*}

\begin{figure*}
	\centering
	\subfloat{\includegraphics[width=3.7in]{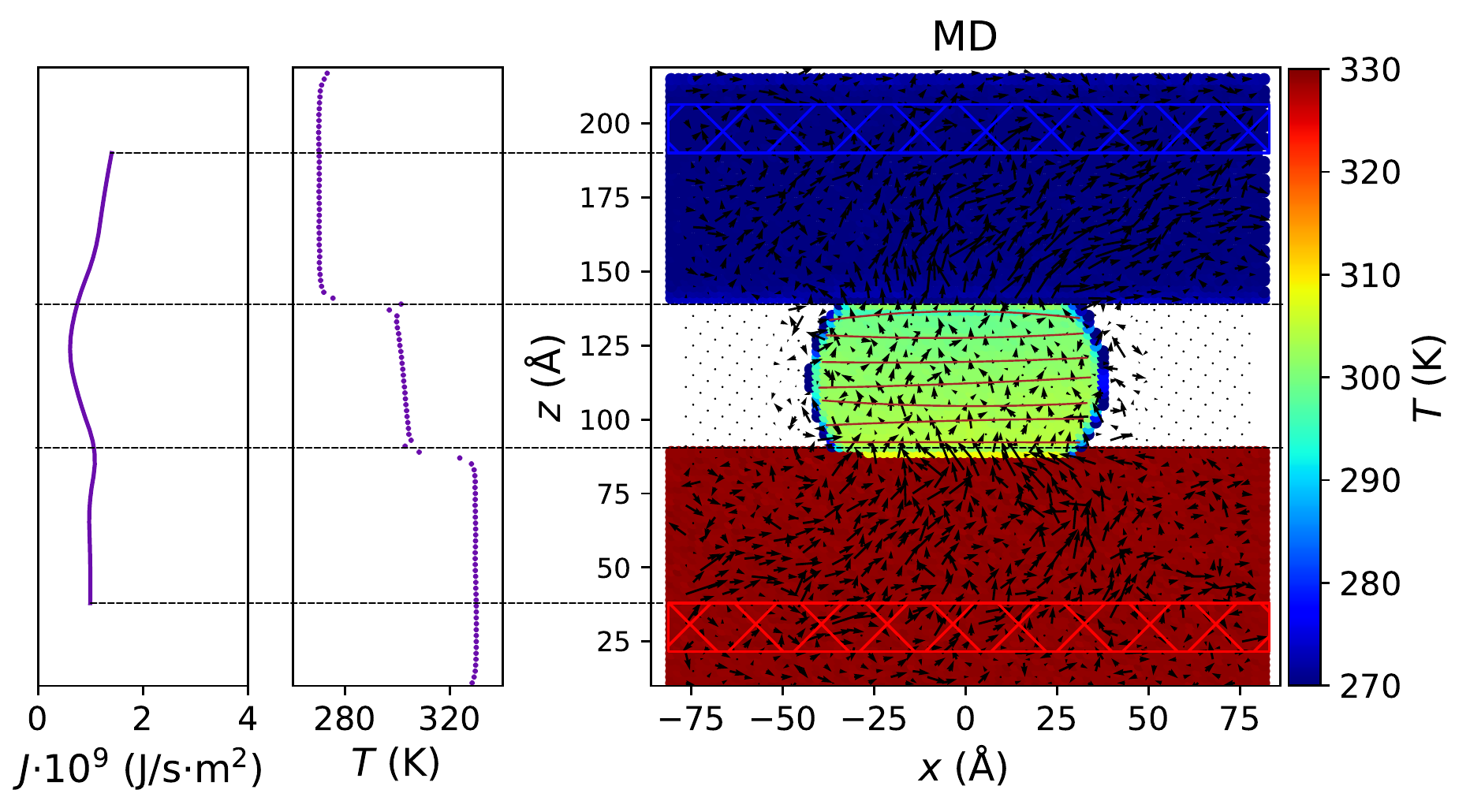}} 
	\subfloat{\includegraphics[width = 3.7in]{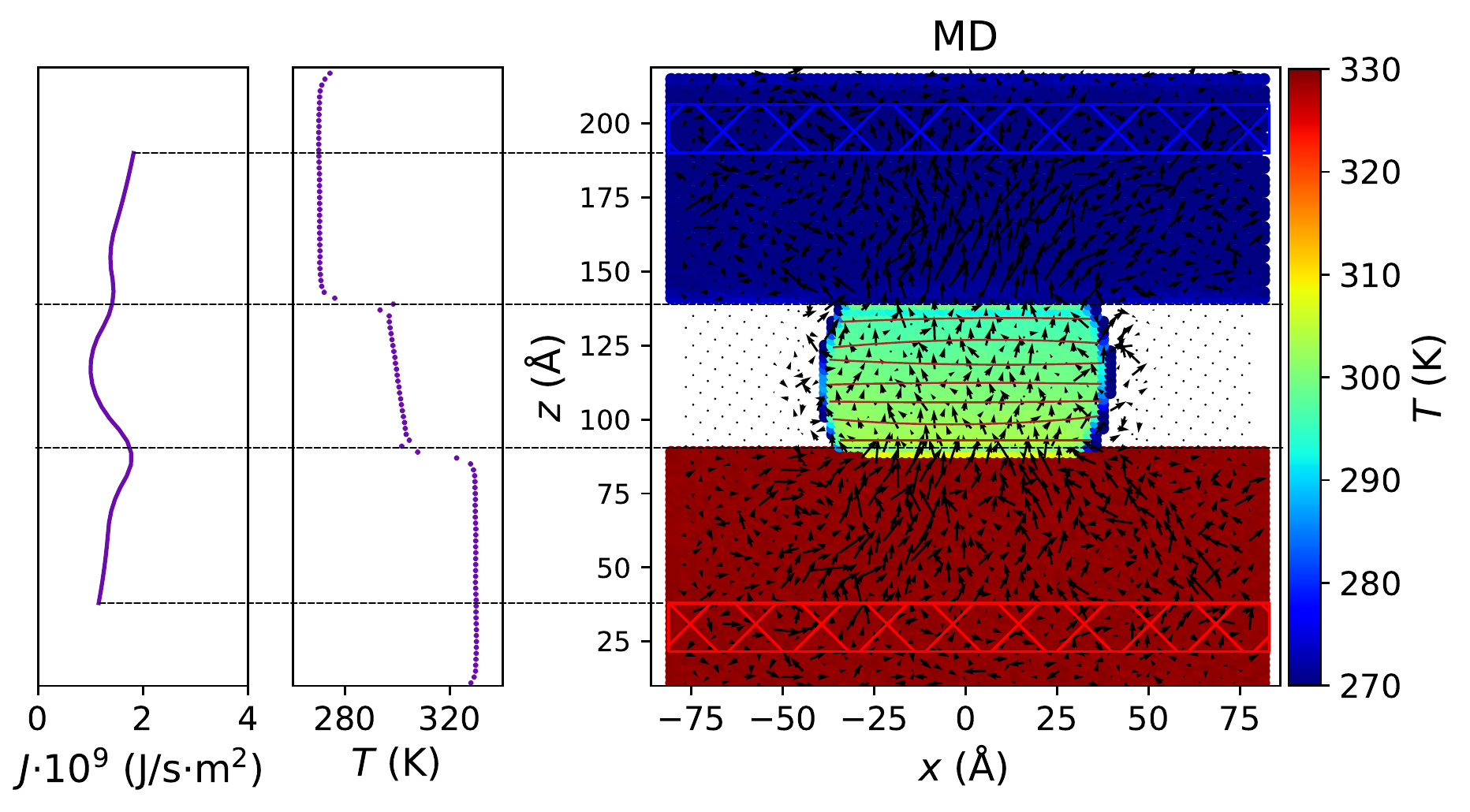}}\\
	\subfloat{\includegraphics[width = 3.7in]{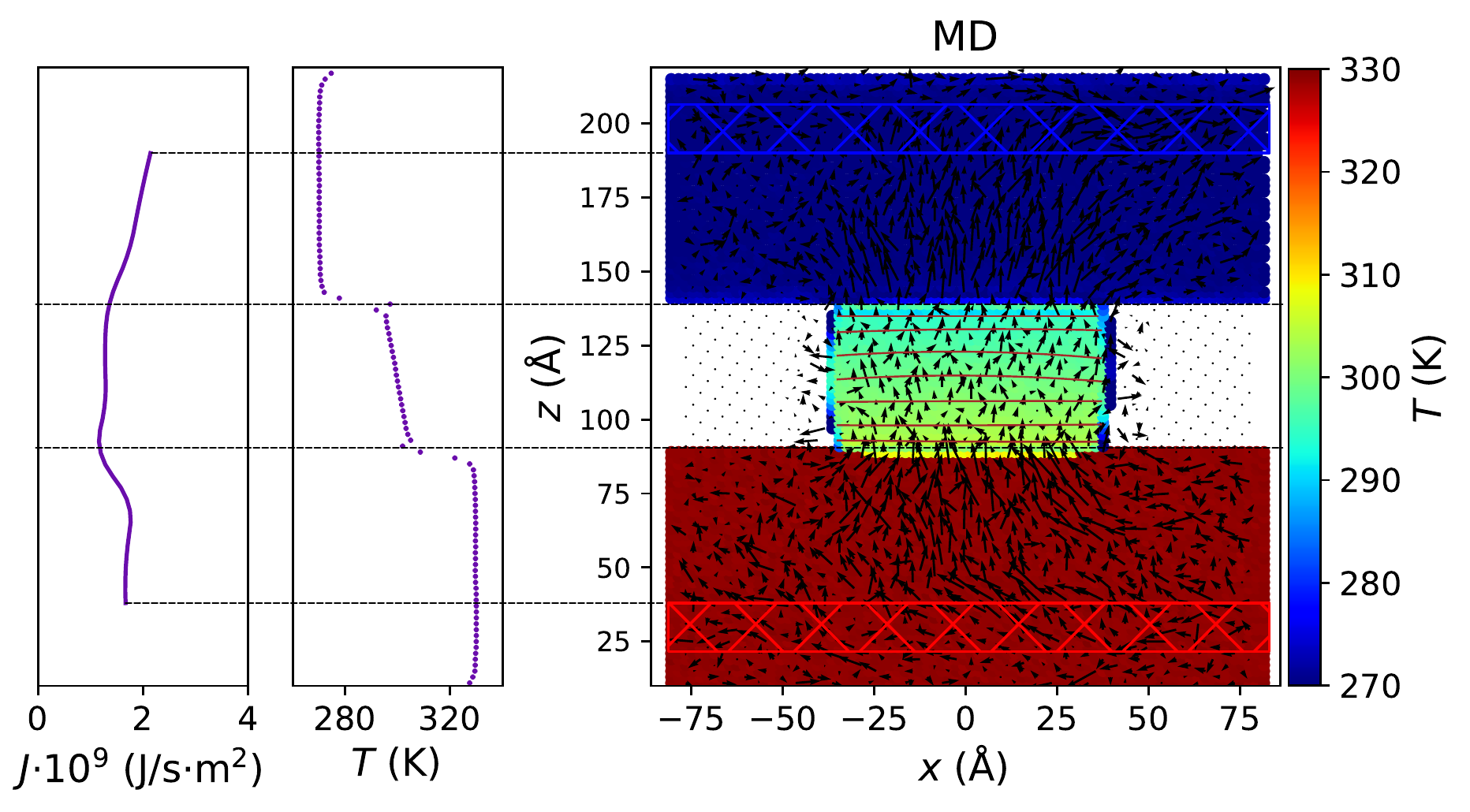}} 
	\subfloat{\includegraphics[width = 3.7in]{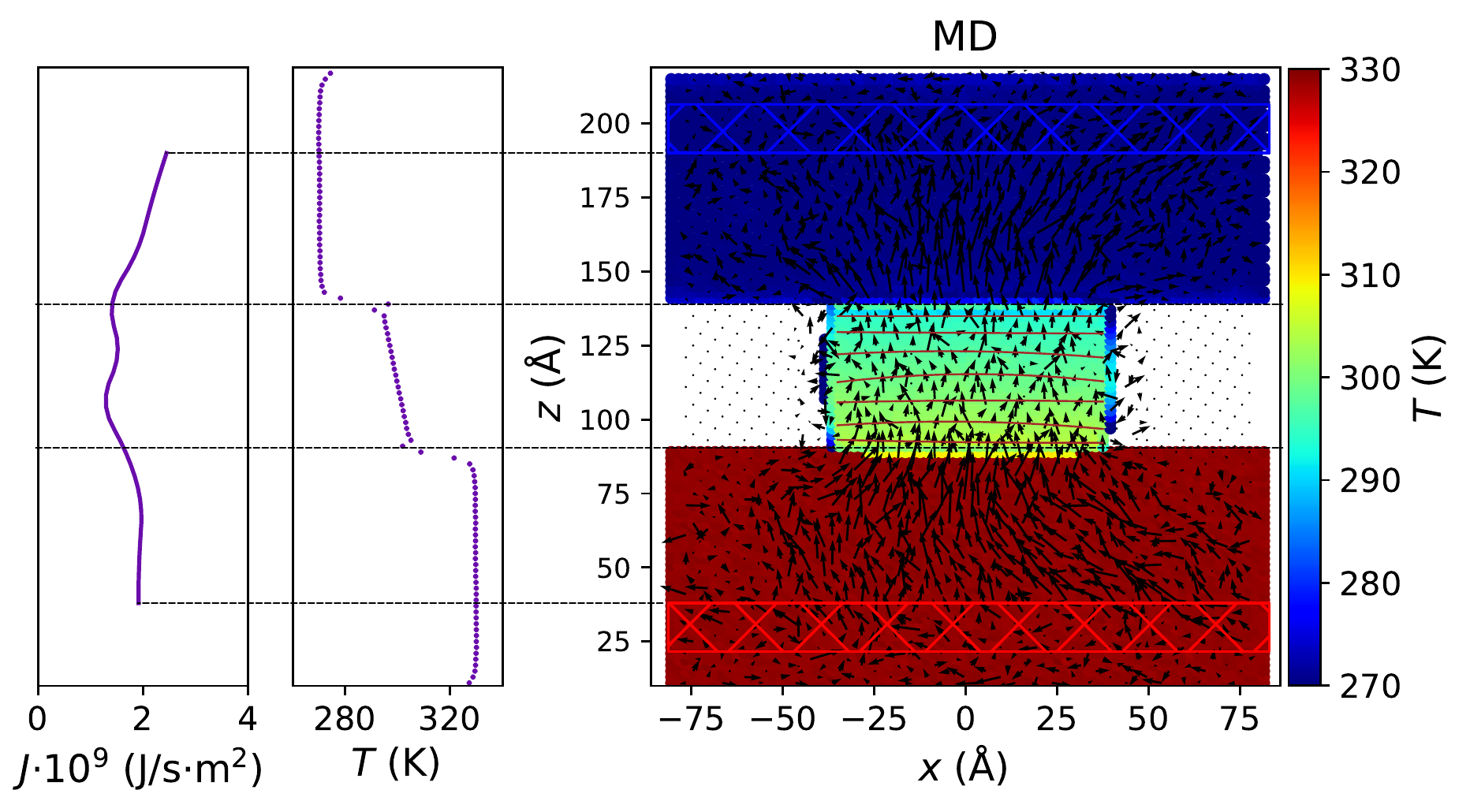}}\\
	\subfloat{\includegraphics[width = 3.7in]{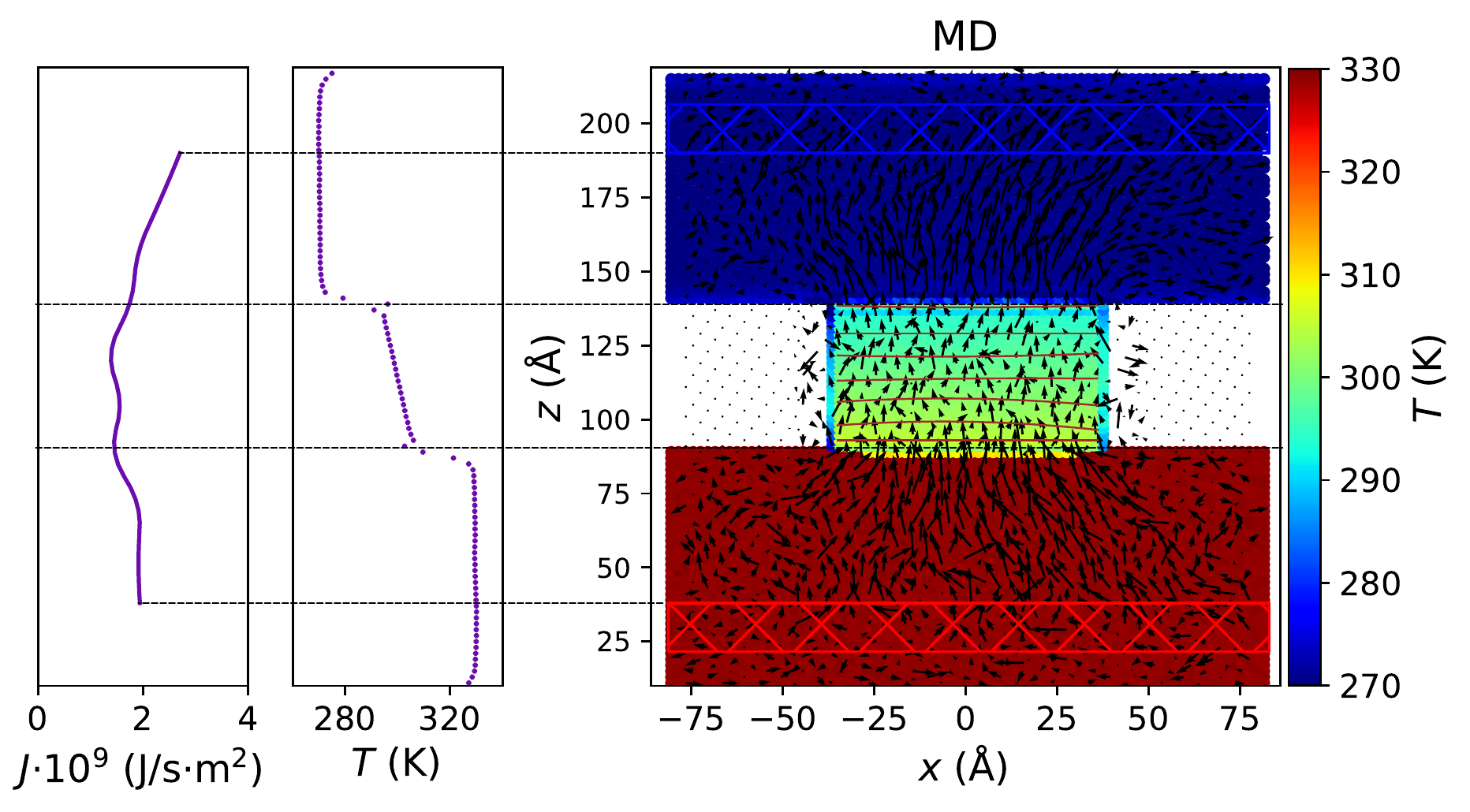}}
	\subfloat{\includegraphics[width = 3.7in]{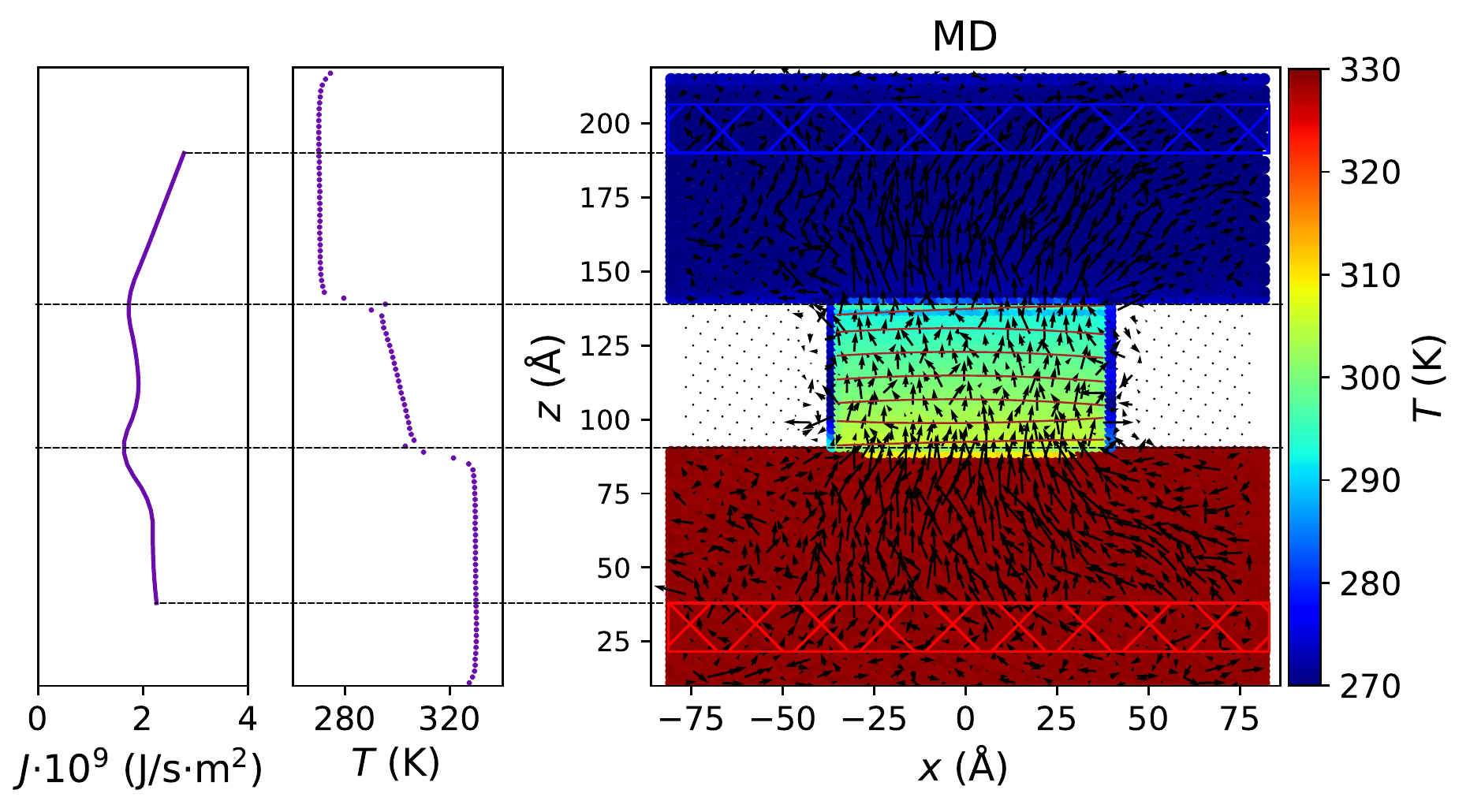}} \\
	\subfloat{\includegraphics[width = 3.7in]{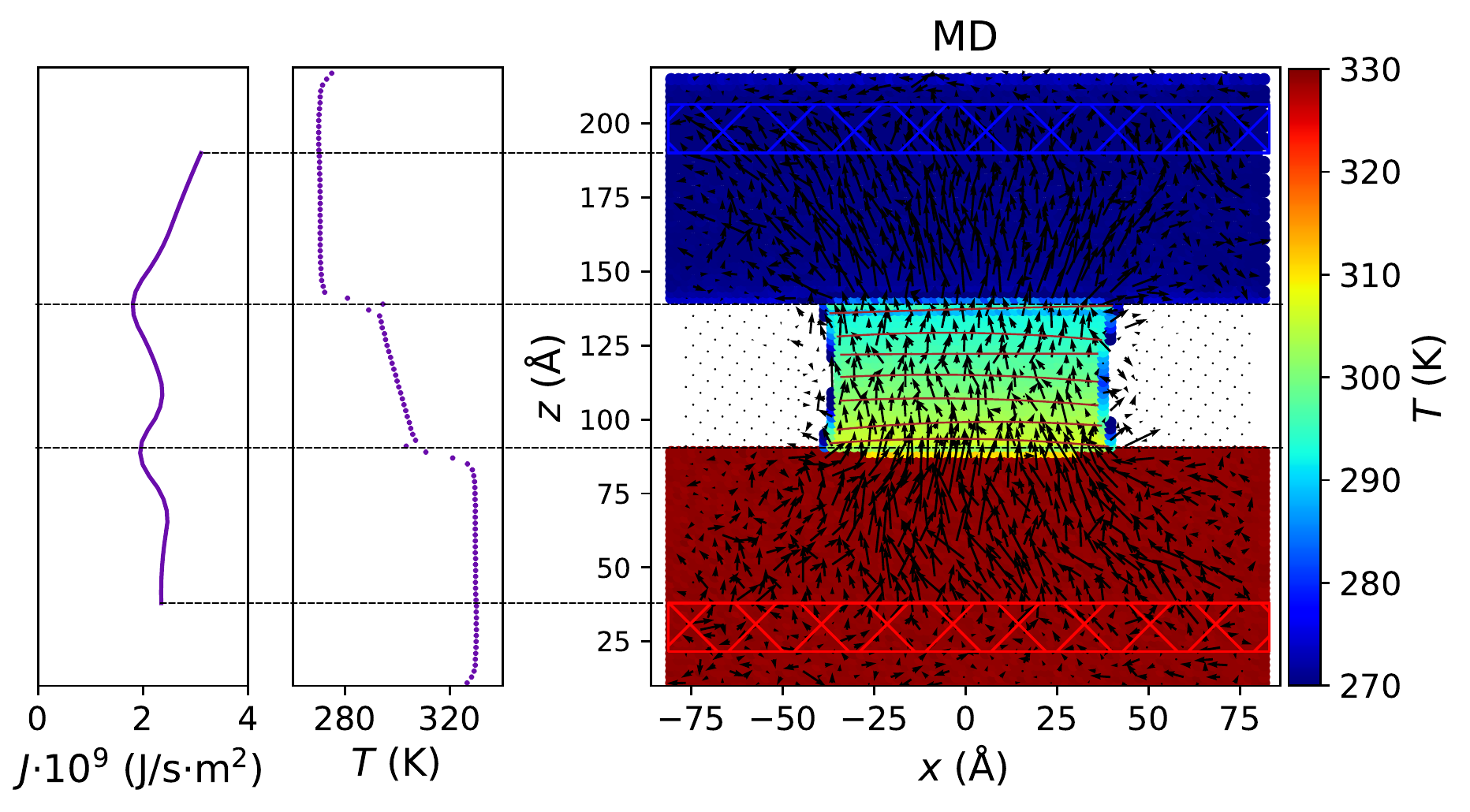}}
	\subfloat{\includegraphics[width = 3.7in]{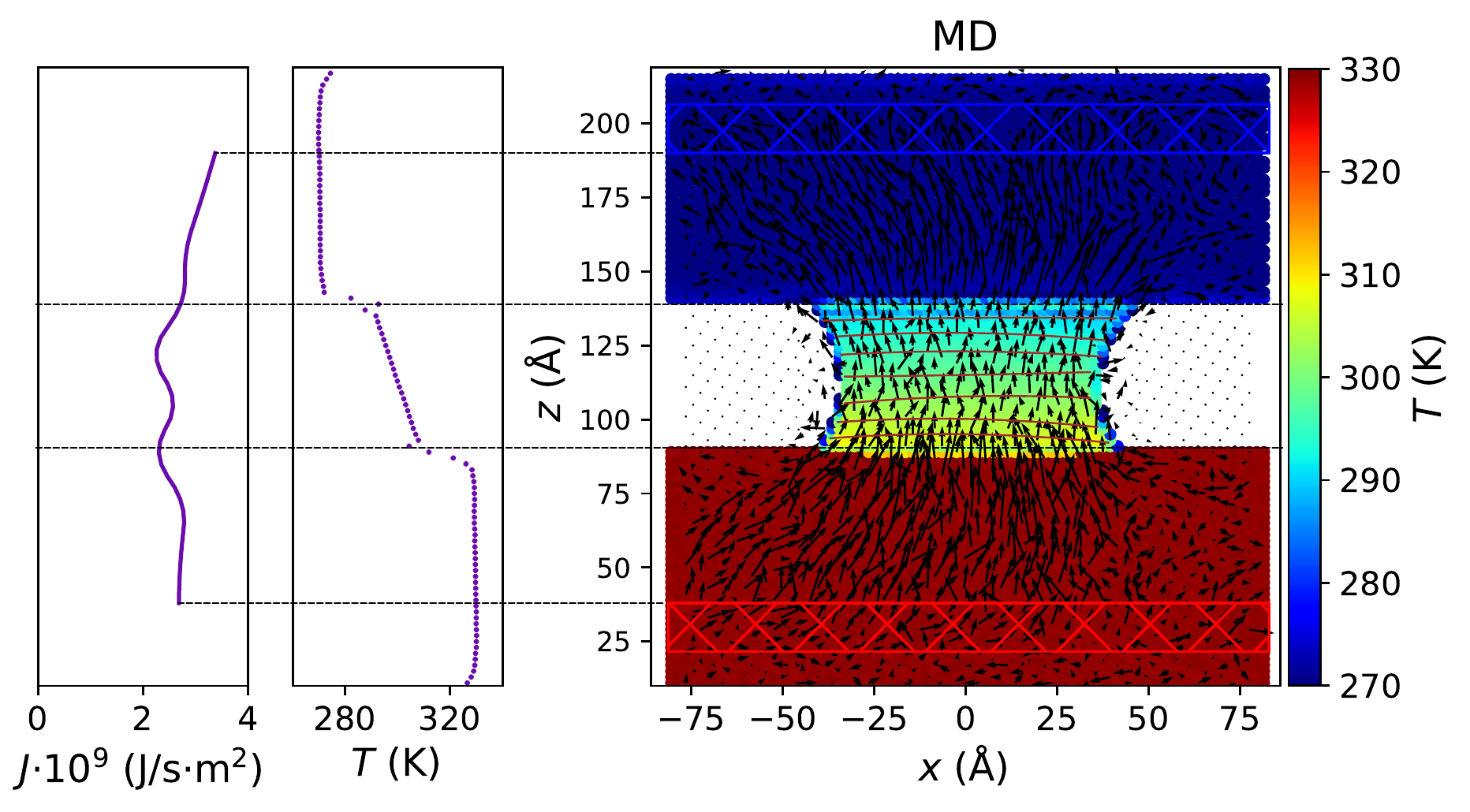}} 
	\caption{\label{fig:6_2} Temperature profiles of the meniscus with size $14\text{a}_0$, ($\varepsilon$ = 10 - 21 meV) with heat fluxes distributions obtained from MD simulations.}
	
\end{figure*}


\begin{figure*}
	\centering
	\subfloat{\includegraphics[width=2.4in]{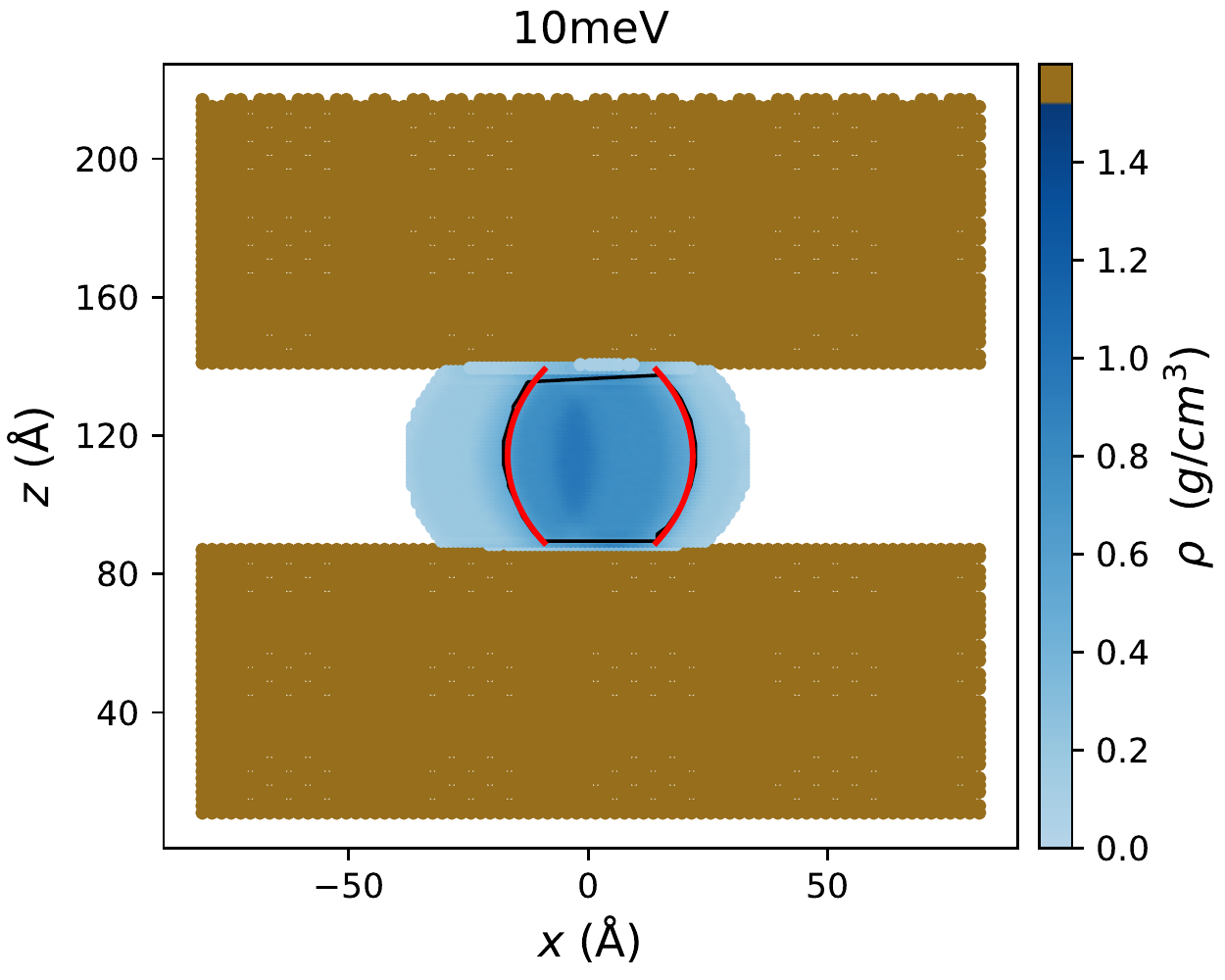}} 
	\subfloat{\includegraphics[width=2.4in]{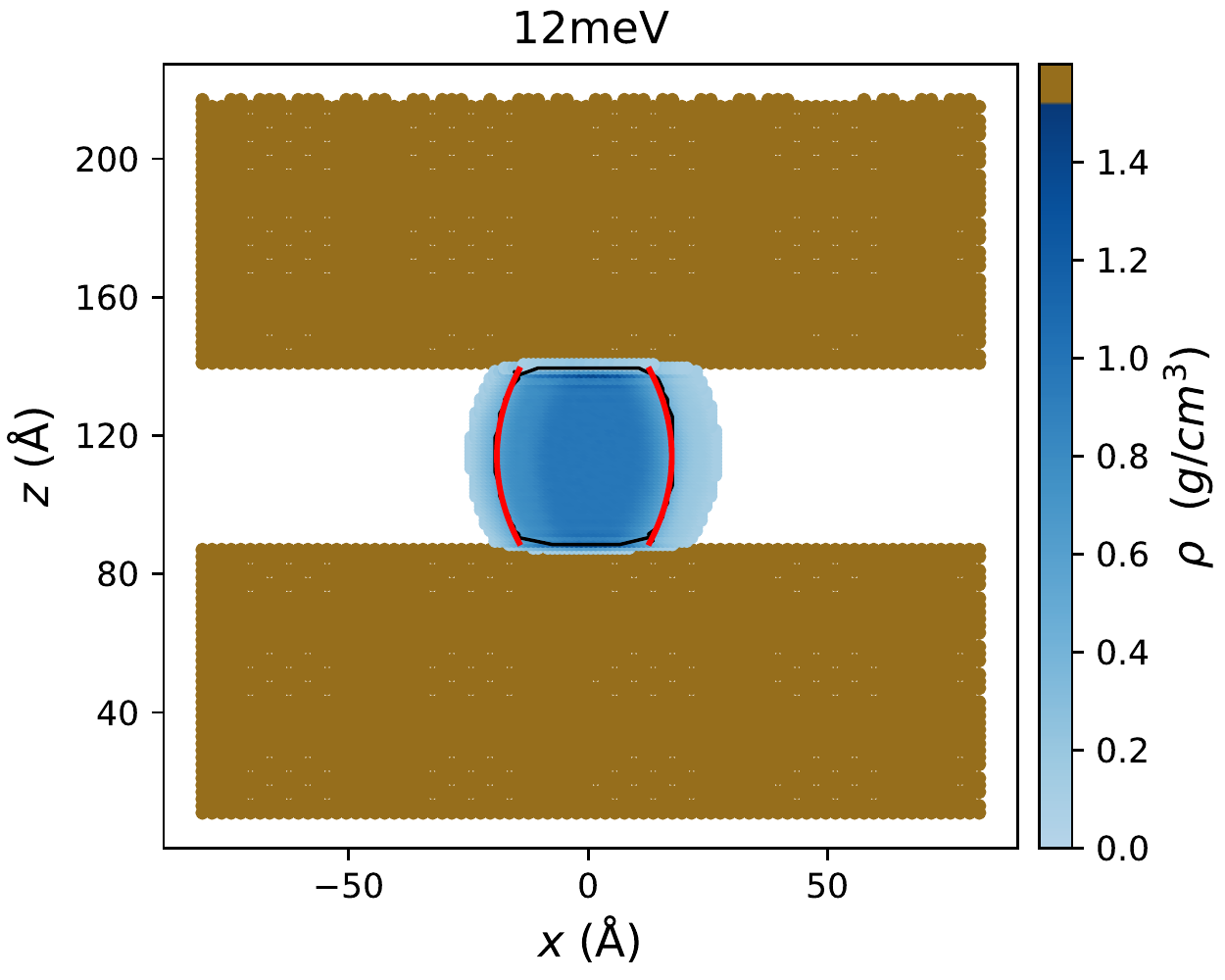}}\\
	\subfloat{\includegraphics[width = 2.4in]{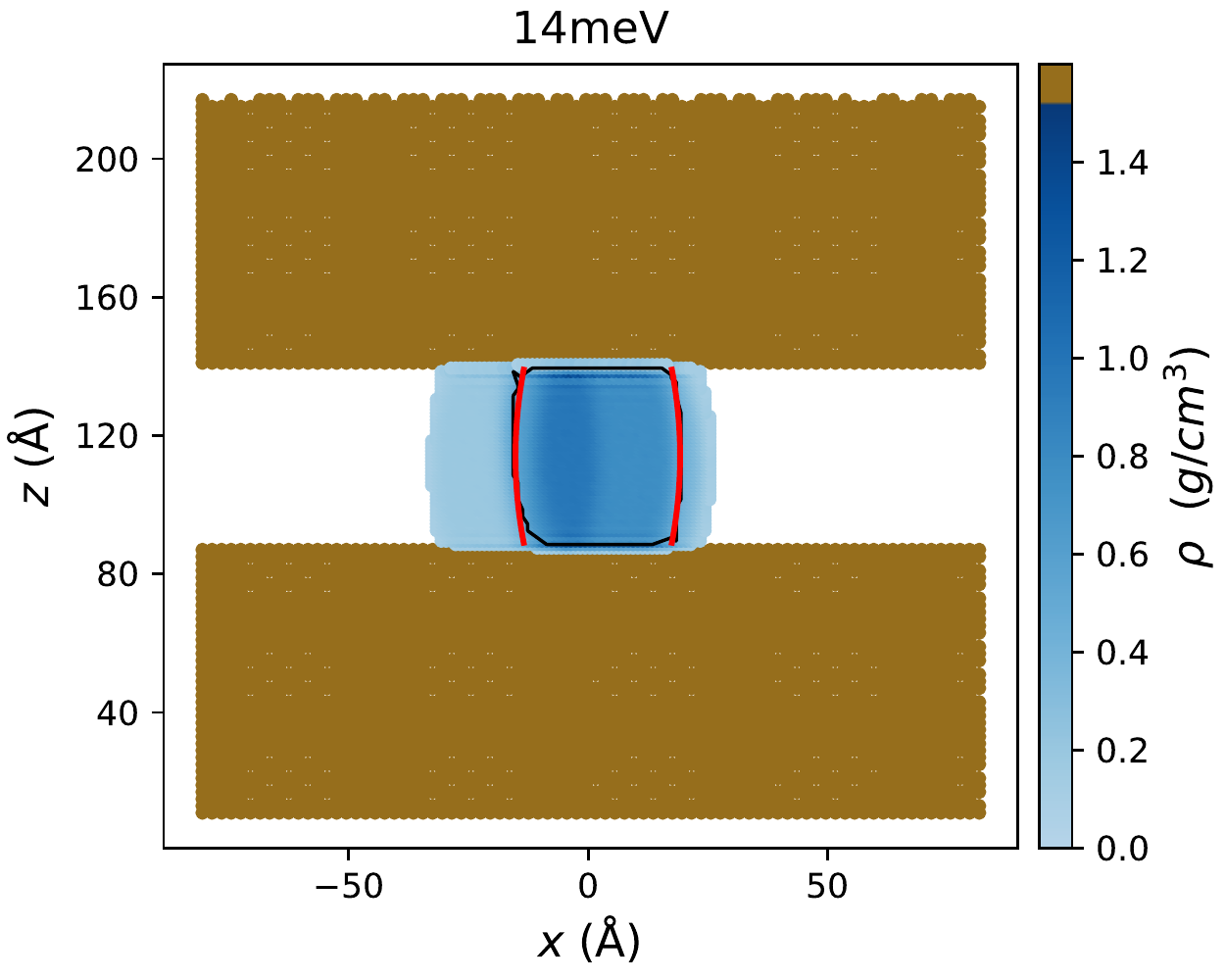}} 
	\subfloat{\includegraphics[width = 2.4in]{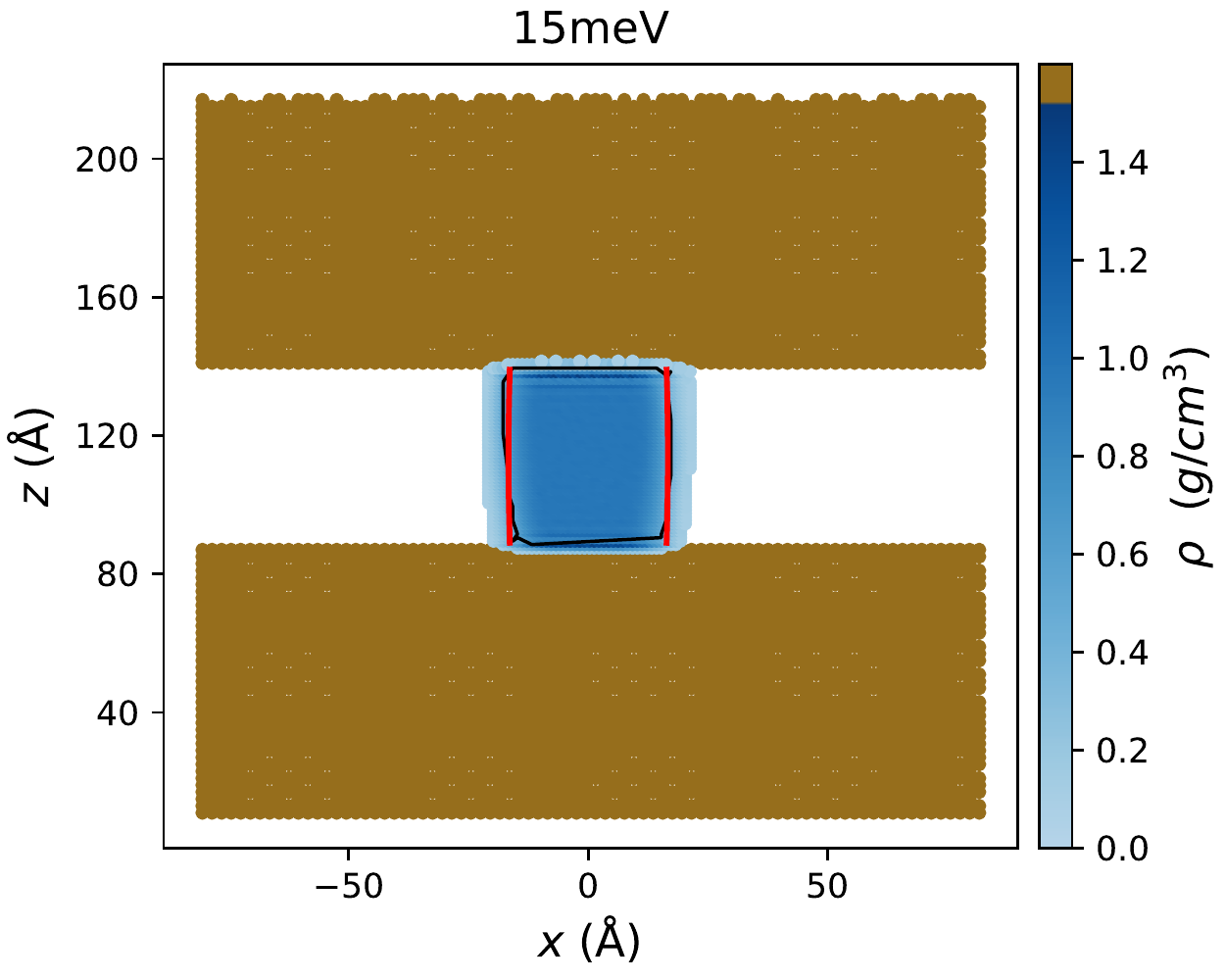}}\\
	\subfloat{\includegraphics[width = 2.4in]{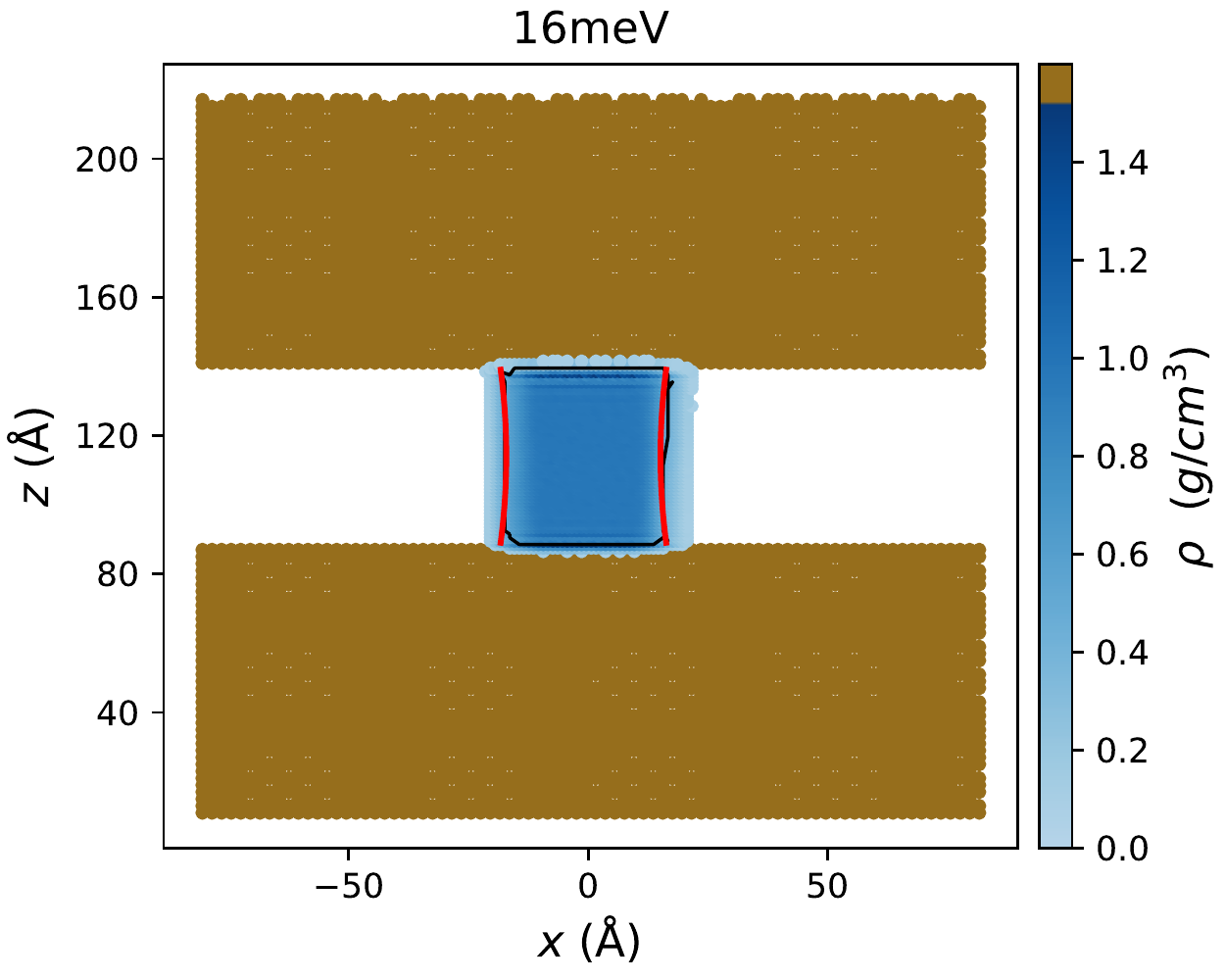}}
	\subfloat{\includegraphics[width = 2.4in]{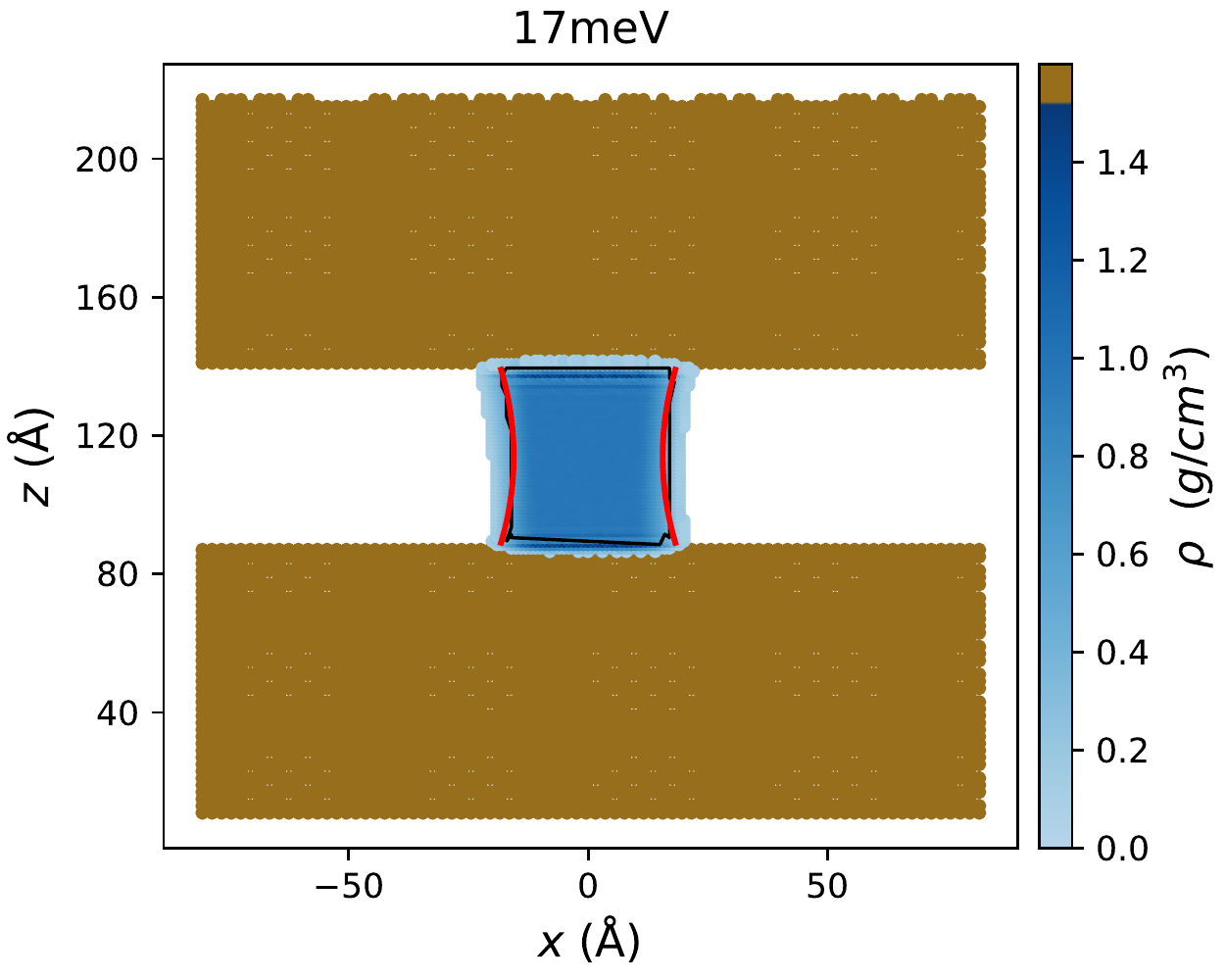}} \\
	\subfloat{\includegraphics[width = 2.4in]{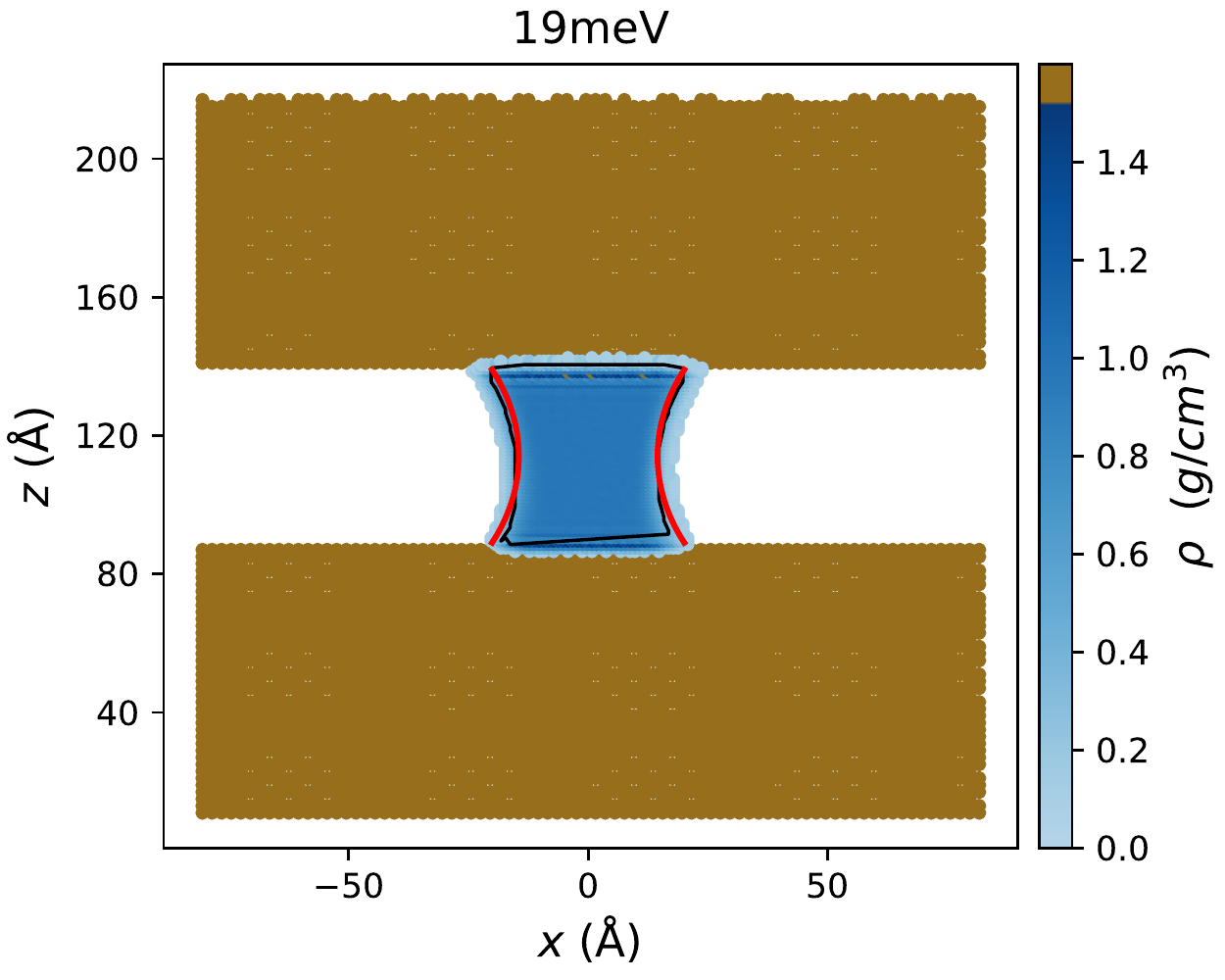}}
	\subfloat{\includegraphics[width = 2.4in]{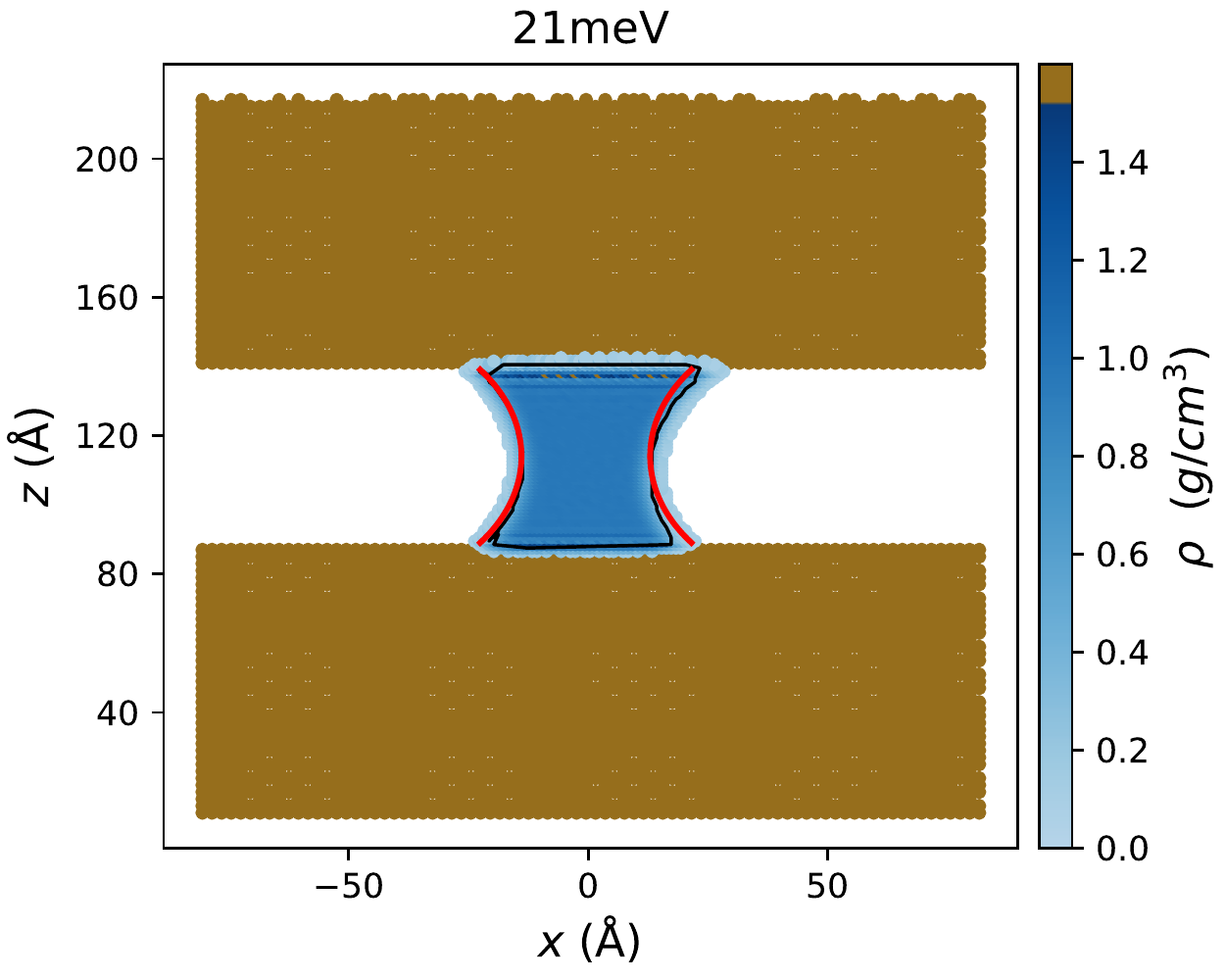}} 
	\caption{\label{fig:8} The density profiles of meniscus with size $6\text{a}_0$ for different parametrization, ($\varepsilon$ = 10 - 21 meV). Values of density are shown in colorbar. The black solid line connects the points where the density of meniscus is equal to the half of the density of confined water. Red lines are the analytically obtained borders of droplet.}
	
\end{figure*}

\begin{figure*}
	\centering
	\subfloat{\includegraphics[width=2.4in]{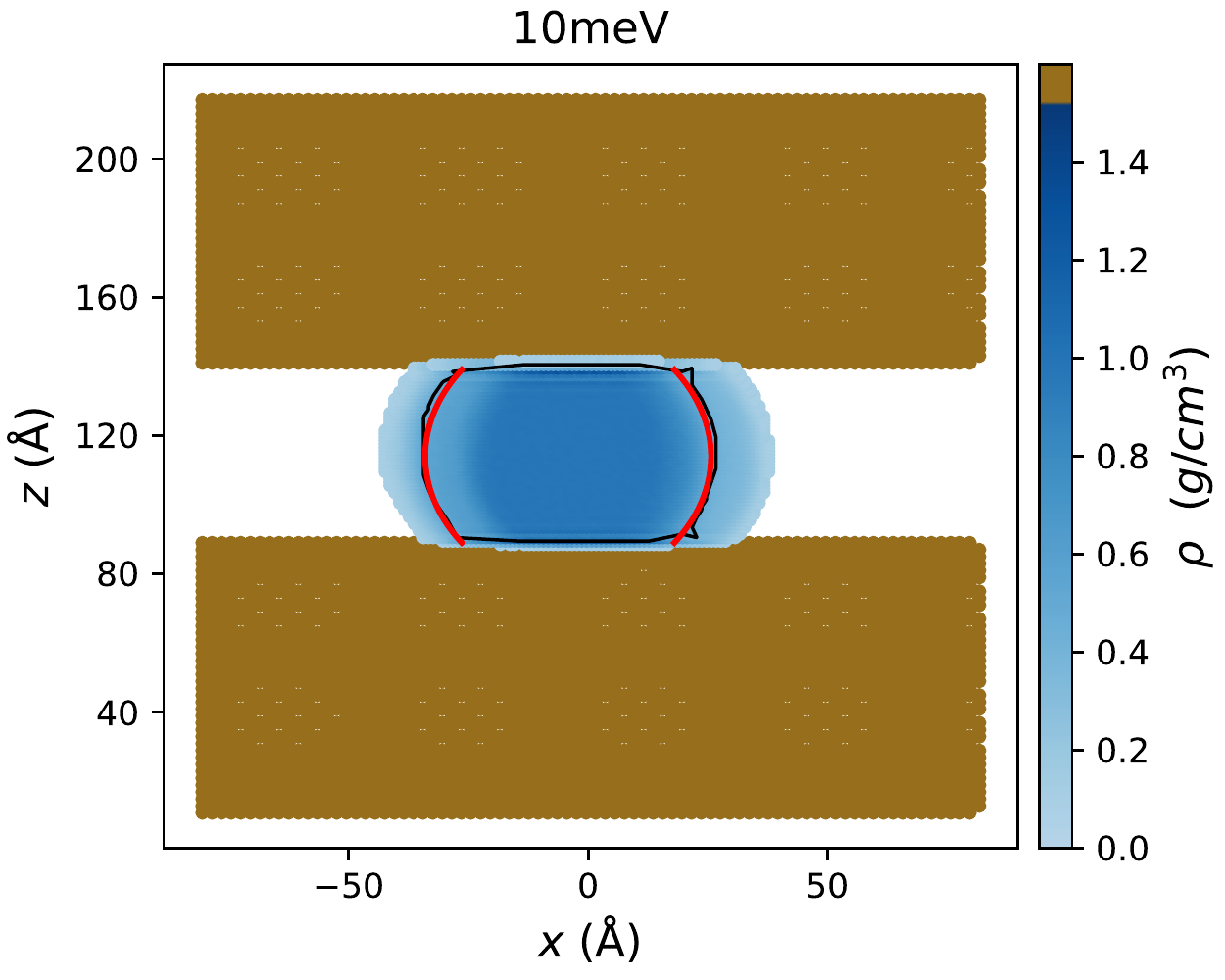}} 
	\subfloat{\includegraphics[width=2.4in]{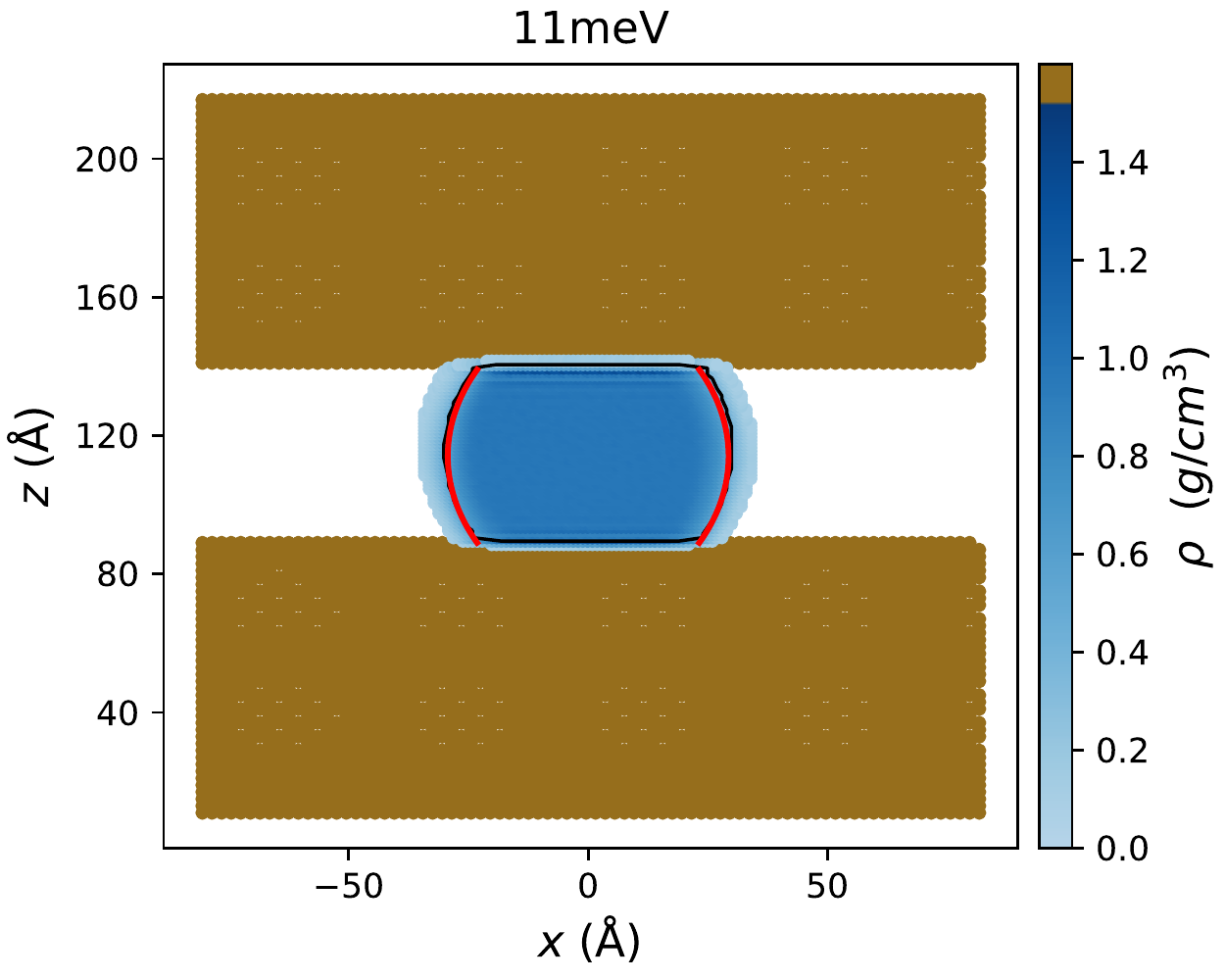}}
	\subfloat{\includegraphics[width=2.4in]{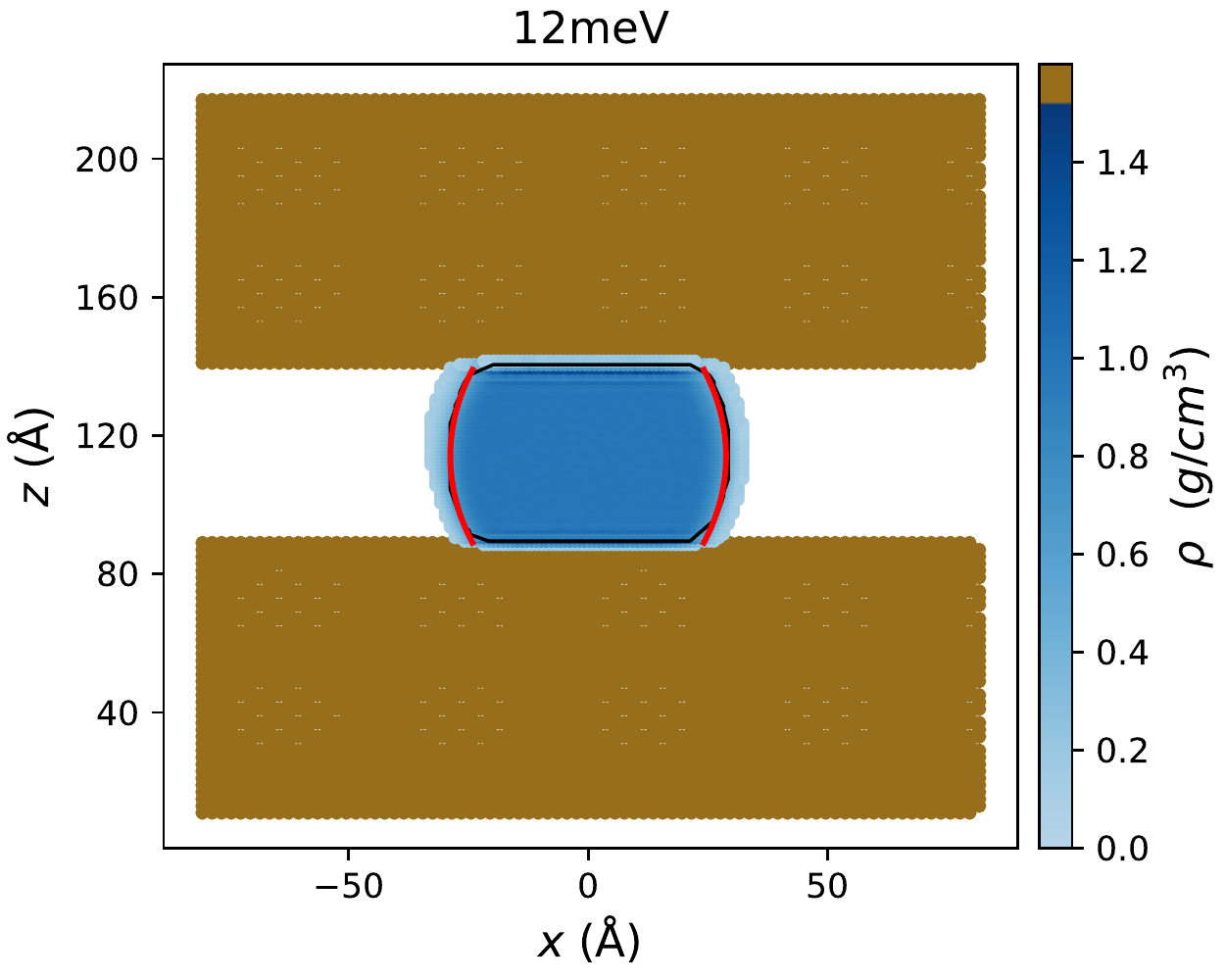}}\\
	\subfloat{\includegraphics[width=2.4in]{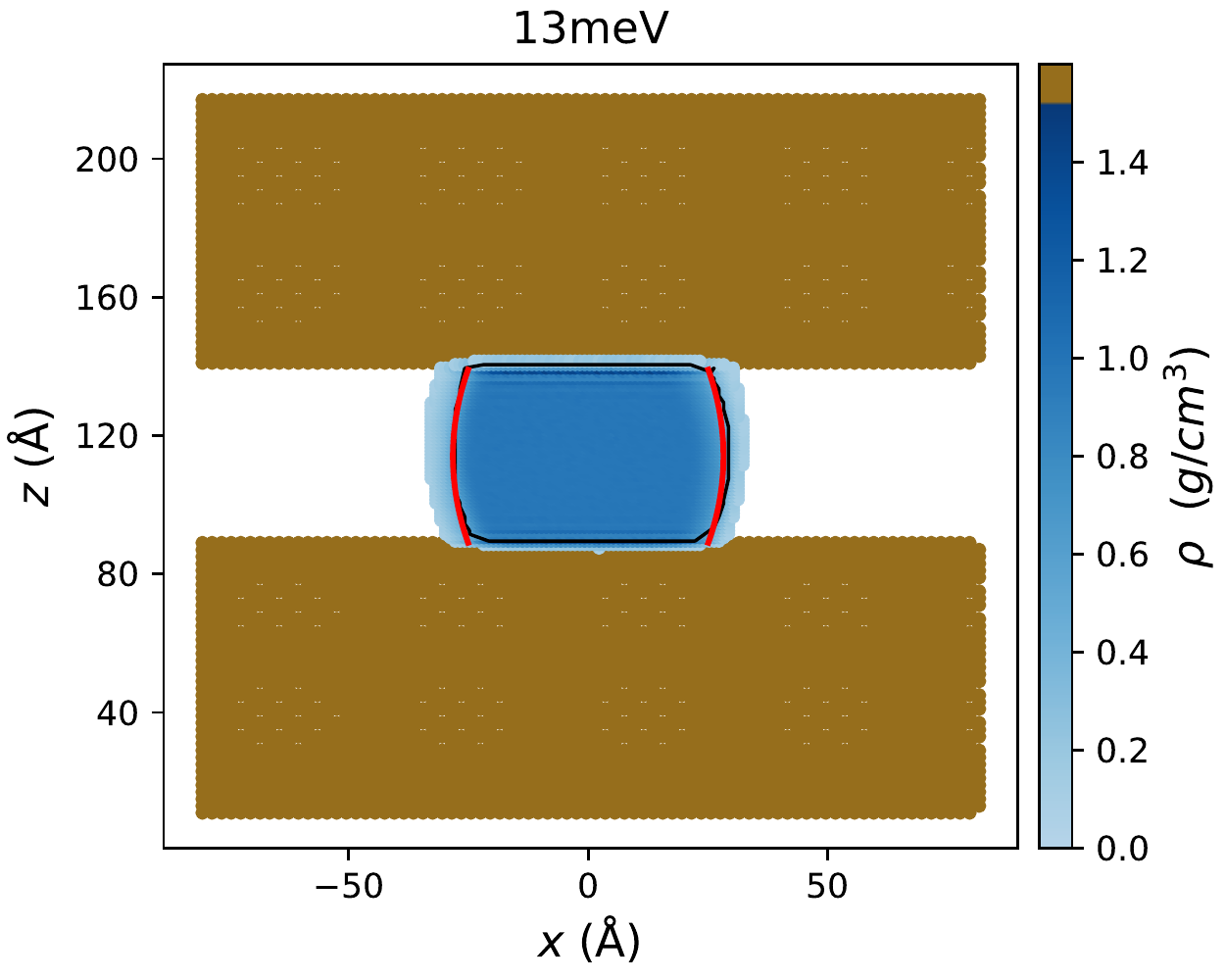}} 
	\subfloat{\includegraphics[width = 2.4in]{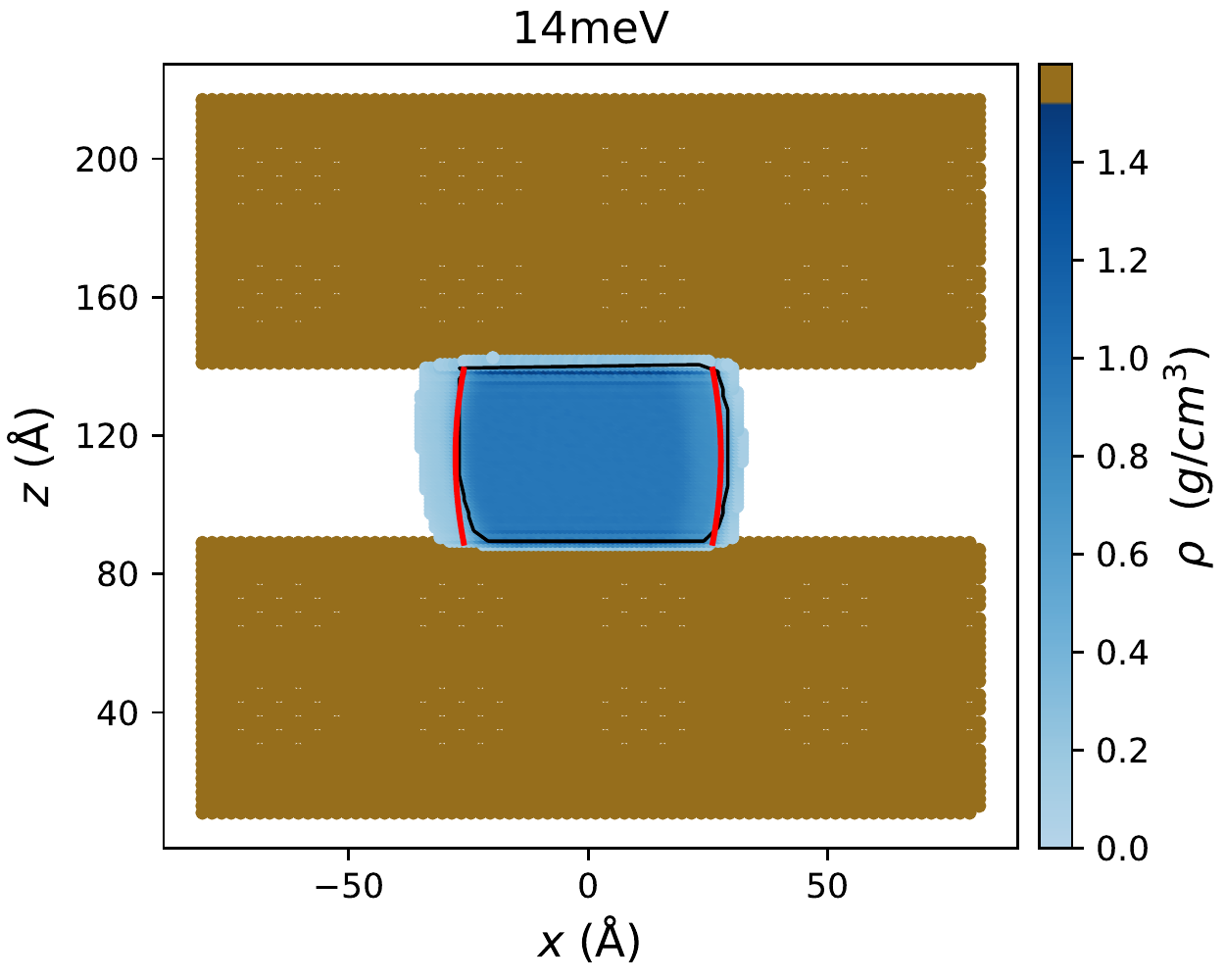}} 
	\subfloat{\includegraphics[width = 2.4in]{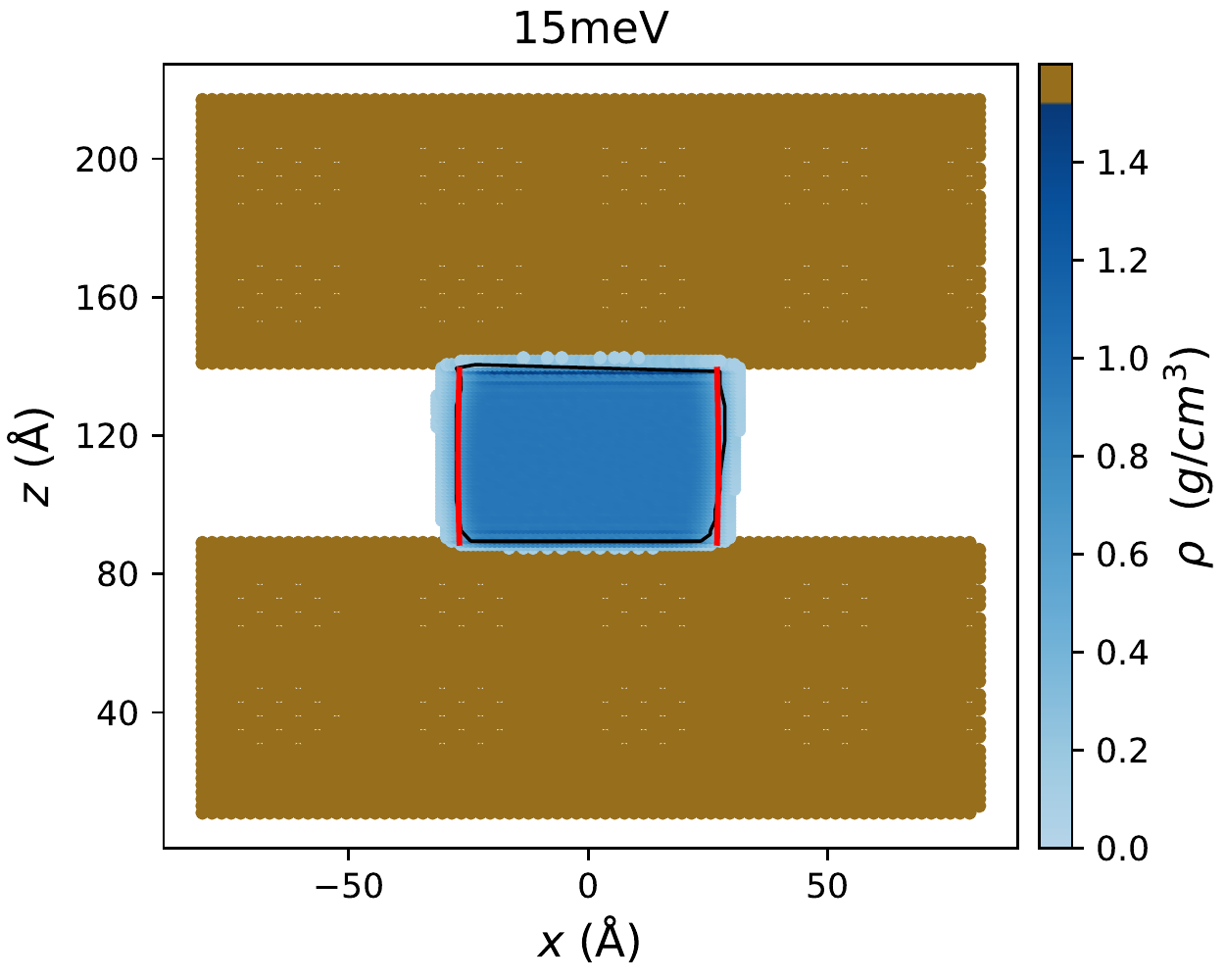}}\\
	\subfloat{\includegraphics[width = 2.4in]{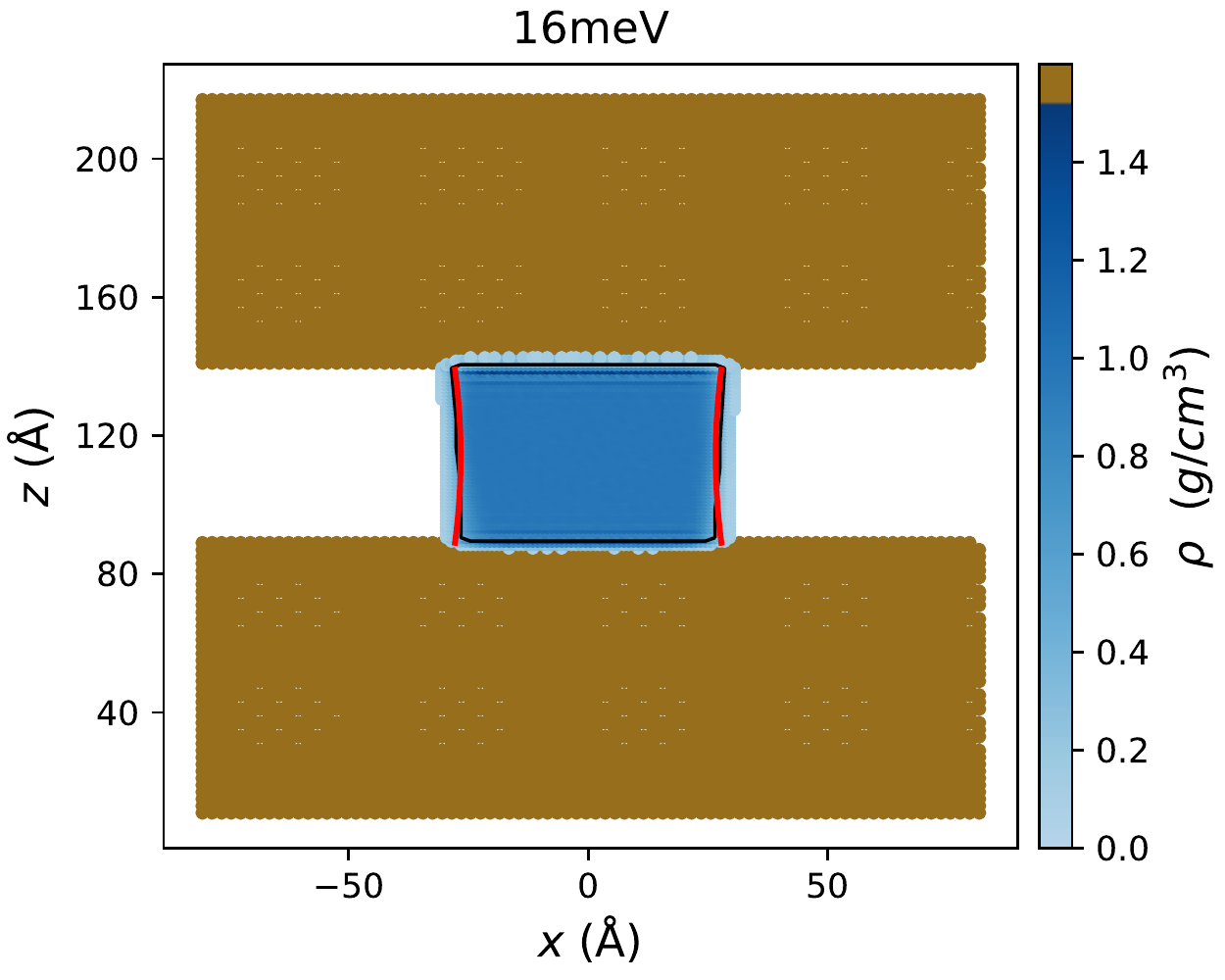}}
	\subfloat{\includegraphics[width = 2.4in]{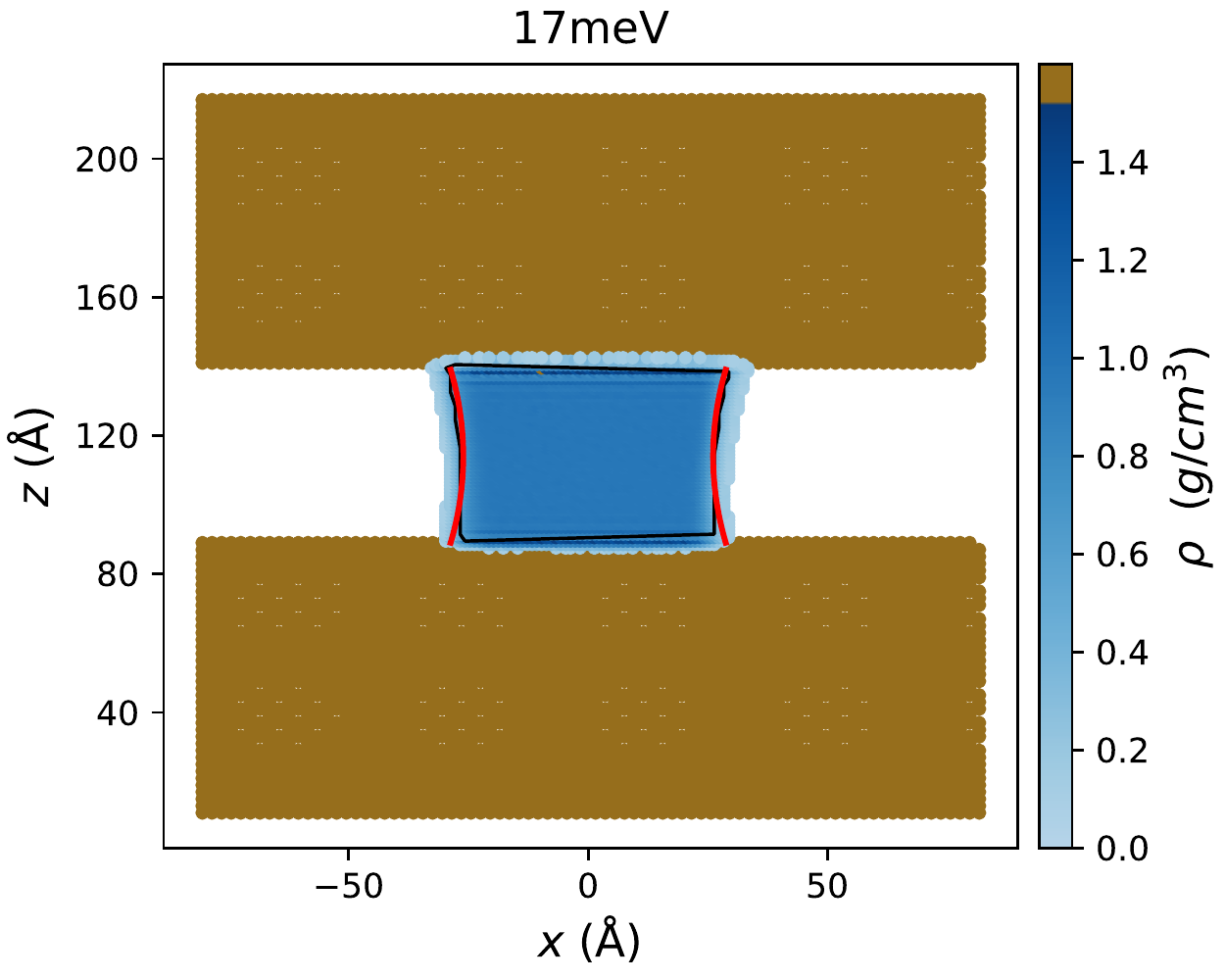}} 
	\subfloat{\includegraphics[width = 2.4in]{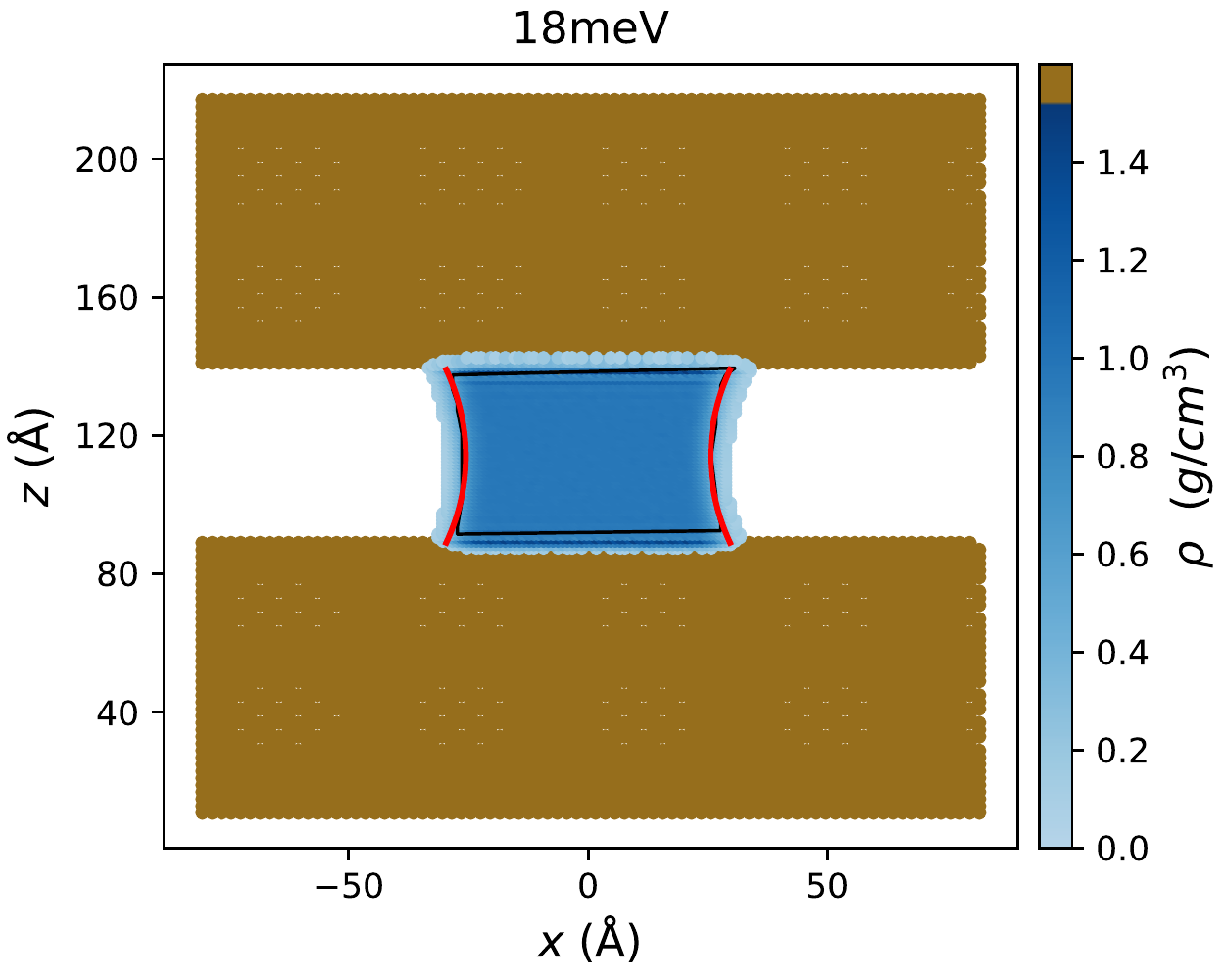}} \\
	\subfloat{\includegraphics[width = 2.4in]{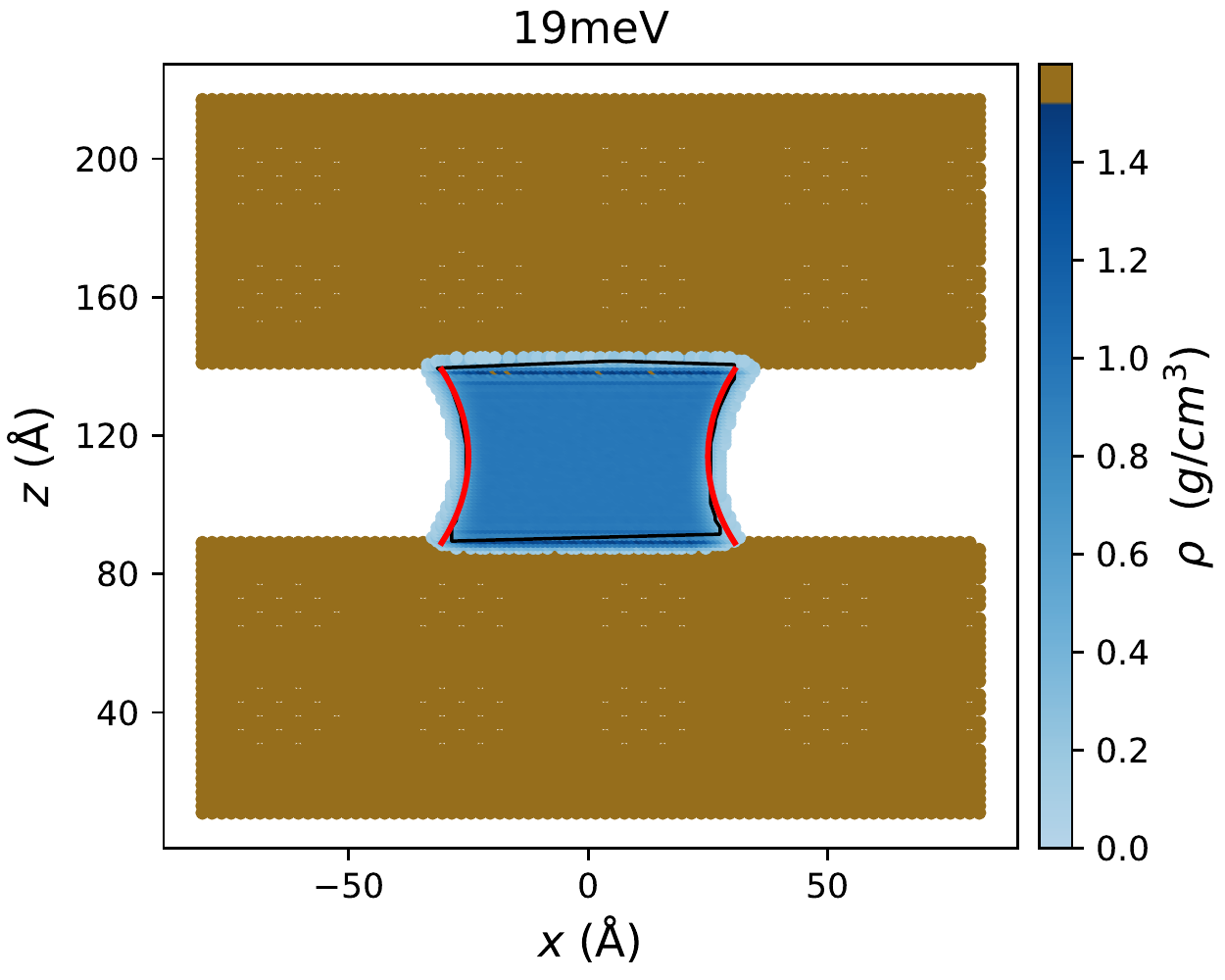}}
	\subfloat{\includegraphics[width = 2.4in]{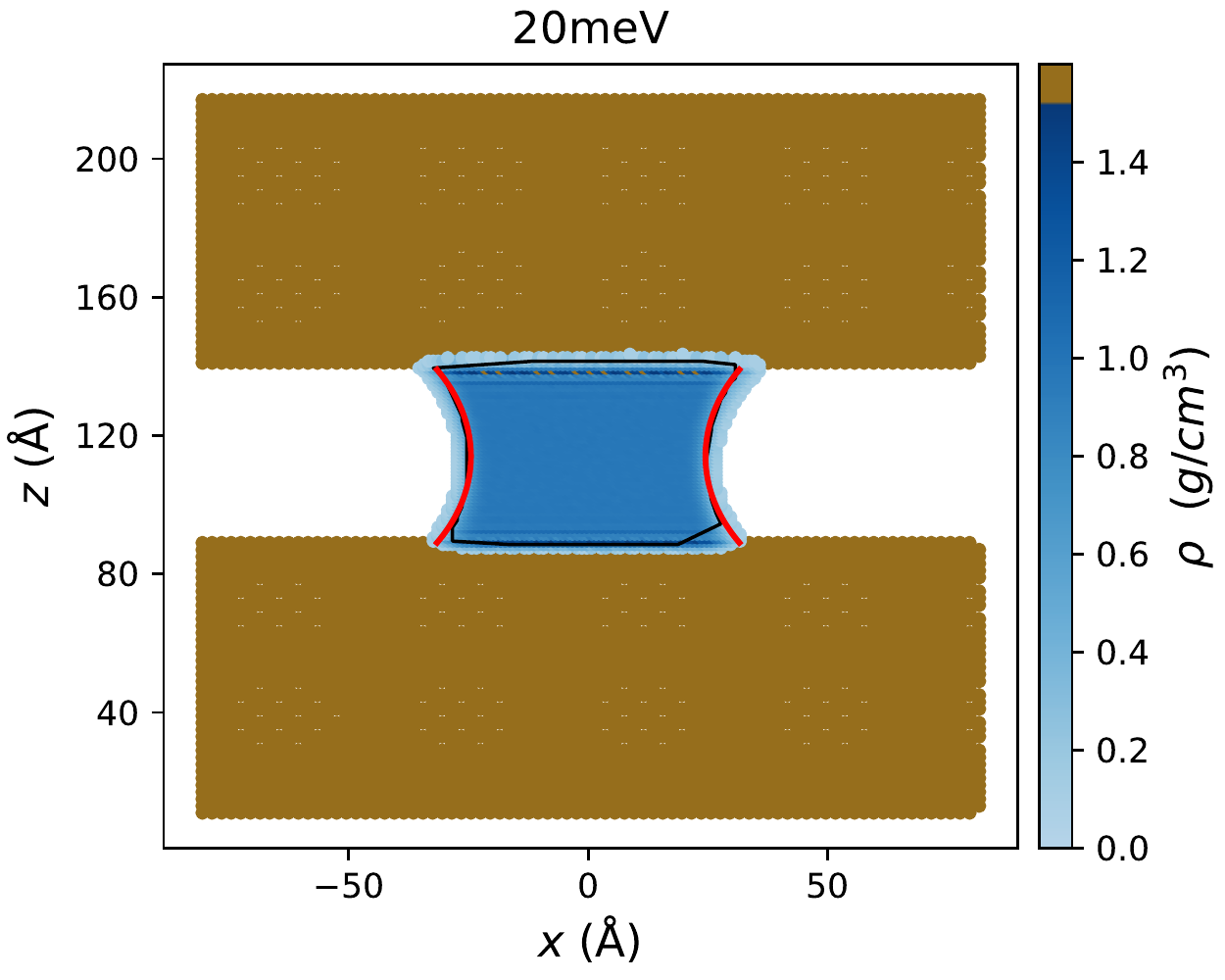}}
	\subfloat{\includegraphics[width = 2.4in]{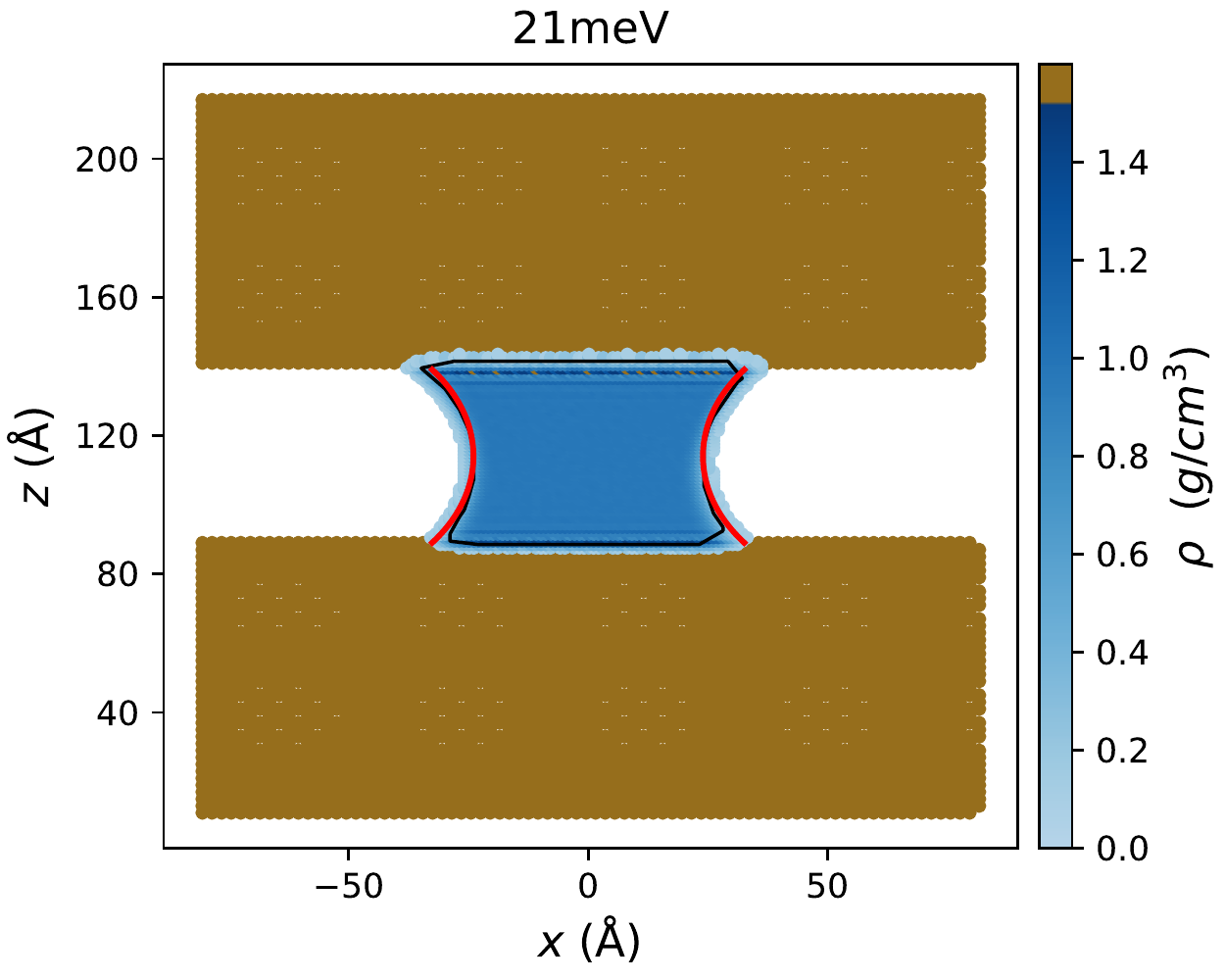}} 
	\caption{\label{fig:8_1} The density profiles of meniscus with size $10\text{a}_0$ for different parametrization, ($\varepsilon$ = 10 - 21 meV). Values of density are shown in colorbar. The black solid line connects the points where the density of meniscus is equal to the half of the density of confined water. Red lines are the analytically obtained borders of droplet.}
	
\end{figure*}

\begin{figure*}
	\centering
	\subfloat{\includegraphics[width=2.4in]{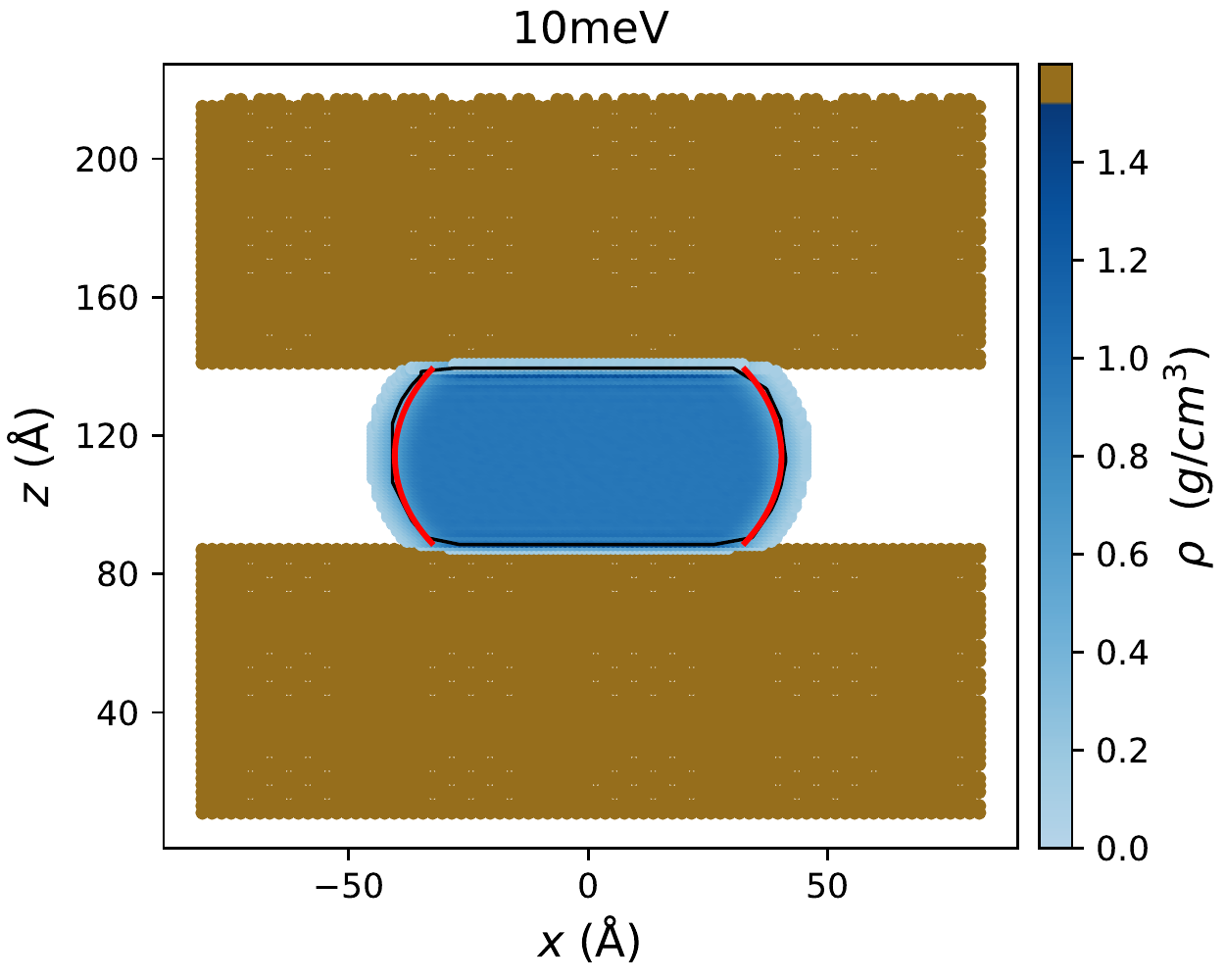}} 
	\subfloat{\includegraphics[width=2.4in]{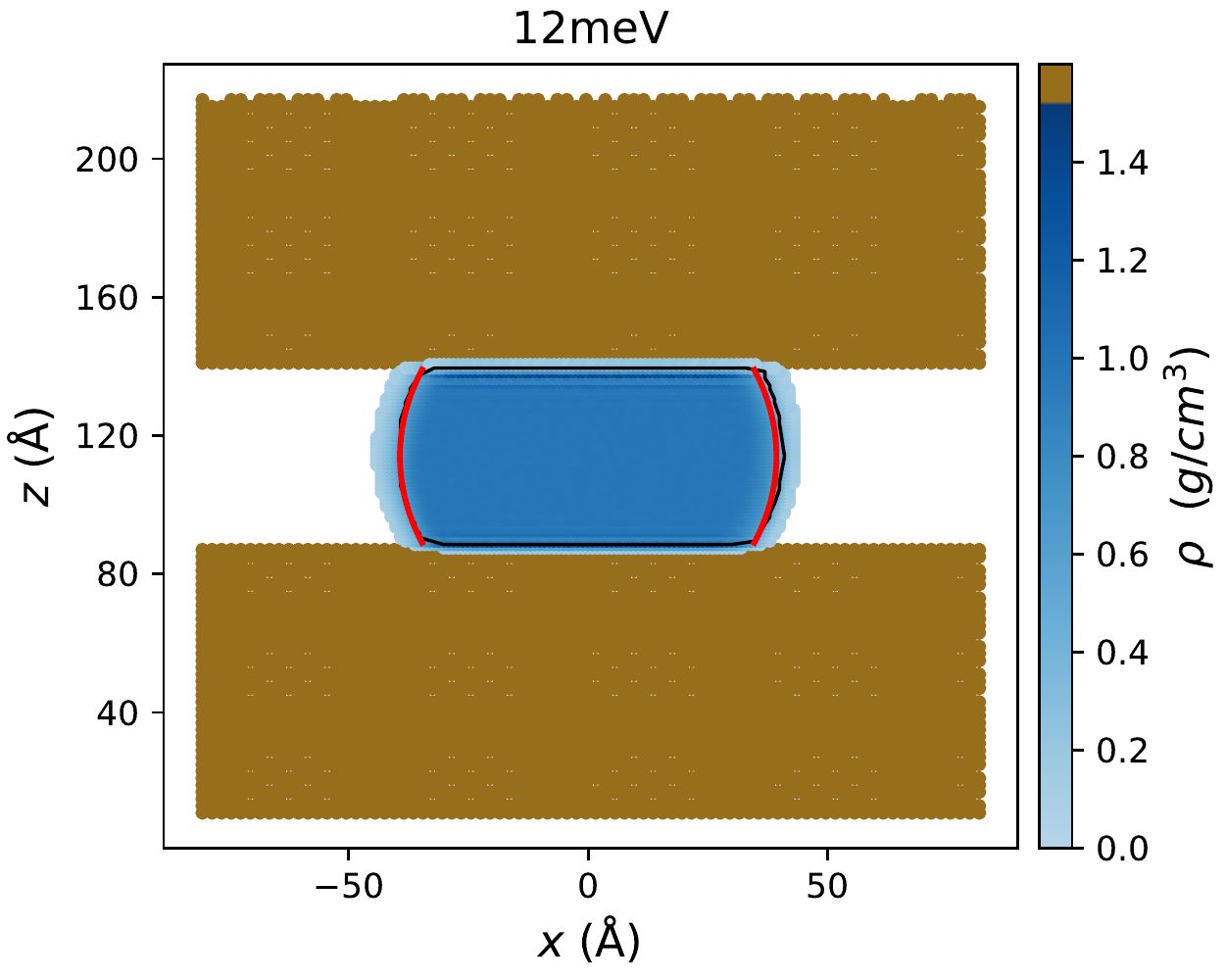}}\\
	\subfloat{\includegraphics[width = 2.4in]{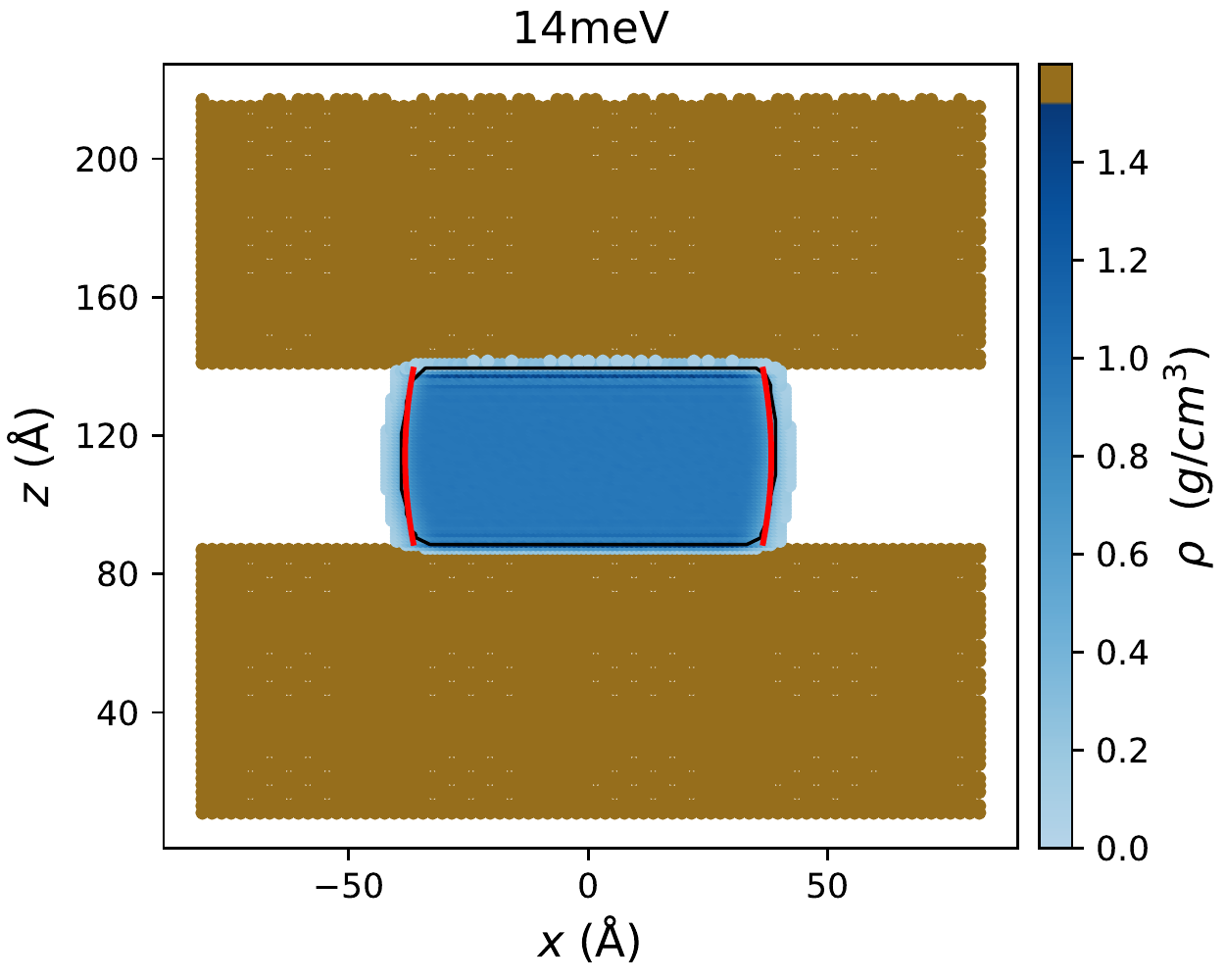}} 
	\subfloat{\includegraphics[width = 2.4in]{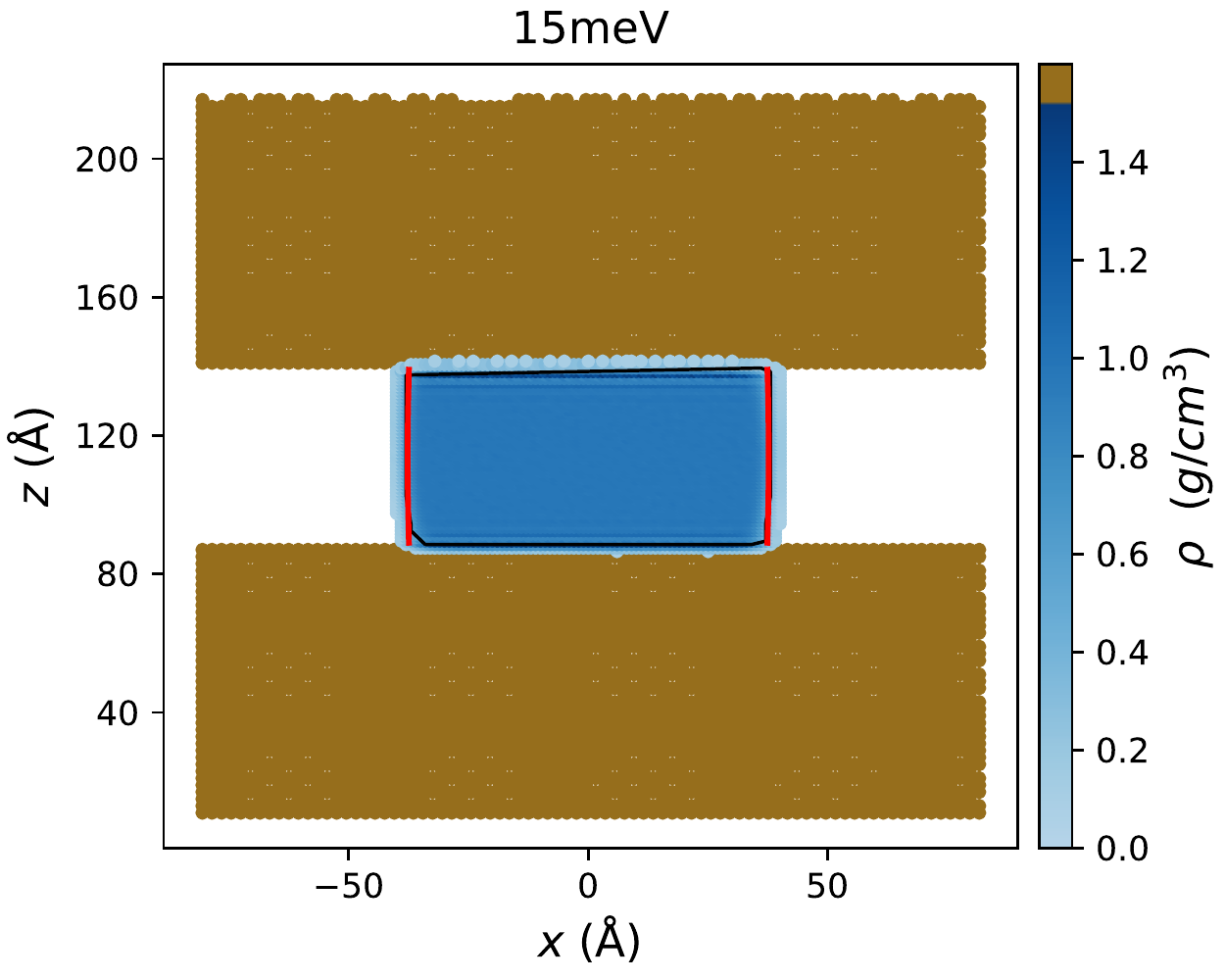}}\\
	\subfloat{\includegraphics[width = 2.4in]{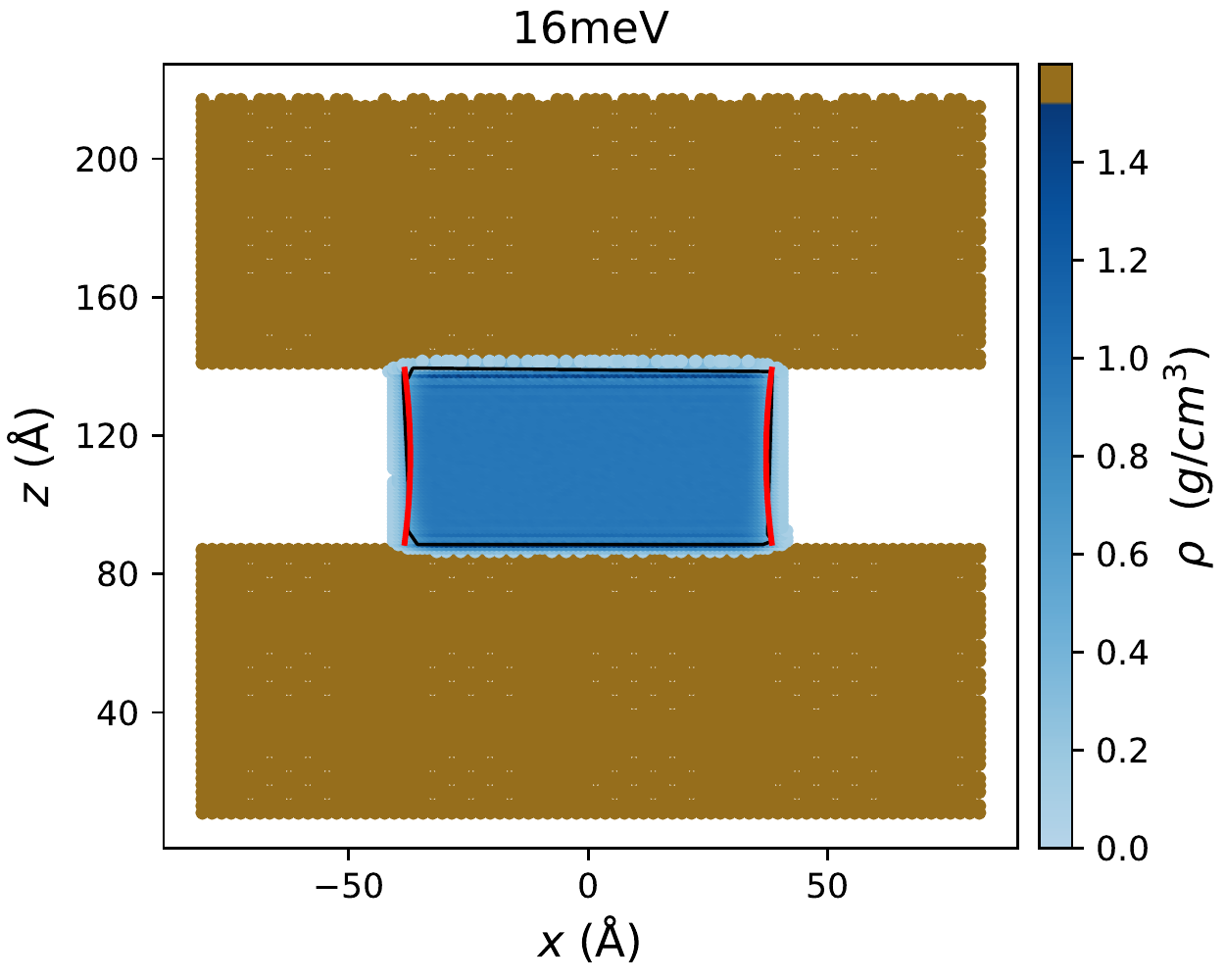}}
	\subfloat{\includegraphics[width = 2.4in]{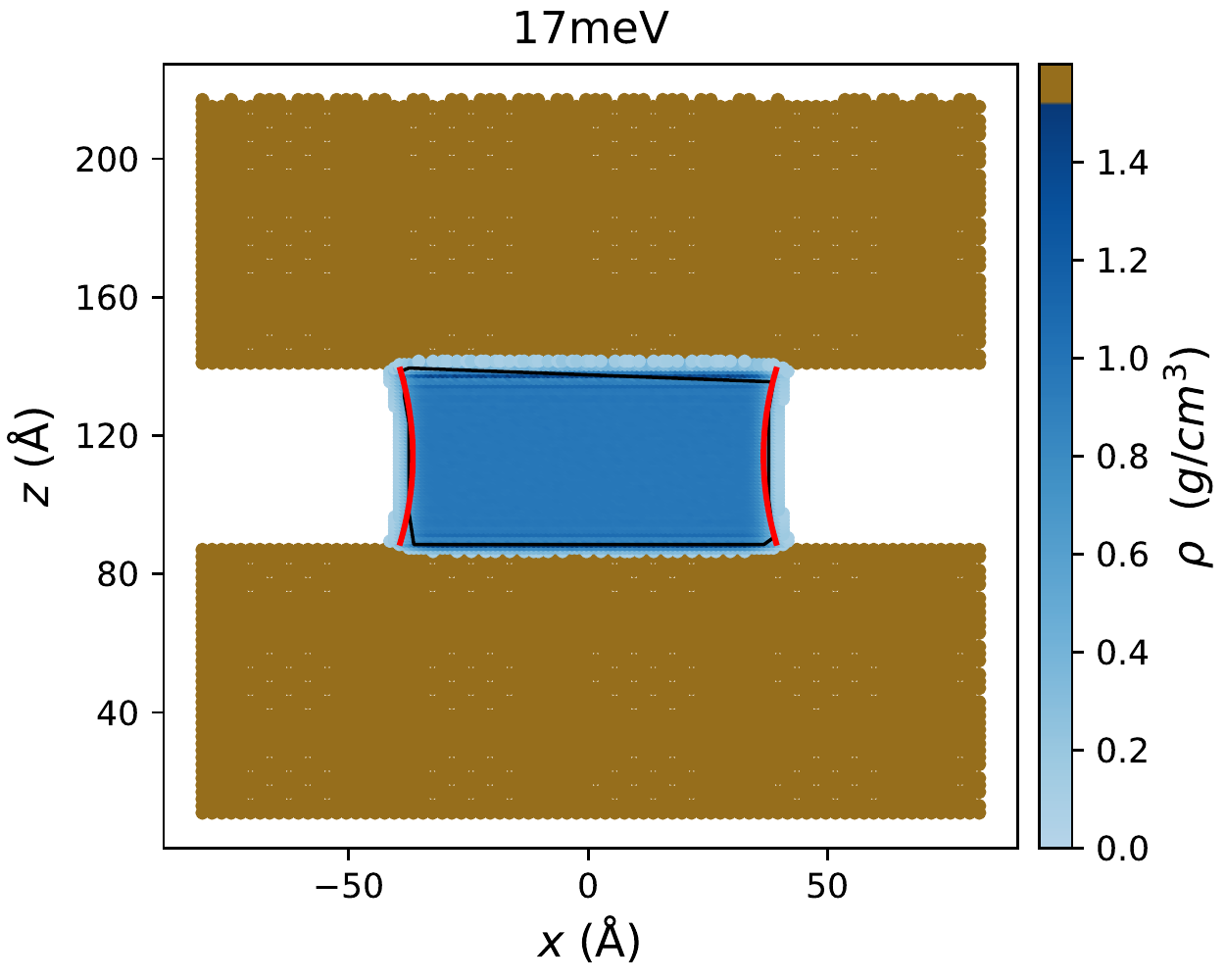}} \\
	\subfloat{\includegraphics[width = 2.4in]{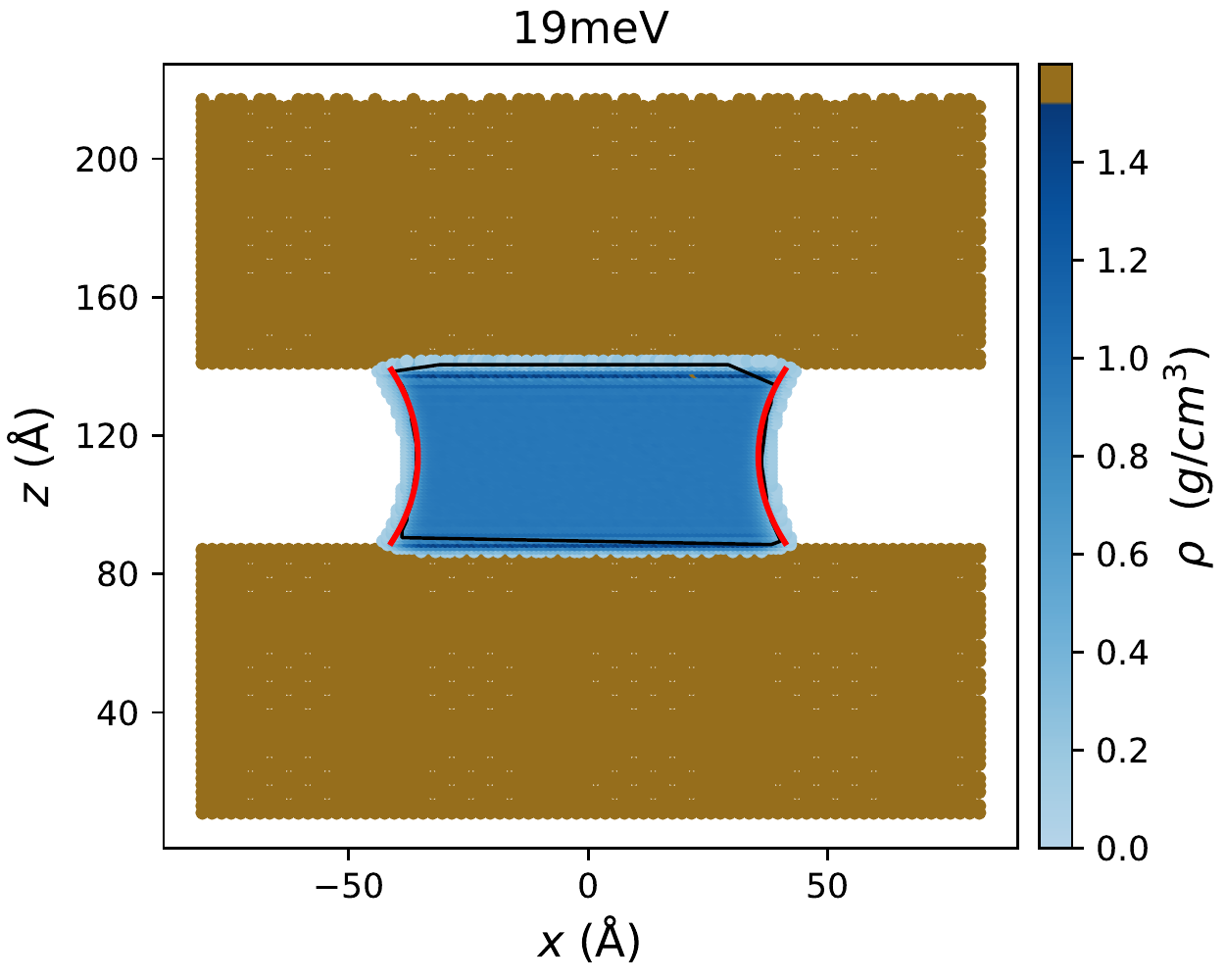}}
	\subfloat{\includegraphics[width = 2.4in]{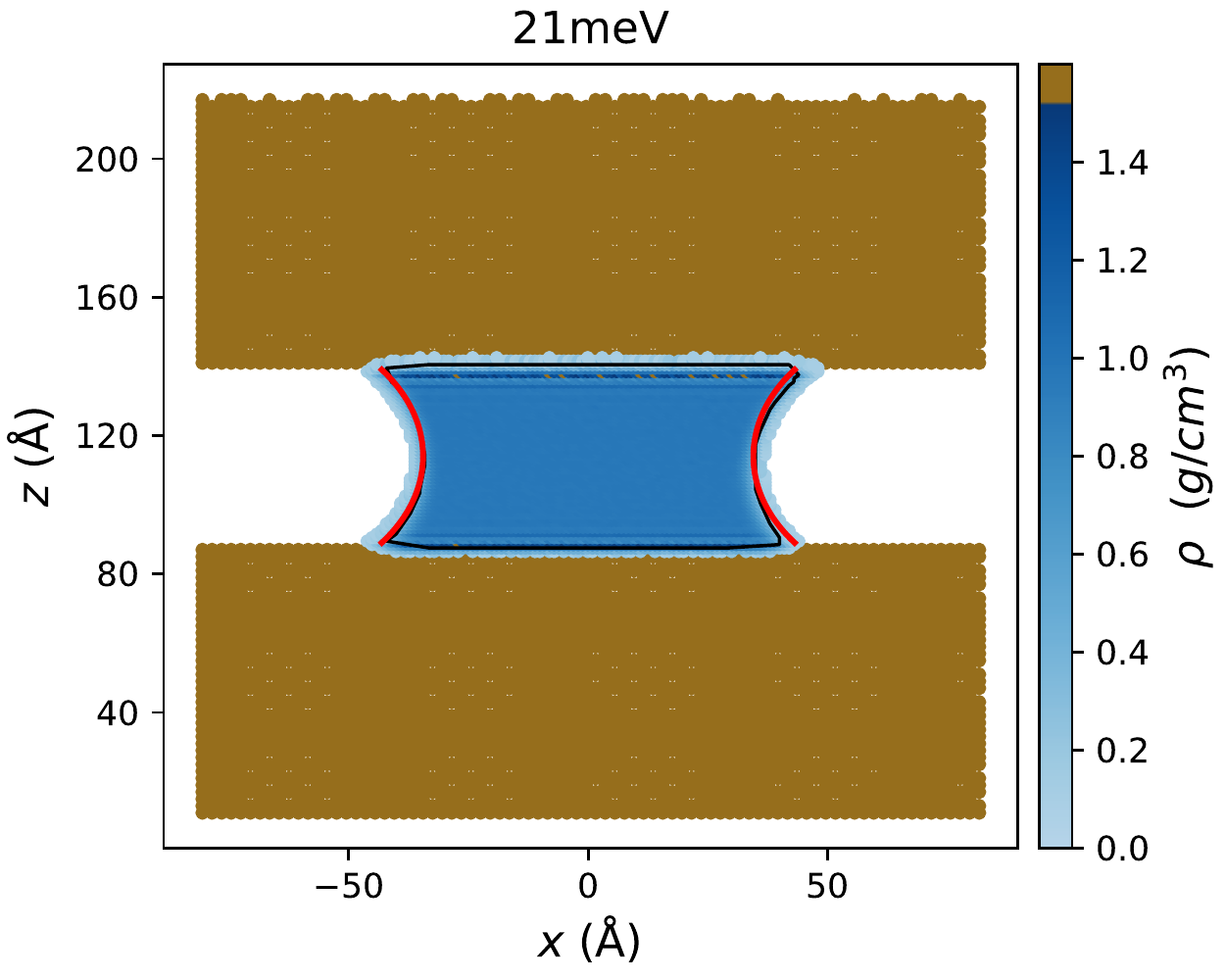}} \\
	\caption{\label{fig:8_2} The density profiles of meniscus with size $14\text{a}_0$ for different parametrization, ($\varepsilon$ = 10 - 21 meV). Values of density are shown in colorbar. The black solid line connects the points where the density of meniscus is equal to the half of the density of confined water. Red lines are the analytically obtained borders of droplet.}
	
\end{figure*}

\nocite{*}

\end{document}